\documentclass[twocolumn,aps,prd,floatfix,preprintnumbers,a4paper,nofootinbib,superscriptaddress,10pt]{revtex4-1} % linenumbers

\usepackage{lipsum}
\usepackage{multirow}
\usepackage{import}
\usepackage[titletoc,title]{appendix}

\usepackage{graphicx}

\usepackage{amsmath}
\usepackage{amssymb}
\usepackage{amsfonts}
\usepackage{mathtools}
\usepackage{bm}

\usepackage[utf8]{inputenc}
\usepackage{newtxtext}

\usepackage[table]{xcolor}
\usepackage{url}
\usepackage[colorlinks, pdfborder={0 0 0}]{hyperref}
\definecolor{LinkColor}{rgb}{0.75, 0, 0}
\definecolor{CiteColor}{rgb}{0.75, 0, 0}
\definecolor{UrlColor}{rgb}{0, 0, 0.75}
\hypersetup{linkcolor=LinkColor}
\hypersetup{citecolor=CiteColor}
\hypersetup{urlcolor=UrlColor}

\usepackage{rotating}
\usepackage{booktabs}

\usepackage{dcolumn}
\newcolumntype{d}{D{.}{.}{-1}}
\newcolumntype{b}[1]{D{(}{(}{#1}}
\newcommand\mc[1]{\multicolumn{1}{l}{#1}}

\newcommand{\Cardiff}{School of Physics and Astronomy, Cardiff University, Queens Buildings, Cardiff, CF24 3AA, United Kingdom}
\newcommand{\MITKavli}{MIT-Kavli Institute for Astrophysics and Space Research and LIGO Laboratory, 77 Massachusetts Avenue, 37-664H, Cambridge, MA 02139, USA}
\newcommand{\Zurich}{Physik-Institut, Universit\"at Z\"urich, Winterthurerstrasse 190, 8057 Z\"urich, Switzerland}
\newcommand{\Lisbon}{CENTRA, Departamento de F\`isica, Instituto Superior T\'ecnico IST, Universidade de Lisboa UL, Avenida Rovisco Pais 1, 1049-001 Lisboa, Portugal}
\newcommand{\Amsterdam}{Institute for High-Energy Physics, University of Amsterdam, Science Park 904, 1098 XH Amsterdam, The Netherlands}
\newcommand{\KCL}{King's  College  London,  Strand,  London  WC2R  2LS,  United Kingdom}
\newcommand{\Nikhef}{Nikhef, Science Park 105, 1098 XG Amsterdam, The Netherlands}
\newcommand{\GRASP}{Institute for Gravitational and Subatomic Physics (GRASP), 
Utrecht University, Princetonplein 1, 3584 CC Utrecht, The Netherlands}
\newcommand{\ICL}{Imperial College London, South Kensington Campus, London SW7 2AZ, United Kingdom}
\newcommand{\Portsmouth}{University of Portsmouth, Portsmouth, PO1 3FX, United Kingdom}
\newcommand{\Sinica}{Institute of Physics, Academia Sinica, Taipei 115201, Taiwan}
\newcommand{\Caltech}{Theoretical Astrophysics Group, California Institute of Technology, Pasadena, CA 91125, USA}

\usepackage[acronym,shortcuts]{glossaries}

\newacronym{bbh}{BBH}{binary black hole}
\newacronym{rit}{RIT}{Rochester Institute of Technology}
\newacronym{spec}{SpEC}{Spectral Einstein Code}

\begin{document}

% ***************************************** %
\def\gw#1{gravitational wave#1}
% ***************************************** %
\def\nr#1{numerical relativity
 (NR)#1\gdef\nr{NR}}
 % ***************************************** %
\def\bh#1{black-hole
 (BH)#1\gdef\bh{BH}}
 % ***************************************** %
 \def\bbh#1{binary black hole#1
  (BBH#1)\gdef\bbh{BBH}}
% ***************************************** %
 \def\qnm#1{Quasinormal Mode
    (QNM)#1\gdef\qnm{QNM}}
% ***************************************** %
\def\oed#1{optimal emission direction#1}
% ***************************************** %
\def\pn#1{post-Newtonian (PN)#1\gdef\pn{PN}}
% ***************************************** %
\def\imr#1{inspiral-merger-ringdown (IMR)#1\gdef\imr{IMR}}
% ***************************************** %
   \def\eob#1{effective-one-body
      (EOB)#1\gdef\eob{EOB}}
% ***************************************** % 

\title{A catalogue of precessing black-hole-binary numerical-relativity simulations}

% #9 - Author list

\author{Eleanor Hamilton}
\affiliation{\Cardiff}
\affiliation{\Zurich}
\author{Edward Fauchon-Jones}
\affiliation{\Cardiff}
\affiliation{\ICL}
\author{Mark Hannam}
\affiliation{\Cardiff}
\author{Charlie Hoy}
\affiliation{\Cardiff}
\affiliation{\Portsmouth}
\author{Chinmay Kalaghatgi}
\affiliation{\Cardiff}
\affiliation{\Nikhef}
\affiliation{\GRASP}
\affiliation{\Amsterdam}
\author{Lionel London}
\affiliation{\Cardiff}
\affiliation{\Amsterdam}
\affiliation{\KCL}
\affiliation{\MITKavli}
\author{Jonathan E. Thompson}
\affiliation{\Cardiff}
\affiliation{\Caltech}
\author{Dave Yeeles}
\affiliation{\Cardiff}
\affiliation{\Sinica}
\author{Shrobana Ghosh}
\affiliation{\Cardiff}
\author{Sebastian Khan}
\affiliation{\Cardiff}
\author{Panagiota Kolitsidou}
\affiliation{\Cardiff}
\author{Alex Vano-Vinuales}
\affiliation{\Cardiff}
\affiliation{\Lisbon}

\begin{abstract}
We present a public catalogue of numerical-relativity binary-black-hole simulations. The catalogue 
contains datasets from 80 distinct configurations of precessing binary-black-hole systems, with mass ratios up to $m_2/m_1 = 8$, dimensionless spin
magnitudes on the larger black hole up to $|\vec{S}_2|/m_2^2 = 0.8$ (the small black hole is non-spinning), and a range of five values of spin 
misalignment for each mass-ratio/spin combination. We discuss the physical properties of the configurations in our catalogue, and assess the 
accuracy of the initial configuration of each simulation and of the gravitational waveforms. We perform a careful analysis of the errors due to the 
finite resolution of our simulations and the finite distance from the source at which we extract the waveform data and provide a conservative 
estimate of the mismatch accuracy. We find that the upper limit on the mismatch uncertainty of our waveforms is $0.4\%$. In doing this
we present a consistent approach to combining mismatch uncertainties from multiple error sources. 
We compare this release to previous catalogues and discuss how these new simulations complement the existing public datasets. In particular, 
this is the first catalogue to uniformly cover this parameter space of single-spin binaries and 
there was previously only sparse coverage of the precessing-binary parameter space for mass ratios $\gtrsim 5$. 
We discuss applications of these new data, and the most urgent directions for future simulation work.
The public dataset can be accessed online at \texttt{https://data.cardiffgravity.org/bam-catalogue/}.

\end{abstract}

\date{\today}

\maketitle

\section{Introduction}
\label{sec:introduction}

After several decades of research to solve the binary-black-hole (BBH) problem, the first \nr{} BBH simulations through one orbit, merger and ringdown 
were produced in 2005~\cite{Pretorius:2005gq,Campanelli:2005dd,Baker:2005vv}. Since then many independent numerical relativity 
codes~\cite{Brugmann:2008zz, Husa:2007hp, Scheel:2006gg, Hemberger:2012jz, Herrmann:2007cwl, Zlochower:2005bj, Sperhake:2006cy, Loffler:2011ay} have been developed to simulate BBH systems for many orbits and added support for more complex configurations such as 
extremely high mass ratios and highly spinning black holes~\cite{Lousto:2010ut, Sperhake:2011ik, Scheel:2014ina}. 

The data products from \nr{} have been crucial for the field of gravitational-wave astronomy, including as input to develop approximate \gw{} models~\cite{Husa:2015iqa, Khan:2015jqa, Pratten:2020fqn, Garcia-Quiros:2020qpx, Estelles:2020twz, Hamilton:2021pkf, Buonanno:2009qa, Taracchini:2013rva, Pan:2013rra, Bohe:2016gbl, Babak:2016tgq, Cotesta:2018fcv, Blackman:2017dfb, Blackman:2017pcm, Varma:2019csw}, to calculate remnant properties of binary mergers~\cite{Gonzalez:2006md, Gonzalez:2007hi, Herrmann:2007ex, Campanelli:2007cga, Campanelli:2007ew, Kesden:2008ga, Lousto:2012gt, Healy:2014yta, Healy:2016lce, Healy:2018swt}, and used directly for \gw{} injection studies~\cite{Aylott:2009ya, LIGOScientific:2014oec}.
These \nr{}-dependent tools have played a central role in the direct detection of gravitational waves and the measurement of their source properties over the last six years~\cite{LIGOScientific:2018mvr, LIGOScientific:2020ibl, LIGOScientific:2021djp}.

Gravitational wave observations during the first three LIGO-Virgo-Kagra (LVK) observing runs~\cite{LIGOScientific:2016aoc, LIGOScientific:2016dsl, LIGOScientific:2016sjg, LIGOScientific:2017bnn, LIGOScientific:2017ycc, LIGOScientific:2017vox, LIGOScientific:2018mvr, LIGOScientific:2020aai, LIGOScientific:2020stg, LIGOScientific:2020zkf, LIGOScientific:2020ibl, LIGOScientific:2021qlt, LIGOScientific:2021usb, LIGOScientific:2021djp} have relied on theoretical models from three families; Phenom, SEOBNR and NRSurrogate. 
The Phenom and SEOBNR families use \nr{} waveforms to inform the merger-ringdown part of the model by calibrating a theoretically-motivated ansatz to the numerical data; the NRSurrogate models are constructed entirely from NR input. 
The simplest \gw{} models calibrated to \nr{} data are aligned-spin models~\cite{Husa:2015iqa, Khan:2015jqa, Taracchini:2013rva}, which 
capture the most important features of the waveform but require \nr{} waveforms that cover only a three-dimensional parameter space.
More recent aligned spin models have benefitted from further calibration to expanded data sets~\cite{Pratten:2020fqn, Garcia-Quiros:2020qpx, Estelles:2020twz, Bohe:2016gbl, Cotesta:2018fcv}. Many other subsequent models have been based on these aligned-spin models, thus indirectly benefitting from calibration to \nr{} data~\cite{London:2017bcn, PhysRevD.100.024059, Khan:2019kot, Thompson:2020nei, Pratten:2020ceb, Estelles:2021gvs, Ossokine:2020kjp, Matas:2020wab}. Recent expansions of \nr{} catalogues to cover the precessing parameter space have enabled the construction of the first generic-spin models calibrated entirely to \nr{} data~\cite{Varma:2019csw}.

\nr{} has been useful beyond modelling. \nr{} waveform injections have been used in several studies, including to assess the presence of systematic bias in waveform models~\cite{Abbott:2016wiq, Purrer:2019jcp}, and to estimate intermediate mass black hole binary merger rates~\cite{Salemi:2019ovz}. \nr{} waveforms have also been used for direct comparisons and parameter estimation of \gw{} observations~\cite{Abbott:2016apu, Lange:2017wki}. Further, \nr{} data can be used in the construction of fits to predict the remnant properties of a BBH merger, namely the final mass and spin as well as the gravitational recoil~\cite{Rezzolla:2007rz, Rezzolla:2007rd, Tichy:2007hk, Barausse:2009uz, Barausse:2012qz, Lousto:2012su, Jimenez-Forteza:2016oae, Varma:2018aht, Zappa:2019ntl}. 
These fits have a number of applications, such as tests of general relativity~\cite{Ghosh:2017gfp, LIGOScientific:2016lio, LIGOScientific:2020tif, LIGOScientific:2021sio}. 

Several large catalogues of BBH \nr{} simulations exist~\cite{Mroue:2013xna, Boyle:2019kee, Healy:2017psd, Healy:2019jyf, Healy:2020vre, Healy:2022wdn, Jani:2016wkt}. 
A quasi-circular BBH is described by 8 intrinsic parameters; the masses of each of the black holes $m_1$ and $m_2$ and their respective spins $\mathbf{S}_1$ and $\mathbf{S}_2$. The total mass $M=m_1+m_2$ sets the overall frequency scale and can be factored out. We therefore choose to set $M=1$. The dimensionless spin is defined as $\vec{\chi}_i=\mathbf{S}_i/m_i^2$.
The majority of simulations contained within these catalogues cover the precessing parameter space up to mass ratio 
$q=m_2/m_1=8$ and dimensionless spin magnitudes $\chi \leqslant 0.8$. However, the existence of simulations beyond $q=4$ of sufficient length and accuracy to be useful in the construction of gravitational wave models is fairly sparse. 
There therefore exists no broad systematic covering of the precessing parameter space up to $q=8$ with \nr{} simulations. One purpose of the current
catalogue is to provide a systematic covering of that parameter space. 

The primary objective of this catalogue was to support the development of a new precessing phenomenological model that is calibrated to numerical relativity waveforms~\cite{Hamilton:2021pkf}. 
Experience with producing previous phenomenological models suggests that we do not require an extremely dense sampling of the parameter space to produce a reasonably accurate model~\cite{Husa:2015iqa, Khan:2015jqa}. For the first catalogue used to inform the first precessing Phenom model we therefore chose no more than five points in each parameter direction. This choice was found to be sufficient: the {\tt PhenomPNR} model of the dominant
contribution to the signal (the $(2,2)$ multipoles in the co-precessing frame)~\cite{Hamilton:2021pkf}, constructed from 19 aligned-spin waveforms and 
40 precessing-binary waveforms, is of comparable accuracy to the equivalent contributions to the NRSurrogate model~\cite{Varma:2019csw}, which was constructed from 
more than 1000 simulations over a smaller volume of parameter space.
(It remains to be seen how many NR simulations are required to accurately include two-spin effects, higher multipoles, and mode asymmetries.) 
Secondary objectives were to contribute data that is useful to the waveform modelling community and to provide processed datasets that are 
appropriate for parameter estimation studies \cite{Abbott:2016wiq,Salemi:2019ovz}.

Our catalogue contains datasets from 80 different configurations of precessing BBH systems. These configurations cover four mass ratios $q = m_2/m_1 \in \{1, 2, 4, 8\}$ at four different spin magnitudes $\chi_2 = |\vec{S}_2|/m_2^2 \in \{0.2, 0.4, 0.6, 0.8\}$ each at five different spin vectors such that the angle between the orbital angular momentum and spin vector of the larger black hole is one of $\{30^{\circ}, 60^{\circ}, 90^{\circ}, 120^{\circ}, 150^{\circ}\}$.
The configurations are specified at a reference orbital frequency.
The catalogue can be accessed online at \texttt{https://data.cardiffgravity.org/bam-catalogue/}.

In the following section we briefly summarise the methods used by the BAM code \cite{Brugmann:2008zz,Husa:2007hp} to perform numerical simulations of BBH systems and describe the workflow we use to produce low eccentricity initial data. In Sec.~\ref{sec: properties} we provide a description of the properties of the simulations contained within the catalogue. In Sec.~\ref{sec: four waveforms} we perform a waveform accuracy analysis to validate the catalogue. 
Finally we conclude with Sec.~\ref{sec:comparison} where we discuss what regions of parameter space and how the catalogue can be used to contribute to the continuing advance of gravitational wave data analysis.

\section{Summary of Methods}	
\label{sec:methods}

\subsection{Simulation method}

The simulations in this catalogue were produced using BAM \cite{Brugmann:2008zz, Husa:2007hp}, a moving-box-based mesh-refinement 
numerical-relativity code that solves the 
$3+1$ decomposed Einstein equations.  Specifically, for the simulations in this catalogue we evolve Bowen-York 
wormhole data~\cite{PhysRevD.21.2047,Brandt:1997tf,PhysRevD.70.064011} via the $\chi$ variant of the moving-puncture treatment~\cite{Campanelli:2005dd,Baker:2005vv} of
the BSSN formulation~\cite{Baumgarte:1998te,Shibata:1995we}.  
Spatial derivatives are approximated algebraically through sixth-order finite differencing in the bulk, which in turn 
are evolved in time through fourth-order Runge-Kutta time stepping; see Refs.~\cite{Brugmann:2008zz, Husa:2007hp} for full details on the 
treatment of boundaries, buffer zones, advection derivatives, and numerical dissipation. Finally, the gravitational wave content of the system is 
extracted at some finite distance using the 
Newman-Penrose scalar $\Psi_4$~\cite{Newman:1961qr} following the procedure outlined in Ref.~\cite{Baker:2001sf}.
Following these references, $\Psi_4$ is decomposed via the spherical multipolar decomposition
\begin{align} 
   \Psi_4\left(\Theta,\Phi\right) = {}& \sum_{\ell=2}^{\infty} \sum_{m=\ell}^{\ell}
   \psi_{4,\ell m} \, ^{-2}Y_{\ell m}\left(\Theta,\Phi\right),
\end{align}
where $\psi_{4,\ell m}$ are the individual multipole moments, $^{-2}Y_{\ell m}$ are the spin-weighted spherical harmonics over the unit-sphere defined by $\Theta$ and $\Phi$.
The individual multipole moments can be written
\begin{align} 
   \psi_{4,\ell m} = {}& A_{\ell m} e^{-im\phi_{\ell m}},
\end{align}
where $A_{\ell m}$ and $\phi_{\ell m}$ are the amplitude and phase of the multipoles respectively.
The quantity $\Psi_4$ can be converted into the \gw{} observable known as the strain $h$ via a double time integral. This is generally the more useful quantity to consider for applications of the numerical data, such as waveform modelling, since it is the quantity measured by \gw{} detectors and so forms the starting point for all \gw{} astronomy. Where the strain is required, we obtain it in the frequency domain by dividing the frequency domain $\Psi_4$ data by $\omega^2$, where $\omega$ is the angular Fourier frequency~\cite{Reisswig:2010di}.

The numerical domain consists of nested Cartesian grids of successively finer spacing, nested in the sense that the grid at each level $n$ is 
encompassed by 
that of level $n-1$. The grid spacing $d_n$ for each refinement level $n$ follows the scaling,
	\begin{equation}
	d_n = \tfrac{d_0}{2^n} \qquad \{n \in \mathbb{Z}^+\} ,
	\end{equation}
where $d_0$ is the spacing on the coarsest level. The coarsest levels (largest boxes) encompass both black holes and are fixed, 
while for the finer levels (smaller boxes) there is a box around each black hole, and these boxes move with the punctures. The boxes are initially 
specified as cubes, where the user provides the number of points along one side for each level, i.e., if $N_n$ is the number of points in each direction
on level $n$, then the user specifies a list of $N_i = \{N_0, N_1, ..., N_{\rm l_{\rm max}}\}$, where $l_{\rm max}$ is the finest level.
During evolution the code will dynamically adjust the number of points, in particular near merger when individual boxes around each puncture 
will be merged when they are about to overlap. In addition, $l_{\rm max}$ differs for each puncture so that $d_{{\rm lmax},i}/m_i$ is
approximately the same for each puncture. See Ref.~\cite{Brugmann:2008zz} for more details on the {\tt BAM} grid structure, and Ref.~\cite{Purrer:2012wy}
for typical choices for numbers of levels and relative box sizes. Sec.~\ref{subsec:grids} provides further details of the choices we made for the grid configurations. 
The values $N_i$, $l_{\rm max}$ for each puncture, and the coarsest grid spacing $d_0$ are all provided in the public data release. 

The temporal resolution is subject to a Berger-Oliger refinement scheme in which the spacing between successive time steps halves with 
each successive level.  
The finest level consists of two grids, one centred on each puncture, an arrangement that is maintained as we move up the levels so long as the grids 
are not so large that they would overlap. These nested grids around each individual puncture move with the punctures as they orbit. Beyond this level 
the two grids are replaced by a single grid that encompasses all of the moving boxes and is centered on the origin.
It was found in Ref.~\cite{Brugmann:2008zz} that the Berger-Oliger 
timestepping becomes unstable on the coarsest non-moving grids, and for these we revert to a single time step specified by a Courant factor of 0.25 applied to the finest
such grid. The details are the same as those used in Refs.~\cite{Brugmann:2008zz, Husa:2007hp}, except for the use of 0.25 rather than 0.5 for the Courant factor, 
which is necessary to sufficiently reduce the timestepping error in long simulations. 

A pseudo-spectral elliptic solver is used to calculate binary wormhole initial data \cite{PhysRevD.70.064011}, with eccentricity reduced to $< 2 \times 10^{-3}$ 
through a series of manual iterations of the linear momenta of the punctures in the initial parameters. This process is described in more detail in Sec.~\ref{subsec:gen-initial-data}.

\subsection{Simulation workflow}

\subsubsection{Initial data construction}
\label{subsec:gen-initial-data}

We wish our simulations to begin 
at a user specified reference orbital frequency $M \Omega_\mathrm{orb}$ with spin vector $\bm{S} \equiv \bm{S}_2$ on the larger black hole 
(which we designate the secondary). The orientation of $\bm{S}_2$ can be defined by the angle 
$\theta_\text{LS} = \arccos(\hat{\bm{L}}_\mathrm{N} \cdot \hat{\bm{S}}_2)$ between the spin vector $\bm{S}_2$ and Newtonian orbital angular 
momentum vector $\bm{L}_\mathrm{N}$, and the angle $\phi_\mathrm{rS} = \arccos(\hat{\bm{r}} \cdot \hat{\bm{S}}_{2 \perp})$ 
between the projection of the spin vector on to the orbital plane $\bm{S}_{2 \perp}$ and the separation vector $\bm{r}$ from the larger black hole 
to the smaller. 
The positions and momenta of the black holes consistent with these constraints must then be determined at $M \Omega_\mathrm{orb}$, 
approximately chosen to 
minimise eccentricity. Bowen-York wormhole data can then be generated from these parameters. The main task of initial data construction is 
therefore reduced to 
identifying the appropriate black-hole positions and momenta at $M \Omega_\mathrm{orb}$. Two methods were used for the simulations 
in this catalogue.

For simulations with $\chi_2 \in \{0.4, 0.8\}$ the initial data parameters were determined by adapting the method used in previous 
work~\cite{Hannam:2010ec, Schmidt:2012rh, Husa:2015iqa}. For this method the physical parameters of the system $(q, \bm{S}_1, \bm{S}_2)$ are specified at a much larger 
separation $D_{\mathrm{start}}$ than the \nr{} simulations will start at. The \eob{} equations of motion are then evolved up to $M \Omega_\mathrm{orb}$ and the parameters at 
this frequency are used as input to a Bowen-York initial data solver. However for precessing systems this method does not allow the user to specify the exact system configuration 
$(q, \bm{S}_1, \bm{S}_2)$ at $M \Omega_\mathrm{orb}$. During the course of inspiral from $D_{\text{start}}$ to $M \Omega_\mathrm{orb}$ for the single-spin precessing systems 
in this catalogue, the angle $\theta_{\rm LS}$ can be seen to vary no more than $\sim 1^{\circ}$, while $\phi_\mathrm{rS}$ increases continuously (see Fig.~3 and Fig.~4 in~\cite{Schmidt:2014iyl}). 
To achieve a specific choice of $(\theta_{\rm LS}, \phi_\mathrm{rS})$ at a prescribed value of $M \Omega_\mathrm{orb}$, the method was extended with an iterative refinement of the angle 
$\phi_\mathrm{rS}$ at $D_{\mathrm{start}}$ until the parameters at $M \Omega_\mathrm{orb}$ are within a suitable tolerance of our desired values. Full details of this adapted method 
are given in Appendix~\ref{app:initial-data-method}.

The simulations  $\chi_2 \in \{0.2, 0.6\}$ were performed later, and were able to make use of a more recent method to produce low-eccentricity initial parameters, as
described in Ref.~\cite{Ramos-Buades:2018azo}. 
This method provides a post-Newtonian estimate of low-eccentricity parameters at a prescribed orbital frequency, making it possible to specify $(\theta_{\rm LS}, \phi_\mathrm{rS})$
without the need for any iterative steps. 
This method also supports additional iteration steps to further reduce the eccentricity based on \nr{} dynamics, however this additional iteration was not used for the simulations 
in this catalogue. We instead relied upon the manual perturbation approach outlined in Sec.~\ref{subsec:gen-initial-data} to reduce eccentricity when using either approach to obtain the 
initial data.

For all the \nr{} configurations described in this work the azimuthal angle for the spin vector $\mathbf{S}_2$ placed on the larger secondary component black hole was chosen to be $\phi = 0^{\circ}$ at $M \Omega_\mathrm{orb}$. 

While the initial data parameters generated in Sec.~\ref{subsec:gen-initial-data} will lead to low eccentricity simulations, in general this will not be low enough to satisfy our definition of a quasi-circular binary. We placed an upper limit on the eccentricity at $2\times10^{-3}$, based on the observation in Ref.~\cite{Purrer:2012wy} that the 
puncture dynamics do not give reliable eccentricity estimates below this value, due to gauge effects. A standard iterative method to further reduce eccentricity is to 
perform a low resolution simulation for $\sim$1000$M$, estimate the eccentricity,
and make iterative small perturbations to the momenta of the component black holes~\cite{Hannam:2010ec, Purrer:2012wy}. 
The method employed to estimate the eccentricity is described below.
For most of the simulations in this work a perturbation of 0.1-0.8\% is applied to the magnitude of the momenta. This is normally sufficient to reduce the eccentricity below the desired threshold. However in cases where this is not sufficient the radial component of the momenta is also reduced by 25-75\%. The eccentricity reduction procedure is performed using low resolution simulations in order to reduce both the computing resources and wall time required. Once initial data parameters are found that yield a sufficiently low eccentricity then a high resolution production simulation is performed using the same parameters. The higher resolution simulations tend to have higher eccentricity than the associated low resolution simulations. Consequently, a number of the simulations presented in this paper have eccentricities marginally above the $2\times10^{-3}$ threshold.

There are two different ways that eccentricity is estimated for the simulations in this work. For the shorter iterative eccentricity reduction simulations where the merger time is not known, the puncture separation $D$ is fit using a quadratic function with data typically in the range $[200, 700]M$ 
similar to the method described in Ref.~\cite{Husa:2007rh}. The eccentricity is then estimated by the maximum absolute relative difference between the fit and the data. For production simulations the eccentricity is estimated using a fit that also incorporates the merger time~\cite{Husa:2007rh}.

In our production simulations we in general find that true eccentricity differs from that calculated in our lower-resolution eccentricity-reduction simulations. 
For a few of the cases in this catalogue  the eccentricity of the lower resolution simulation was below our $2\times10^{-3}$ threshold, but the eccentricity of the production 
simulation exceeded it, as can be seen in Tab.~\ref{table:metadata}. Nonetheless, only a handful of cases have eccentricities above $3\times10^{-3}$, and only one
is close to $4\times10^{-3}$ (\texttt{CF\_8}).

\subsubsection{Grid configurations}
\label{subsec:grids}

The simulations performed for this catalogue are all computationally expensive, requiring $O(10^5)$ CPU hours for each production run, 
and we do not have the luxury of exhaustive experiments to identify a choice of 
numerical grids that provides a good balance between computational efficiency and physical accuracy. In 3D simulations of this scale 
it is impractical to perform standard convergence tests where the grid spacing $d_0$ is halved between successive runs, and indeed clean 
convergence has rarely been observed in binary simulations with any code, and even given promising convergence results for one binary 
configuration, there exists no robust algorithm to determine the resolution requirements to guarantee clean convergence for a second 
configuration. Sec.~\ref{sec: four waveforms} presents a convergence study of several of our configurations. 
In this section we discuss
the heuristic requirements we place on our grid configurations, based on past experience with BAM binary simulations. 

Our first requirement is that the width of the smallest moving box following each component black hole should be between 1.2 and 1.5 times the maximum effective coordinate 
diameter of the apparent horizon of its respective black hole before merger. This requirement is achieved by changing the values of the grid spacing $d_0$ on the coarsest level, 
and the finest level that exists for the larger black hole. The number of grid points $N_L$ on the finest level can also be used to adjust the size of the finest box around the black
hole, if necessary, but we find in most cases that adjusting $d_0$ is sufficient.

The second requirement is to have at least ten grid points per wavelength of the $(4,4)$ multipole moment 
on the level where \gw{}s are extracted. 
The maximum frequency is estimated by doubling the $(2,2)$ ringdown frequency calculated by the aligned spin \gw{} model 
\texttt{PhenomD}~\cite{Husa:2015iqa, Khan:2015jqa}, using the parameters $(q, 0, \chi_2)$. In precessing configurations the ringdown frequency will always be lower
than this estimate, and therefore this provides a conservative estimate of the resolution requirements.
One could use a more accurate estimate of the ringdown frequency for each precessing configuration using, for example, the method described in 
Refs.~\cite{Hannam:2013oca,Schmidt:2014iyl,London:2018nxs}, but for this work we found no need to do this. 
The required grid spacing on level $n$ where \gw{}s are extracted is then approximated as $d_n \leq 1/(20f_\mathrm{RD})$. 
This requirement is achieved by changing the values of the finest grid spacing $d_0$. If this requirement cannot be satisfied on level $n$ and level $n$ is not the last fixed box level, then the number of grid points $N_{n+1}$ on level $n+1$ is increased until the box size $N_{n+1} \times h_{n+1}$ is large enough to support \gw{} extraction at the radius required. The use of much larger numbers of points on the wave-extraction level means that the wave-extraction resolution requirements are also a strong determinant of the 
overall computational cost, along with the resolution requirements local to the black hole. 

For most configurations both of the requirements outlined in the preceding two paragraphs
 can be satisfied. However it is not always possible to satisfy both requirements, and for such cases the smallest box sizes and extraction level grid spacing are balanced to achieve the best possible result.

\section{Simulation properties}
\label{sec: properties}

In this section we discuss the properties of the simulations in our catalogue. We first motivate our coverage
of the single-spin parameter space, our choice of starting frequency for each binary, and our procedure to 
estimate initial black-hole momenta and spins to achieve quasi-circular inspiral with a prescribed spin orientation. 
We then discuss in detail the accuracy with which our desired configurations are achieved, in particular the
accuracy of our specification of the black-hole masses and spins, and the spin orientations. Finally, we summarise
the properties of the remnant black holes.

\subsection{Simulation configurations}

\begin{table*}[htbp]
    \begin{tabular*}{\textwidth}{lllb{6}b{7}*{8}{l}r}
\toprule[0.16em]
Name & $q$ & $\chi$ & \mc{$\theta_{\mathrm{LS}} \left(^\circ\right)$} & \mc{$\chi_{\mathrm{eff}}$} & $\chi_{\rm p}$ & $D/M$ & $e$ & $M\Omega_{\mathrm{orb}}$ & $t_{\rm M}$ & $N_{\mathrm{orb}}$ & $M_{f}$ & $\chi_{f}$ & $v_R$ \\
& & & & & & & \scriptsize{$\left(\times 10^{-3}\right)$} & \scriptsize{$\left(\times 10^{-2}\right)$} & & & & & \scriptsize{$(\mathrm{kms}^{-1})$} \\
\midrule
\texttt{CF\_1} & $1$ & 0.2 (0.200) & 30.0 \, (29.9) & 0.087 \, (0.087) & $0.100$ $(0.100)$ & $11.6$ $(11.4)$ & $1.327$ & $2.28$ & $1710$ & $9.61$ & $0.949$ & $0.713$ & $188$ \\
\texttt{CF\_2} & $1$ & $0.2$ $(0.200)$ & 60.0 \, (59.9) & 0.050 \, (0.050) & $0.173$ $(0.173)$ & $11.6$ $(11.4)$ & $0.931$ & $2.28$ & $1685$ & $9.43$ & $0.950$ & $0.703$ & $96$ \\
\texttt{CF\_3} & $1$ & 0.2 (0.200) & 90.0 \, (89.8) & 0.000 \, (0.000) & $0.200$ $(0.200)$ & $11.6$ $(11.4)$ & $1.084$ & $2.27$ & $1667$ & $9.25$ & $0.951$ & $0.688$ & $176$ \\
\texttt{CF\_4} & $1$ & 0.2 (0.200) & 120.0 \, (119.8) & -0.050 \, (-0.050) & $0.173$ $(0.173)$ & $11.6$ $(11.4)$ & $0.759$ & $2.28$ & $1628$ & $9.01$ & $0.953$ & $0.672$ & $329$ \\
\texttt{CF\_5} & $1$ & 0.2 (0.200) & 150.0 \, (149.9) & -0.087 \, (-0.086) & $0.100$ $(0.100)$ & $11.6$ $(11.4)$ & $0.723$ & $2.27$ & $1610$ & $8.86$ & $0.953$ & $0.661$ & $147$ \\
\texttt{CF\_6} & $1$ & 0.4 (0.400) & 30.0 \, (29.9) & 0.173 \, (0.174) & $0.200$ $(0.200)$ & $11.6$ $(11.3)$ & $1.227$ & $2.28$ & $1768$ & $10.0$ & $0.946$ & $0.740$ & $120$ \\
\texttt{CF\_7} & $1$ & 0.4 (0.400) & 60.0 \, (59.9) & 0.100 \, (0.101) & $0.346$ $(0.346)$ & $11.6$ $(11.3)$ & $2.478$ & $2.28$ & $1723$ & $9.69$ & $0.948$ & $0.720$ & $735$ \\
\texttt{CF\_8} & $1$ & 0.4 (0.400) & 90.0 \, (89.9) & 0.000 \, (0.001) & $0.400$ $(0.400)$ & $11.5$ $(11.3)$ & $3.898$ & $2.30$ & $1626$ & $9.11$ & $0.951$ & $0.691$ & $757$ \\
\texttt{CF\_9} & $1$ & $0.4$ $(0.400)$ & 120.0 \, (119.9) & -0.100 \, (-0.099) & $0.346$ $(0.347)$ & $11.6$ $(11.3)$ & $3.213$ & $2.29$ & $1577$ & $8.72$ & $0.954$ & $0.660$ & $93$ \\
\texttt{CF\_10} & $1$ & 0.4 (0.400) & 150.0 \, (149.9) & -0.173 \, (-0.173) & $0.200$ $(0.201)$ & $11.6$ $(11.4)$ & $2.675$ & $2.29$ & $1542$ & $8.45$ & $0.956$ & $0.634$ & $324$ \\
\texttt{CF\_11} & $1$ & 0.6 (0.600) & 30.0 \, (29.9) & 0.260 \, (0.260) & $0.300$ $(0.299)$ & $10.0$ $(9.79)$ & $1.269$ & $2.79$ & $1064$ & $7.34$ & $0.943$ & $0.767$ & $189$ \\
\texttt{CF\_12} & $1$ & 0.6 (0.601) & 60.0 \, (59.8) & 0.150 \, (0.151) & $0.520$ $(0.519)$ & $11.6$ $(11.3)$ & $1.446$ & $2.29$ & $1737$ & $9.81$ & $0.946$ & $0.740$ & $977$ \\
\texttt{CF\_13} & $1$ & 0.6 (0.600) & 90.1 \, (89.8) & 0.000 \, (0.001) & $0.600$ $(0.600)$ & $11.5$ $(11.3)$ & $1.261$ & $2.30$ & $1625$ & $9.11$ & $0.950$ & $0.697$ & $1170$ \\
\texttt{CF\_14} & $1$ & $0.6$ $(0.600)$ & 120.0 \, (119.8) & -0.150 \, (-0.149) & $0.519$ $(0.521)$ & $11.7$ $(11.4)$ & $1.200$ & $2.27$ & $1589$ & $8.65$ & $0.955$ & $0.651$ & $98$ \\
\texttt{CF\_15} & $1$ & 0.6 (0.600) & 150.0 \, (149.9) & -0.260 \, (-0.259) & $0.300$ $(0.301)$ & $11.6$ $(11.4)$ & $1.761$ & $2.29$ & $1480$ & $8.03$ & $0.958$ & $0.609$ & $159$ \\
\texttt{CF\_16} & $1$ & 0.8 (0.801) & 30.0 \, (29.9) & 0.346 \, (0.347) & $0.400$ $(0.399)$ & $11.6$ $(11.3)$ & $2.958$ & $2.28$ & $1902$ & $10.9$ & $0.939$ & $0.792$ & $874$ \\
\texttt{CF\_17} & $1$ & 0.8 (0.801) & 60.0 \, (59.8) & 0.200 \, (0.202) & $0.693$ $(0.692)$ & $11.6$ $(11.4)$ & $2.691$ & $2.27$ & $1832$ & $10.3$ & $0.943$ & $0.758$ & $1690$ \\
\texttt{CF\_18} & $1$ & 0.8 (0.801) & 90.1 \, (89.8) & 0.000 \, (0.002) & $0.800$ $(0.801)$ & $11.6$ $(11.3)$ & $1.027$ & $2.29$ & $1644$ & $9.17$ & $0.950$ & $0.707$ & $1220$ \\
\texttt{CF\_19} & $1$ & 0.8 (0.801) & 120.1 \, (119.8) & -0.200 \, (-0.198) & $0.692$ $(0.696)$ & $11.6$ $(11.3)$ & $1.402$ & $2.30$ & $1503$ & $8.26$ & $0.955$ & $0.641$ & $1110$ \\
\texttt{CF\_20} & $1$ & 0.8 (0.801) & 150.0 \, (149.9) & -0.347 \, (-0.346) & $0.399$ $(0.403)$ & $11.6$ $(11.3)$ & $0.552$ & $2.32$ & $1374$ & $7.48$ & $0.959$ & $0.584$ & $394$ \\
\midrule
\texttt{CF\_21} & $2$ & 0.2 (0.200) & 30.0 \, (29.9) & 0.115 \, (0.116) & $0.100$ $(0.100)$ & $11.6$ $(11.4)$ & $1.718$ & $2.27$ & $1934$ & $10.6$ & $0.958$ & $0.680$ & $114$ \\
\texttt{CF\_22} & $2$ & 0.2 (0.200) & 60.0 \, (59.9) & 0.067 \, (0.067) & $0.173$ $(0.173)$ & $11.6$ $(11.4)$ & $1.257$ & $2.28$ & $1887$ & $10.3$ & $0.959$ & $0.659$ & $374$ \\
\texttt{CF\_23} & $2$ & 0.2 (0.200) & 90.0 \, (89.8) & 0.000 \, (0.000) & $0.200$ $(0.200)$ & $11.6$ $(11.4)$ & $0.655$ & $2.28$ & $1823$ & $9.84$ & $0.961$ & $0.629$ & $207$ \\
\texttt{CF\_24} & $2$ & 0.2 (0.200) & 120.0 \, (119.8) & -0.067 \, (-0.066) & $0.173$ $(0.174)$ & $11.6$ $(11.4)$ & $1.159$ & $2.29$ & $1759$ & $9.42$ & $0.963$ & $0.596$ & $287$ \\
\texttt{CF\_25} & $2$ & 0.2 (0.200) & 150.0 \, (149.9) & -0.116 \, (-0.115) & $0.100$ $(0.100)$ & $11.6$ $(11.4)$ & $0.981$ & $2.29$ & $1713$ & $9.13$ & $0.964$ & $0.569$ & $227$ \\
\texttt{CF\_26} & $2$ & 0.4 (0.400) & 30.0 \, (29.9) & 0.231 \, (0.231) & $0.200$ $(0.200)$ & $11.6$ $(11.4)$ & $1.237$ & $2.29$ & $2005$ & $11.2$ & $0.954$ & $0.737$ & $209$ \\
\texttt{CF\_27} & $2$ & 0.4 (0.400) & 60.1 \, (59.9) & 0.133 \, (0.134) & $0.347$ $(0.346)$ & $11.6$ $(11.5)$ & $2.131$ & $2.26$ & $1979$ & $10.8$ & $0.956$ & $0.700$ & $713$ \\
\texttt{CF\_28} & $2$ & 0.4 (0.400) & 90.1 \, (89.9) & -0.001 \, (0.001) & $0.400$ $(0.400)$ & $11.6$ $(11.4)$ & $0.580$ & $2.28$ & $1838$ & $9.91$ & $0.961$ & $0.646$ & $169$ \\
\texttt{CF\_29} & $2$ & 0.4 (0.400) & 120.1 \, (119.8) & -0.134 \, (-0.133) & $0.346$ $(0.347)$ & $11.6$ $(11.4)$ & $2.226$ & $2.30$ & $1692$ & $9.02$ & $0.964$ & $0.577$ & $609$ \\
\texttt{CF\_30} & $2$ & 0.4 (0.400) & 150.1 \, (149.9) & -0.231 \, (-0.231) & $0.200$ $(0.201)$ & $11.6$ $(11.4)$ & $1.436$ & $2.29$ & $1626$ & $8.53$ & $0.966$ & $0.518$ & $270$ \\
\texttt{CF\_31} & $2$ & 0.6 (0.601) & 30.1 \, (29.9) & 0.346 \, (0.347) & $0.301$ $(0.300)$ & $11.5$ $(11.4)$ & $1.726$ & $2.28$ & $2118$ & $12.0$ & $0.948$ & $0.795$ & $154$ \\
\texttt{CF\_32} & $2$ & 0.6 (0.600) & 60.1 \, (59.8) & 0.199 \, (0.201) & $0.520$ $(0.519)$ & $11.5$ $(11.3)$ & $0.876$ & $2.30$ & $1950$ & $10.9$ & $0.952$ & $0.746$ & $1280$ \\
\texttt{CF\_33} & $2$ & 0.6 (0.601) & 90.1 \, (89.8) & -0.001 \, (0.002) & $0.600$ $(0.601)$ & $11.7$ $(11.5)$ & $0.681$ & $2.25$ & $1908$ & $10.2$ & $0.958$ & $0.669$ & $1270$ \\
\texttt{CF\_34} & $2$ & 0.6 (0.601) & 120.1 \, (119.8) & -0.201 \, (-0.199) & $0.519$ $(0.522)$ & $11.5$ $(11.3)$ & $1.531$ & $2.31$ & $1610$ & $8.57$ & $0.965$ & $0.571$ & $390$ \\
\texttt{CF\_35} & $2$ & 0.6 (0.600) & 150.1 \, (149.9) & -0.347 \, (-0.346) & $0.299$ $(0.301)$ & $11.9$ $(11.6)$ & $0.525$ & $2.24$ & $1657$ & $8.39$ & $0.968$ & $0.471$ & $395$ \\
\texttt{CF\_36} & $2$ & 0.8 (0.802) & 30.1 \, (29.9) & 0.461 \, (0.463) & $0.401$ $(0.399)$ & $11.5$ $(11.2)$ & $2.096$ & $2.31$ & $2156$ & $12.4$ & $0.940$ & $0.851$ & $866$ \\
\texttt{CF\_37} & $2$ & 0.8 (0.802) & 60.1 \, (59.7) & 0.265 \, (0.269) & $0.694$ $(0.692)$ & $11.5$ $(11.3)$ & $2.046$ & $2.30$ & $2014$ & $11.4$ & $0.948$ & $0.800$ & $903$ \\
\texttt{CF\_38} & $2$ & 0.8 (0.802) & 90.2 \, (89.7) & -0.002 \, (0.003) & $0.800$ $(0.802)$ & $11.6$ $(11.4)$ & $3.134$ & $2.28$ & $1844$ & $10.0$ & $0.958$ & $0.707$ & $720$ \\
\texttt{CF\_39} & $2$ & 0.8 (0.802) & 120.2 \, (119.7) & -0.268 \, (-0.264) & $0.691$ $(0.696)$ & $11.5$ $(11.3)$ & $0.869$ & $2.32$ & $1554$ & $8.29$ & $0.964$ & $0.571$ & $1180$ \\
\texttt{CF\_40} & $2$ & 0.8 (0.802) & 150.1 \, (149.9) & -0.462 \, (-0.462) & $0.398$ $(0.403)$ & $11.6$ $(11.3)$ & $0.801$ & $2.32$ & $1408$ & $7.26$ & $0.969$ & $0.431$ & $443$ \\
\bottomrule[0.16em]
\end{tabular*}

    \caption{Initial data parameters and relaxed properties of the precessing BBH configurations in this catalogue with mass ratio 1 or 2. The smaller black hole has no initial spin. The associated properties of the larger black hole are identified with a subscript $2$. The spin magnitude $S_2/m_2^2$, the spin tilt angle $\arccos(\hat{\bm{L}}_{\mathrm{N}} \cdot \hat{\bm{S}}_2)$, the effective spins $\chi_{\mathrm{eff}}$ and $\chi_p$ and the separation $D/M$ are derived from the initial conditions of the simulations and relaxed times given in brackets. The eccentricity $e$ is estimated over the region $[200, 1000]M$ using the method described in \cite{Husa:2007rh}. The orbital frequency $M\omega_\mathrm{orb}$ is derived from the dynamics at relaxed times. The number of orbits $N_{\mathrm{orb}}$ is from calculated from the relaxed time that $M\omega_\mathrm{orb}$ is reported at until the peak in the $(2,2)$ multipole moment of $\Psi_4$.
%    \mdh{Some table thoughts: (let's talk about these before making change!) We should talk about numbers of digits. For example, $M\Omega_{\rm orb}$ 
%    is specified with three significant digits in the text, but only 2 here. Maybe we use again the $10^{-3}$ thing, as with the eccentricity? And how many 
%    digits can we really believe in the dynamically measured quantities? Do we want the remnant quantities at the far right, since they are the last we 
%    discuss in the text? }
%\EH{Round recoil velocities to nearest whole number.}
    }
	\label{table:metadata}
\end{table*}
\begin{table*}[htbp]
    \begin{tabular*}{\textwidth}{lllb{6}b{7}*{8}{l}r}
\toprule[0.16em]
Name & $q$ & $\chi$ & \mc{$\theta_{\mathrm{LS}} \left(^\circ\right)$} & \mc{$\chi_{\mathrm{eff}}$} & $\chi_{\rm p}$ & $D/M$ & $e$ & $M\Omega_{\mathrm{orb}}$ & $t_{\rm M}$ & $N_{\mathrm{orb}}$ & $M_{f}$ & $\chi_{f}$ & $v_R$ \\
& & & & & & & \scriptsize{$\left(\times 10^{-3}\right)$} & \scriptsize{$\left(\times 10^{-2}\right)$} & & & & & \scriptsize{$(\mathrm{kms}^{-1})$} \\
\midrule
\texttt{CF\_41} & $4$ & 0.2 (0.200) & 30.0 \, (29.9) & 0.139 \, (0.139) & $0.100$ $(0.100)$ & $10.9$ $(10.9)$ & $1.083$ & $2.48$ & $1998$ & $11.4$ & $0.975$ & $0.566$ & $175$ \\
\texttt{CF\_42} & $4$ & 0.2 (0.200) & 60.0 \, (59.8) & 0.080 \, (0.080) & $0.173$ $(0.173)$ & $11.0$ $(10.9)$ & $2.708$ & $2.47$ & $1982$ & $11.1$ & $0.976$ & $0.536$ & $154$ \\
\texttt{CF\_43} & $4$ & 0.2 (0.200) & 90.1 \, (89.9) & 0.000 \, (0.000) & $0.200$ $(0.200)$ & $10.5$ $(10.5)$ & $1.386$ & $2.61$ & $1610$ & $9.34$ & $0.977$ & $0.488$ & $170$ \\
\texttt{CF\_44} & $4$ & 0.2 (0.200) & 120.0 \, (119.8) & -0.080 \, (-0.080) & $0.173$ $(0.173)$ & $11.2$ $(11.2)$ & $1.390$ & $2.38$ & $2004$ & $10.7$ & $0.978$ & $0.433$ & $241$ \\
\texttt{CF\_45} & $4$ & 0.2 (0.200) & 150.0 \, (149.9) & -0.139 \, (-0.138) & $0.100$ $(0.100)$ & $11.3$ $(11.3)$ & $0.701$ & $2.37$ & $1980$ & $10.4$ & $0.979$ & $0.385$ & $192$ \\
\texttt{CF\_46} & $4$ & 0.4 (0.400) & 30.1 \, (29.9) & 0.277 \, (0.277) & $0.201$ $(0.201)$ & $10.6$ $(10.6)$ & $1.501$ & $2.58$ & $1947$ & $11.7$ & $0.972$ & $0.664$ & $234$ \\
\texttt{CF\_47} & $4$ & 0.4 (0.400) & 60.2 \, (59.9) & 0.159 \, (0.160) & $0.347$ $(0.347)$ & $10.7$ $(10.7)$ & $1.424$ & $2.53$ & $1929$ & $11.2$ & $0.974$ & $0.616$ & $164$ \\
\texttt{CF\_48} & $4$ & 0.4 (0.400) & 90.2 \, (89.9) & -0.001 \, (0.000) & $0.400$ $(0.400)$ & $11.6$ $(11.5)$ & $1.608$ & $2.28$ & $2378$ & $12.4$ & $0.977$ & $0.533$ & $448$ \\
\texttt{CF\_49} & $4$ & 0.4 (0.400) & 120.2 \, (119.9) & -0.161 \, (-0.160) & $0.346$ $(0.346)$ & $11.6$ $(11.5)$ & $0.758$ & $2.30$ & $2148$ & $11.0$ & $0.980$ & $0.424$ & $336$ \\
\texttt{CF\_50} & $4$ & 0.4 (0.400) & 150.1 \, (149.9) & -0.277 \, (-0.277) & $0.199$ $(0.200)$ & $11.5$ $(11.5)$ & $1.761$ & $2.32$ & $1954$ & $9.87$ & $0.981$ & $0.313$ & $272$ \\
\texttt{CF\_51} & $4$ & 0.6 (0.600) & 30.1 \, (29.9) & 0.415 \, (0.415) & $0.301$ $(0.301)$ & $10.4$ $(10.3)$ & $1.644$ & $2.63$ & $1982$ & $12.5$ & $0.967$ & $0.762$ & $413$ \\
\texttt{CF\_52} & $4$ & 0.6 (0.601) & 60.2 \, (59.8) & 0.238 \, (0.239) & $0.521$ $(0.521)$ & $10.8$ $(10.7)$ & $1.817$ & $2.52$ & $2054$ & $12.1$ & $0.970$ & $0.704$ & $782$ \\
\texttt{CF\_53} & $4$ & 0.6 (0.602) & 90.3 \, (89.8) & -0.002 \, (-0.001) & $0.600$ $(0.602)$ & $11.0$ $(10.9)$ & $2.232$ & $2.46$ & $1935$ & $10.8$ & $0.975$ & $0.600$ & $764$ \\
\texttt{CF\_54} & $4$ & 0.6 (0.601) & 120.3 \, (119.8) & -0.242 \, (-0.241) & $0.518$ $(0.520)$ & $11.4$ $(11.4)$ & $0.438$ & $2.34$ & $1949$ & $10.0$ & $0.980$ & $0.451$ & $524$ \\
\texttt{CF\_55} & $4$ & 0.6 (0.600) & 150.2 \, (149.8) & -0.416 \, (-0.416) & $0.298$ $(0.300)$ & $10.1$ $(9.88)$ & $3.600$ & $2.88$ & $945$ & $5.45$ & $0.983$ & $0.270$ & $296$ \\
\texttt{CF\_56} & $4$ & 0.8 (0.801) & 30.2 \, (29.9) & 0.553 \, (0.554) & $0.402$ $(0.402)$ & $10.1$ $(10.0)$ & $1.638$ & $2.72$ & $1951$ & $13.1$ & $0.959$ & $0.859$ & $722$ \\
\texttt{CF\_57} & $4$ & 0.8 (0.802) & 60.4 \, (59.7) & 0.317 \, (0.318) & $0.695$ $(0.696)$ & $10.5$ $(10.4)$ & $0.751$ & $2.60$ & $1968$ & $12.2$ & $0.965$ & $0.800$ & $1150$ \\
\texttt{CF\_58} & $4$ & 0.8 (0.802) & 90.4 \, (89.7) & -0.004 \, (-0.002) & $0.800$ $(0.802)$ & $11.5$ $(11.4)$ & $1.675$ & $2.31$ & $2342$ & $12.3$ & $0.973$ & $0.684$ & $1160$ \\
\texttt{CF\_59} & $4$ & 0.8 (0.802) & 120.4 \, (119.7) & -0.324 \, (-0.323) & $0.690$ $(0.693)$ & $11.6$ $(11.5)$ & $1.225$ & $2.30$ & $1984$ & $9.92$ & $0.979$ & $0.506$ & $710$ \\
\texttt{CF\_60} & $4$ & 0.8 (0.801) & 150.2 \, (149.8) & -0.555 \, (-0.556) & $0.397$ $(0.399)$ & $11.6$ $(11.5)$ & $0.565$ & $2.31$ & $1690$ & $8.18$ & $0.983$ & $0.271$ & $383$ \\
\midrule
\texttt{CF\_61} & $8$ & $0.2$ $(0.200)$ & 30.0 \, (29.9) & 0.154 \, (0.154) & $0.100$ $(0.100)$ & $9.84$ $(9.96)$ & $1.359$ & $2.86$ & $2006$ & $12.4$ & $0.988$ & $0.437$ & $81$ \\
\texttt{CF\_62} & $8$ & $0.2$ $(0.200)$ & 60.0 \, (59.8) & 0.089 \, (0.089) & $0.173$ $(0.173)$ & $9.95$ $(10.1)$ & $0.656$ & $2.82$ & $2008$ & $12.1$ & $0.988$ & $0.402$ & $87$ \\
\texttt{CF\_63} & $8$ & $0.2$ $(0.200)$ & 90.0 \, (89.7) & 0.000 \, (0.000) & $0.200$ $(0.200)$ & $10.1$ $(10.2)$ & $1.147$ & $2.77$ & $1989$ & $11.7$ & $0.989$ & $0.345$ & $98$ \\
\texttt{CF\_64} & $8$ & $0.2$ $(0.200)$ & 120.0 \, (119.8) & -0.089 \, (-0.089) & $0.173$ $(0.174)$ & $10.3$ $(10.4)$ & $0.739$ & $2.70$ & $2001$ & $11.3$ & $0.990$ & $0.273$ & $100$ \\
\texttt{CF\_65} & $8$ & $0.2$ $(0.200)$ & 150.0 \, (149.9) & -0.154 \, (-0.154) & $0.100$ $(0.100)$ & $10.4$ $(10.5)$ & $1.909$ & $2.67$ & $1987$ & $11.0$ & $0.990$ & $0.199$ & $100$ \\
\texttt{CF\_66} & $8$ & 0.4 (0.400) & 30.2 \, (29.9) & 0.307 \, (0.309) & $0.201$ $(0.199)$ & $9.54$ $(9.67)$ & $1.360$ & $2.96$ & $2022$ & $13.3$ & $0.986$ & $0.573$ & $118$ \\
\texttt{CF\_67} & $8$ & $0.4$ $(0.400)$ & 60.3 \, (59.9) & 0.176 \, (0.178) & $0.347$ $(0.346)$ & $9.69$ $(9.84)$ & $1.921$ & $2.90$ & $1960$ & $12.4$ & $0.987$ & $0.525$ & $83$ \\
\texttt{CF\_68} & $8$ & 0.4 (0.400) & 90.3 \, (89.8) & -0.002 \, (0.000) & $0.400$ $(0.400)$ & $10.1$ $(10.2)$ & $1.222$ & $2.76$ & $2007$ & $11.8$ & $0.988$ & $0.439$ & $193$ \\
\texttt{CF\_69} & $8$ & 0.4 (0.400) & 120.3 \, (119.8) & -0.179 \, (-0.178) & $0.345$ $(0.346)$ & $10.5$ $(10.6)$ & $1.402$ & $2.65$ & $2022$ & $11.1$ & $0.990$ & $0.318$ & $166$ \\
\texttt{CF\_70} & $8$ & 0.4 (0.400) & 150.1 \, (149.9) & -0.308 \, (-0.308) & $0.199$ $(0.200)$ & $10.8$ $(10.8)$ & $1.578$ & $2.56$ & $2040$ & $10.7$ & $0.991$ & $0.169$ & $103$ \\
\texttt{CF\_71} & $8$ & 0.6 (0.601) & 30.2 \, (29.9) & 0.461 \, (0.463) & $0.302$ $(0.299)$ & $9.28$ $(9.34)$ & $2.038$ & $3.07$ & $2040$ & $14.3$ & $0.983$ & $0.711$ & $161$ \\
\texttt{CF\_72} & $8$ & 0.6 (0.601) & 60.4 \, (59.8) & 0.263 \, (0.268) & $0.522$ $(0.519)$ & $9.60$ $(9.71)$ & $2.220$ & $2.92$ & $2046$ & $13.3$ & $0.985$ & $0.658$ & $330$ \\
\texttt{CF\_73} & $8$ & 0.6 (0.601) & 90.5 \, (89.8) & -0.004 \, (0.001) & $0.600$ $(0.601)$ & $10.0$ $(10.1)$ & $2.370$ & $2.78$ & $1990$ & $11.8$ & $0.988$ & $0.561$ & $320$ \\
\texttt{CF\_74} & $8$ & 0.6 (0.600) & 120.4 \, (119.7) & -0.270 \, (-0.271) & $0.518$ $(0.517)$ & $10.6$ $(10.7)$ & $1.100$ & $2.60$ & $2008$ & $10.8$ & $0.990$ & $0.420$ & $124$ \\
\texttt{CF\_75} & $8$ & 0.6 (0.601) & 150.2 \, (149.9) & -0.463 \, (-0.462) & $0.298$ $(0.301)$ & $10.6$ $(10.6)$ & $0.482$ & $2.64$ & $1637$ & $8.57$ & $0.991$ & $0.247$ & $116$ \\
\texttt{CF\_76} & $8$ & 0.8 (0.802) & 30.3 \, (29.8) & 0.614 \, (0.619) & $0.404$ $(0.397)$ & $8.96$ $(8.97)$ & $1.145$ & $3.20$ & $2044$ & $15.5$ & $0.977$ & $0.848$ & $210$ \\
\texttt{CF\_77} & $8$ & 0.8 (0.801) & 60.5 \, (59.7) & 0.350 \, (0.362) & $0.697$ $(0.690)$ & $9.37$ $(9.36)$ & $1.334$ & $3.04$ & $2013$ & $13.8$ & $0.982$ & $0.799$ & $193$ \\
\texttt{CF\_78} & $8$ & 0.8 (0.801) & 90.6 \, (89.6) & -0.008 \, (0.003) & $0.800$ $(0.801)$ & $10.0$ $(10.0)$ & $2.868$ & $2.80$ & $2006$ & $11.9$ & $0.987$ & $0.700$ & $363$ \\
\texttt{CF\_79} & $8$ & 0.8 (0.802) & 120.5 \, (119.7) & -0.361 \, (-0.359) & $0.689$ $(0.693)$ & $10.9$ $(10.9)$ & $2.327$ & $2.50$ & $2145$ & $11.1$ & $0.990$ & $0.548$ & $207$ \\
\texttt{CF\_80} & $8$ & 0.8 (0.802) & 150.3 \, (149.8) & -0.618 \, (-0.618) & $0.397$ $(0.398)$ & $11.5$ $(11.4)$ & $0.701$ & $2.36$ & $2169$ & $10.3$ & $0.991$ & $0.371$ & $178$ \\
\bottomrule[0.16em]
\end{tabular*}

    \caption{Initial data parameters and relaxed properties of the precessing BBH configurations in this catalogue with mass ratio 4 or 8. The smaller black hole has no initial spin. The associated properties of the larger black hole are identified with a subscript $2$. The spin magnitude $S_2/m_2^2$, the spin tilt angle $\arccos(\hat{\bm{L}}_{\mathrm{N}} \cdot \hat{\bm{S}}_2)$, the effective spins $\chi_{\mathrm{eff}}$ and $\chi_p$ and the separation $D/M$ are derived from the initial conditions of the simulations and relaxed times given in brackets. The eccentricity $e$ is estimated over the region $[200, 1000]M$ using the method described in \cite{Husa:2007rh}. The orbital frequency $M\omega_\mathrm{orb}$ is derived from the dynamics at relaxed times. The number of orbits $N_{\mathrm{orb}}$ is from calculated from the relaxed time that $M\omega_\mathrm{orb}$ is reported at until the peak in the $(2,2)$ multipole moment of $\Psi_4$.}
	\label{table:metadata-2}
\end{table*}

Our catalogue consists of dynamics and waveform data from \nr{} simulations of 80 binary-black-hole configurations. 
We chose configurations with four mass ratios  $q = m_2/m_1 \in \{1, 2, 4, 8\}$, four values of the dimensionless spin 
on the larger black hole, $\chi_2 = |\bm{S}_2|/m_2^2 \in \{0.2, 0.4, 0.6, 0.8\}$ (the smaller black hole has zero spin), and
five values of the misalignment of the black-hole spin with direction of the Newtonian orbital angular momentum,
$\theta_\text{LS} = \arccos(\hat{\bm{L}}_{\mathrm{N}} \cdot \hat{\bm{S}}_2) \in \{30^{\circ}, 60^{\circ}, 90^{\circ}, 120^{\circ},
150^{\circ}\}$. The configurations with spins $\chi_2 = \{0.4, 0.8\}$ were simulated first and used to produce the first
{\tt PhenomPNR} model, and the later simulations at $\chi_2 = \{0.2, 0.6\}$ were used as independent verification 
waveforms~\cite{Hamilton:2021pkf}. 

For each simulation two mass parameters $m_1, m_2$ were chosen such that $M = m_1 + m_2 = 1$. 
The initial data are iteratively constructed from these parameters such that the ADM mass of each puncture 
equals its respective mass parameter to within 0.02\%~\cite{Brugmann:2008zz}. At
subsequent times the masses of each black hole are recorded as the
apparent-horizon masses $M_{\mathrm{AH},1}, M_{\mathrm{AH},2}$ of each puncture,
which are related to the black hole hole masses $m_1, m_2$ through the
Christodoulou formula~\cite{Christodoulou:1970wf}. This approach agrees well with the ADM mass
of each puncture; the level of agreement is quantified further in Sec.~\ref{sec:masses}. 

In addition to the masses and spins, we must also choose the initial separation of the 
binary. For a binary undergoing non-eccentric inspiral there is a one-to-one correspondence between
the black-hole separation and the orbital frequency, so we may alternatively specify the initial orbital 
frequency, $M \Omega_{\rm orb}$. For this catalogue we prefer to choose $M\Omega_{\rm orb}$, because
our primary purpose is to construct a frequency-domain waveform model, and it would be convenient if
we were able to start to \nr{} tuning at the same frequency for each configuration. This also motivates the
iterative procedure described in Sec.~\ref{subsec:gen-initial-data}, with the goal of finding parameters 
consistent with non-eccentric inspiral for a configuration defined at a specified starting frequency. 
For the first simulations we performed, at mass ratios $q=1$ and $q=2$, we chose $M\Omega_{\rm orb} = 
0.0225$. This value was chosen to produce simulations of $\sim$2000$M$ in length, 
which we expected to be sufficiently accurate for our modelling purposes, based on the experience of
producing the aligned-spin simulations in Refs.~\cite{Husa:2015iqa,Khan:2015jqa}.

The duration of the simulations varies with binary mass ratio and the magnitude of the component of the spin aligned with 
the orbital angular momentum. At leading post-Newtonian order the merger time from a given starting frequency 
scales with $\Delta T \sim 1/\eta$, where $\eta = m_1 m_2 / M^2$ is the 
symmetric mass ratio. Therefore, if simulations at mass-ratios $q=1$ ($\eta = 0.25$) and $q=8$  
($\eta \approx 0.1$) start at the same orbital frequency, the $q=8$ simulation will take roughly 2.5 times
as long to merge. (This is a first-order approximation, and we see in the final results that the variation is not quite
so extreme.) In addition, if the black-hole spin is aligned with the orbital angular momentum, the
binary will inspiral more slowly, and this will also increase the time to merger. Conversely, a spin in the 
opposite direction to the orbital angular momentum will decrease the time to merger. This effect of spin is 
most easily seen in PN calculations, e.g., Refs.~\cite{Cutler:1994ys,Poisson:1995ef}. For example, for our 
$q=2$ configurations with $\chi_2 = 0.8$, where all simulations begin at $M\Omega = 0.0225$, we see
that the $\theta_{\rm LS} = 30^\circ$ configuration merges in $2254M$, while the $\theta_{\rm LS} = 150^\circ$
configuration merges in only $1505M$. Since we do not 
wish to perform expensive tests on the resolution requirements to achieve similar levels of accuracy for
much longer simulations, for mass ratios $q=4$ and $q=8$, we adjust the starting frequency to limit
the time to merger to approximately $2000M$. 

To meet the soft requirement of simulation merger by $2000M$ we estimate the merger time
 using the LALSimulation~\cite{lalsuite} implementation of
\texttt{PhenomD}~\cite{Khan:2015jqa}. This provides a utility function
\texttt{XLALSimIMRPhenomDChirpTime} that calculates the time until the peak in
the (2,2)-strain of a specific system configuration given a starting
\gw{} frequency, which is approximately twice the orbital frequency. 
The starting frequency is optimized using a simple interval
bisection procedure until the peak time is $\sim 2000M$. 
A lower bound on the
orbital frequency is set at 0.0225. The average retarded merger time (calculated as detailed below) 
for the simulations
that required a higher starting orbital frequency was $1983M$, with a
minimum of $1610$, and a maximum of $2169M$. 
While \texttt{XLALSimIMRPhenomDChirpTime} performed sufficiently well, overall it
slightly under estimated the merger time.
One simulation (\texttt{CF\_55}) was mistakenly performed using an increased starting frequency, resulting in a much shorter simulation with a retarded merger time of just $945M$.

The properties of each simulation are presented in Tabs.~\ref{table:metadata} and~\ref{table:metadata-2}.
Each configuration is characterised by its mass ratio $q$, the dimensionless spin magnitude $\chi_2 = S_2/m_2^2$, 
the spin angle $\theta_\text{LS} = \arccos(\hat{\bm{L}_\mathrm{N}} \cdot \hat{\bm{S}}_2)$, the
initial orbital frequency $M\Omega_{\rm orb}$ (or alternatively the initial binary separation $D/M$), and the binary's 
eccentricity, $e$. For the final values reported in Tab.~\ref{table:metadata} and~\ref{table:metadata-2}, the eccentricity $e$ is estimated over the region $[300, 800]M$ using the method described in
\cite{Husa:2007rh}. 
We also show the effective spin parameters $\chi_{\rm eff}$ and $\chi_{\rm p}$. The effective aligned spin
 $\chi_{\rm eff}$ is defined in terms of the individual parallel spin components $\chi_i^\parallel = \boldsymbol{\chi}_i \cdot \hat{\mathbf{L}}_{\rm N}$ 
 as~\cite{Ajith:2009bn}, 
 \begin{equation}
\chi_{\rm eff} = \frac{m_1 \chi_1^\parallel + m_2 \chi_2^\parallel}{M},
\end{equation} and parameterises the dominant 
spin effect on the orbital phasing, as discussed in Refs.~\cite{Cutler:1994ys,Poisson:1995ef,Baird:2012cu}. The effective
precession spin $\chi_{\rm p}$ is defined as~\cite{Schmidt:2014iyl},
\begin{equation} 
   \chi_{\rm p} =  \frac{S_{\rm p}}{m_1^2}, \label{eqn: chi_p}
\end{equation}
where $A_1 S_{\rm p} = \text{max} \left( A_1 S_1^\perp, A_2 S_2^\perp \right)$, $A_1 = 2 + 3 m_2 /(2 m_1)$, and 
$A_2 = 2 + 3 m_1/(2 m_2)$. In a generic two-spin system the dominant precession effect can be approximated by a
single-spin system where the larger black hole has an in-plane spin of $\chi_{\rm p}$, based on the leading-order spin
precession effects~\cite{Apostolatos:1994mx,Kidder:1995zr}. In the single-spin configurations in this catalogue, 
we will always have $\chi_{\rm eff} = m_2 \chi_2 \cos \theta_{\rm LS} / M$ and $\chi_{\rm p} = \chi_2 \sin \theta_{\rm LS}$. 

In Tab.~\ref{table:metadata} we provide the quantities $(\chi, \theta_{\rm LS}, \chi_{\mathrm{eff}}, \chi_{\rm p}, D/M)$ as 
specified in the initial data and (in brackets)
at a \emph{relaxed time}, $t_\mathrm{rel}$. This is the time at which we
estimate that the unphysical junk radiation in the initial data have radiated away, and the GW data can be used for analysis
and modelling. We wish $t_\mathrm{rel}$ to be as early as possible, to maximise the length of the usable waveform.
We choose a relaxed time of, \begin{equation}
t_\mathrm{rel} = t_\mathrm{peak} + 2t_\mathrm{damp},
\end{equation} 
where $t_\mathrm{peak}$ is the time of the peak amplitude of the junk radiation in the $(2,2)$ multipole moment of $\Psi_4$ 
and $t_\mathrm{damp}$ is an estimate of the exponential decay time of the junk radiation, which we estimate as 
$t_\mathrm{damp} = 76 m_2$. The damping time of the $(2,2,0)$ quasi-normal mode for a nonspinning black hole of mass $m$ is approximately $71m$, 
and approximately $83m$ for a black hole with dimensionless spin magnitude 0.8~\cite{PhysRevD.73.064030};
we find that $76m$ is a reasonable choice for all of the configurations in this catalogue. In Tab.~\ref{table:metadata} the initial orbital
frequency $M\Omega_{\rm orb}$ is calculated at $t_\mathrm{rel}$ and the number of orbits $N_{\rm orb}$ is calculated from 
$t_\mathrm{rel}$ to the merger time, which we define to be the time at which the peak in the $\ell=2$ multipole moments of $\Psi_4$ occurs. The retarded merger time, which we denote as $t_{\rm M}$, is given by the difference between the merger time and the tortoise co-ordinate 
\begin{align} 
   r^* = {}& r + 2\ln \left| \frac{r}{2} - 1 \right|,
\end{align}
where $r$ is the distance from the punctures at which the $\Psi_4$ data is extracted. 

The final black hole has a mass of $M_f$, a spin of $\chi_f$ and a recoil velocity $v_R$. We discuss the calculation of these remnant quantities in more detail 
in Sec.~\ref{sec:remnant} below.

\subsection{Initial black-hole masses}
\label{sec:masses}

We estimate the black-hole masses using the ADM mass calculated at the puncture. The black holes are represented in the initial
data as wormholes, and the ADM mass calculated at the second asymptotically flat end of each wormhole provides a good estimate of
that black hole's mass. In the puncture framework, this mass estimate is easy to calculate at each black hole's puncture~\cite{Brandt:1997tf}. 
The ADM puncture mass agrees well with the mass calculated from the area of the apparent horizon in the case of nonspinning black
holes~\cite{Tichy:2003qi}, but becomes less accurate for high spins~\cite{Hannam:2010ec}. 

\begin{figure}[ht!]
\includegraphics[width=\columnwidth]{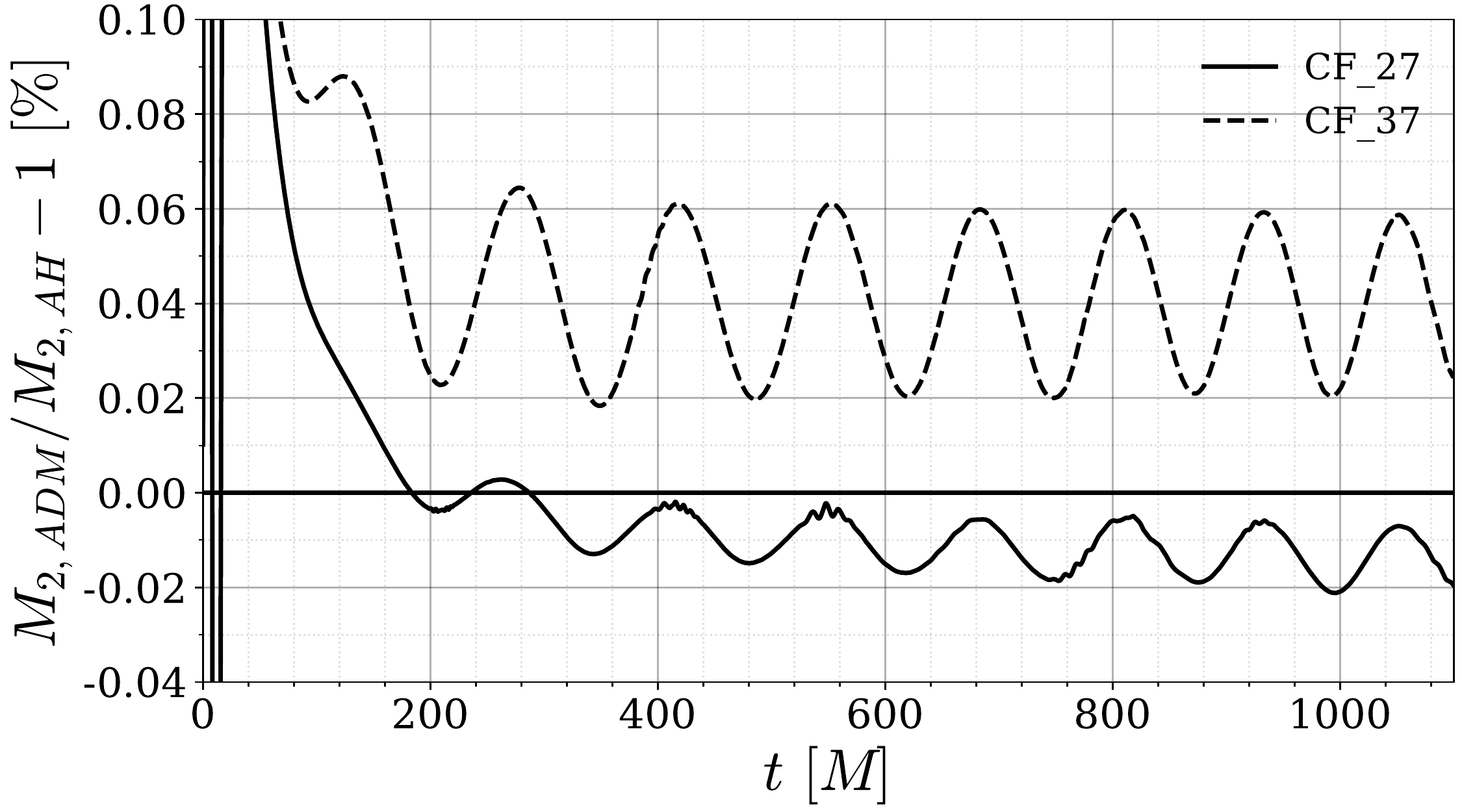}
\caption{Comparison of the relative percentage error between the apparent-horizon mass and the initial data ADM puncture mass for the larger black hole,
as a function of simulation time $t$. 
Both simulations have initial parameters $q = 2$, $\theta_\text{LS} = 60$. The solid black line is for a configuration with dimensionless spin magnitude $\chi_2 = 0.4$ 
and the dashed black line represents $\chi_2 = 0.8$. 
\label{fig:adm-puncture-mass-error}
}
\end{figure}

Figure~\ref{fig:adm-puncture-mass-error} demonstrates the effect spin magnitude has on the initial data ADM puncture mass as a function of time from the start of the simulation. The results
in Appendix~A of Ref.~\cite{Hannam:2010ec} suggest that the error in the mass estimate could be on the order of $\sim$0.5\% for black
holes with spin 0.8. However, the estimates in that paper were made on the initial data. Our results suggest that after the junk radiation has 
left the system (most radiates to infinity, but some falls back into the black hole), the apparent-horizon estimate of the mass is closer to the
original ADM-puncture-mass estimate. We see that for black holes with spins of 0.4 the error due to using the ADM puncture mass is 
on the order of $\sim$0.01\%, while for spins of 0.8 it is $\sim$0.04\%. (The oscillations in Fig.~\ref{fig:adm-puncture-mass-error} are due to 
uncertainties in the apparent-horizon estimate.) From this we conclude that the errors in the mass estimates are negligible.

\subsection{Initial black-hole spins}

The black-hole spin $\mathbf{S}_2$ is specified as part of the Bowen-York extrinsic curvature. 
The main source of uncertainty in the dimensionless spin $\chi_2 = \mathbf{S}_2/m_2^2$ is
the accuracy of the mass: as some of the junk radiation falls into the black hole, the mass increases, and
so $\chi_2$ decreases. However, as we saw previously, the final value of the mass as estimated from 
the area of the apparent horizon agrees well with the nominal value for each configuration. We also 
see in Tab.~\ref{table:metadata} that there is only a small discrepancy in the spin magnitude after the
relaxation time. Since the initial black-hole spins are prescribed analytically in the Bowen-York initial data,
we can reliably estimate the uncertainty in the apparent-horizon measurement of the spin magnitude to be within 
$\sim$0.001.

\begin{figure}[ht!]
\includegraphics[width=\columnwidth]{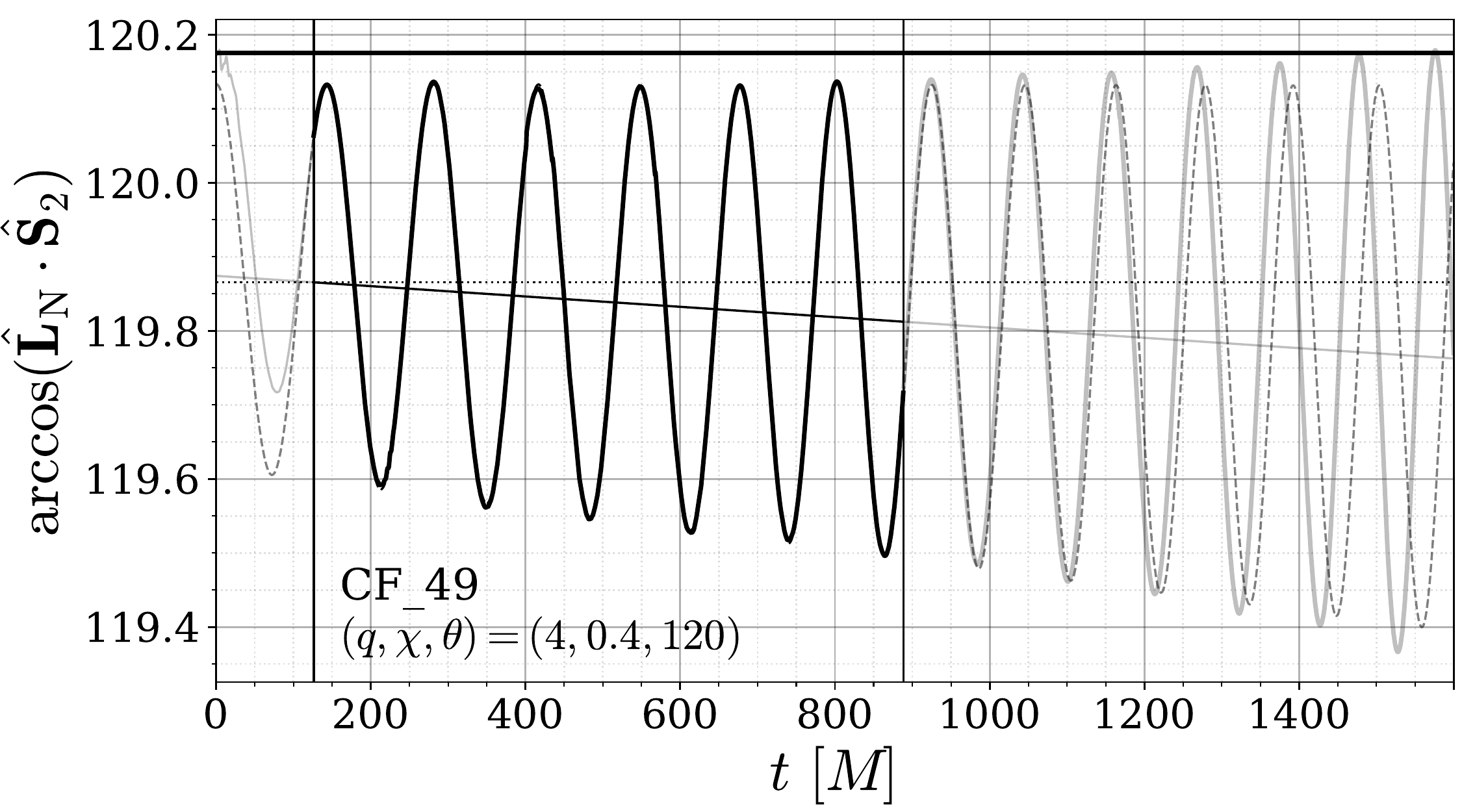}
\caption{Linear trend of the angle $\theta_\text{LS}$ between the Newtonian orbital
angular momentum and spin of the spinning black hole as a function of simulation time $t$. 
$\theta_\text{LS}$ as determined
from \nr{} data is plotted as a thick black line. The initial data value of
$\theta_\text{LS}$ is plotted as a horizontal thick black line. The dashed black line is
a sinusoidal fit using Eq.~\ref{eq:sin-ansatz}. The vertical black lines mark
the fit bounds. The linear part of the sinusoidal fit is plotted as a solid
black line. The dotted black line marks the value of the linear trend line at
the lower bound. Lower opacity lines of the \nr{} data and extrapolated fit are
plotted outside of the fit region. \label{fig:theta-average}}
\end{figure}

During the inspiral $\theta_\text{LS}$ is not constant; it will oscillate, as illustrated for one configuration in 
Fig.~\ref{fig:theta-average}. Ideally, we would set up our simulations so that the mean value of $\theta_{\rm LS}$ 
was equal to our prescribed value at the start frequency. We see in Fig.~\ref{fig:theta-average} the two ways in 
which our data deviate from this ideal. (1) There is an inaccuracy in the initial value of $\theta_{\rm LS}$, which is 
within the tolerance set in our initial-data construction procedure, and (2) this value is at an extremum of the 
oscillations in $\theta_{\rm LS}$, and so the mean will be offset from the target value in the initial-data construction.
We also see that the mean value slowly varies over the course of the simulation, although typically by only a fraction 
of a degree over the entire inspiral. 

In Tabs.~\ref{table:metadata} and ~\ref{table:metadata-2} we report the mean value of $\theta_{\rm LS}$ at the start of the simulation and at the relaxed time. We estimate
the value at the relaxed time by fitting to $\theta_{\rm LS}$ a sinusoidal ansatz of the form,
\begin{equation}
\label{eq:sin-ansatz}
\Big(A_0 + A_1t\Big)\sin\Big(2\pi\big(f_0t+f_1t^2\big) + \varphi\Big) + C_0 + C_1t,
\end{equation}
where $A_0$, $A_1$, $f_0$, $f_1$, $\varphi$, $C_0$ and $C_1$ are all free parameters,
from the relaxed time up to three orbits after the relaxed time.
The value of the linear part of the fit at the relaxed time 
is reported in Tabs.~\ref{table:metadata} and~\ref{table:metadata-2} instead of the pointwise value of the \nr{}
data for $\theta_{\rm LS}$. An example of this fit can be seen in Fig.~\ref{fig:theta-average}. The resulting value of 
$\theta_{\rm LS}$ is used, along with the relaxed-time value of the spin magnitude, to calculate the relaxed-time
values of $\chi_{\rm eff}$ and $\chi_{\rm p}$.

\subsection{Remnant properties}
\label{sec:remnant}

The final black hole that remains after the merger is characterised by its mass, spin, and  recoil. 
We report each of these quantities in Tabs.~\ref{table:metadata} and~\ref{table:metadata-2}.

As with the relaxed-time quantities reported in Tab.~\ref{table:metadata} and~\ref{table:metadata-2}, the mass and spin of the final black hole, $M_f$ and $\chi_f$, 
are calculated from the apparent horizon\cite{Alcubierre:1138167}. 
As a consistency check we also estimate the mass and angular momentum of the final 
spacetime from the gravitational-wave signal. The mass can be calculated by subtracting the radiated energy from the initial ADM mass
of the spacetime. The radiated energy is in turn calculated from the \gw{} signal measured at a series of extraction 
radii~\cite{Brugmann:2008zz} and the 
result extrapolated to infinity. The final mass estimated in this way typically agrees with the horizon measure to within $5 \times 10^{-4}$.
Given the mass, perturbation theory provides a relationship between the black-hole spin and the frequency of the signal multipoles during ringdown~\cite{London:2018nxs,leaver85}. 
We calculate the ringdown frequency of the $(\ell = 2, |m|=2)$ multipoles by taking the Fourier transform of the waveform between 
$10$ and $100M$ after merger~\cite{London_nrutils_2015}, where merger is here defined as the time at which the sum of the square of the $\ell=2$ multipoles is maximised.
The ringdown frequency is then the frequency at which the peak in this frequency domain post-merger waveform occurs~\cite{London:2014cma}. 
This then allows us to make an independent estimate of the final spin. We find that this estimate of the final spin 
typically agrees with the horizon measure to within $5 \times 10^{-3}$. 

We calculated the recoil (or kick) velocity of the final black hole by integrating the radiated linear momentum from the relaxed time 
$t_\text{rel}$ until the end of the simulation. We report here only the magnitude of the the recoil velocity $v_R$ . The linear 
momentum is itself calculated as described in~\cite{Brugmann:2008zz}. 
We used the value of the linear momentum extracted at a distance $90M$ from the source. 
Since the recoil velocity is very sensitive to the in-plane spin directions, this catalogue does not comprehensively explore the range of recoil 
velocities that can be seen for systems with mass ratios up to $q=8$ and dimensionless spin magnitudes up to $\chi=0.8$. However, from 
the values presented in Tabs.~\ref{table:metadata} and~\ref{table:metadata-2}, we can see that the largest magnitude kick velocities tend to
 be seen for systems where the two initial black holes are equal in mass, a general trend that can be seen even in inspiral post-Newtonian estimates~\cite{Kidder:1995zr}, and for full merger calculations in the numerical-relativity recoil studies cited in the Introduction.
 
 \begin{figure*}[htbp]
   \centering
   \includegraphics[width=\textwidth]{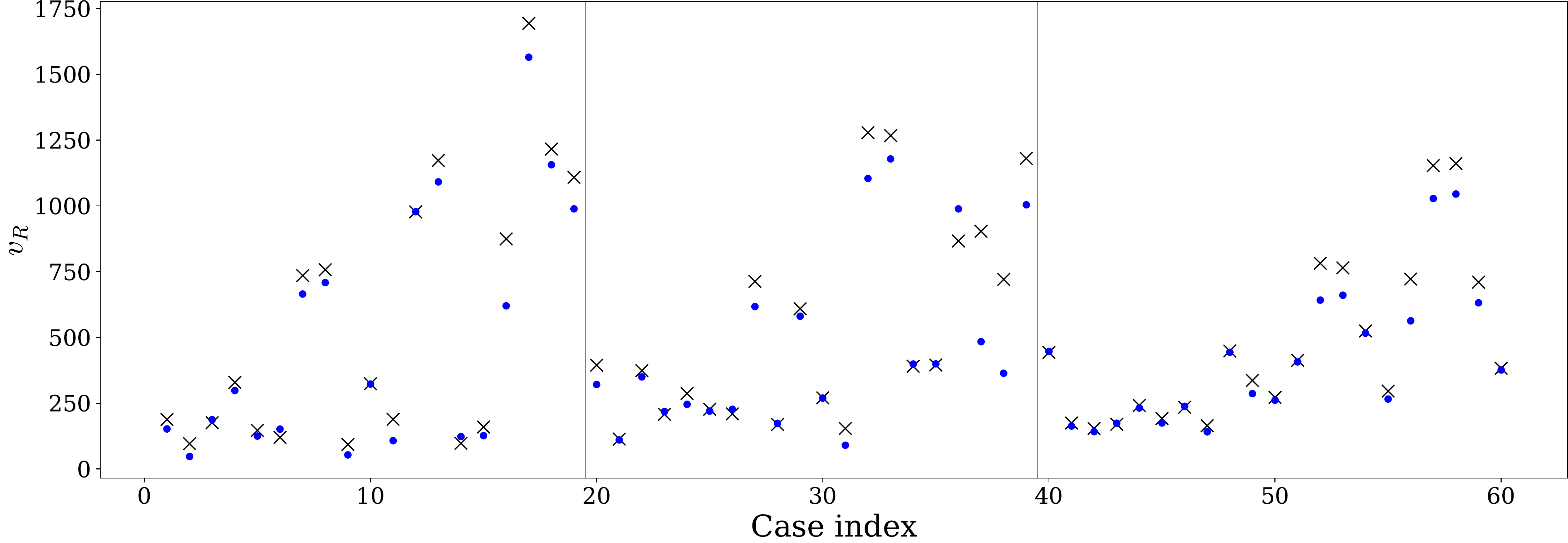}
   \caption{A comparison of the calculation of the recoil velocity from the radiated linear momentum calculated by the \texttt{BAM} code (black crosses) with the prediction by the \texttt{NRSur7dq4} model (blue dots).}
   \label{fig: recoil velocity}
\end{figure*}

We investigated the effect of the finite resolution of the simulations and the radius at which the rate of change of the linear momentum was extracted on the calculation of the final recoil velocity and found it to be negligible. The effect of the extraction radius was found to be less than $10\%$ of the final recoil velocity, while the effect of the resolution was even smaller.

We compared the results of our calculation of the recoil velocity with the prediction given by the NRSurrogate model \texttt{NRSur7dq4}~\cite{Varma:2019csw} for those cases within the catalogue that lie within the calibration region of \texttt{NRSur7dq4} ($q\le4$). To obtain this prediction, we used the value of the black hole spins, rotated into the LAL frame~\cite{Schmidt:2017btt}, and the orbital frequency $100M$ prior to merger. This comparison is shown in Fig.~\ref{fig: recoil velocity}. As can be seen from these results, for most of the cases contained within the catalogue, the calculation from the radiated linear angular momentum agrees well with the prediction by \texttt{NRSur7dq4}. However, in a small number of cases (most notably \texttt{CF\_37} and \texttt{CF\_38}) the two values differ by around $50\%$ of the value calculated from the radiated linear momentum. However, these values remain within the bounds predicted by \texttt{NRSur7dq4} for an equivalent configuration but with a different value for the in-plane spin angle. We therefore do not find these discrepancies too concerning, and we leave determining their exact cause to a future investigation.

\section{Waveform accuracy}\label{sec: four waveforms}

In order to assess the accuracy of the data that comprise this catalogue we studied a subset of four of the configurations described in 
Tabs.~\ref{table:metadata} and \ref{table:metadata-2}. These four configurations are {\tt CF\_47, CF\_59, CF\_66}, and {\tt CF\_80}, with physical 
parameters $\left(q,\chi,\theta_\text{LS}\right) = \left\{ \left(4,0.4,60\right), \left(4,0.8,120\right), \left(8,0.4,30\right), \left(8,0.8,150\right) \right\}$. 
The set of simulations used in the accuracy analysis of the $\left(4,0.4,60\right)$ case were performed with a lower starting frequency of $M\omega_\text{orb}=0.023$ to provide an assessment of the accuracy of a longer simulation.

The two main sources of error in our waveforms are the finite resolution of the simulation and the finite radius at which the data are extracted. In order to assess the effect of the finite resolution, we performed a set of three simulations with low, medium and high resolution for each of the four configurations listed above. We also performed an additional simulation with very high resolution for the $\left(8,0.8,150\right)$ configuration. These resolutions correspond to a number of grid points $N = \{80, 96, 120, 144\}$ in the boxes surrounding the punctures. Typically the width of the 
smallest box around each black hole is on the order of $\sim 2 m /N$, where $m$ is the mass of that black hole; the details of how the
grid is determined for each configuration are given in Sec.~\ref{subsec:grids}.
We extracted the waveform data at $R_\text{ext} = \left\{50,60,70,80,90\right\}M$, which were all on the same refinement level.

In quantifying the error in the waveforms due to these two sources we focus on estimating the mismatch between the medium resolution waveforms extracted at a distance of 90M from the source and the ``true'' waveform at infinitely good resolution and infinitely far from the source. We calculated the convergence order of the BAM code with respect to the resolution and extraction radius then used this to extrapolate the mismatch. We also used Richardson extrapolation to estimate the truncation error due to resolution and extraction radius. 

Mismatches are calculated from a noise-weighted inner product between waveforms, and extrapolate differently to the quantities that 
are usually considered in an convergence analysis, e.g., waveform amplitude and phase. In Sec.~\ref{sec: brief derivation} we 
sketch out how standard numerical convergence properties translate to the waveform mismatch, and provide more detailed derivations 
in Appendix~\ref{sec: mismatch-convergence-relationship}.

\begin{figure*}[ht!]
\includegraphics[width=\textwidth]{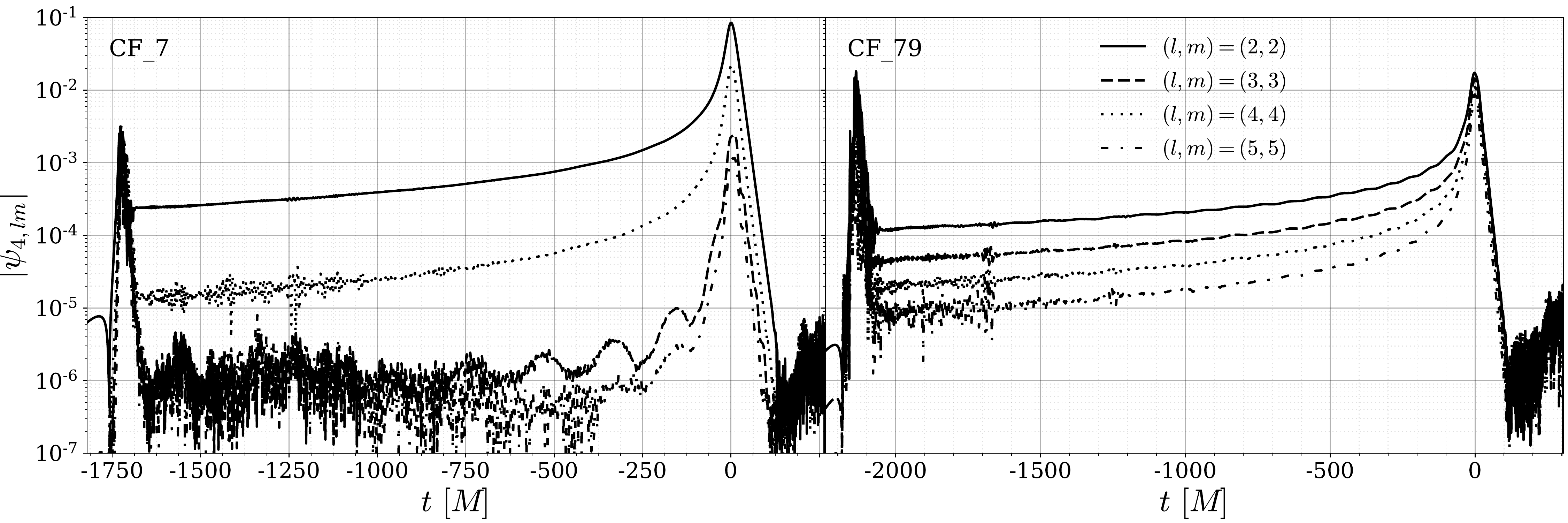}
\caption{Comparison of coprecessing frame time domain amplitudes for the $l=m$ modes for $l \in \{2, 3, 4, 5\}$. The left panel shows 
\texttt{CF\_7} with initial parameters $(q, \chi_2, \theta) = (1, 0.4, 60)$ and the right panel shows \texttt{CF\_79} with parameters 
$(q, \chi_2, \theta) = (8, 0.8, 120)$. \label{fig:data-quality}}
\end{figure*}

\subsection{Data quality}

Our accuracy analysis of the gravitational waveforms calculated from our simulations focuses on the mismatch 
uncertainty as detailed in section~\ref{sec: waveform accuracy mismatches}. 
This is because it is the overall mismatch uncertainty that is most relevant to most gravitational-wave astronomy
applications. We also consider the error in the phase and amplitude of the dominant multipole of the co-precessing waveform in 
section~\ref{sec: cp waveform errors}.
However, we often also wish to know the accuracy of the individual signal multipoles, for example when using
them to construct waveform models, or when using the NR waveforms as proxy signals to test gravitational-wave  
data-analysis pipelines. In this paper we do not perform a separate convergence analysis of the individual multipoles; given that
clean convergence is rare in any binary-black-hole waveforms, even for the dominant multipoles, we do not expect a 
convergence analysis of sub-dominant multipoles to be informative. 

Here we simply note that the phasing accuracy of the waveforms is dominated by the phase accuracy of the inspiral 
dynamics, and this can be assessed through an accuracy analysis of the dominant multipole. (An important exception is
the signal near merger, as discussed in Ref.~\cite{CalderonBustillo:2015lrg}.) For the signal amplitude
we assess the accuracy by the presence of noise in the data. For example, Fig.~\ref{fig:data-quality} compares the relative 
strength of the gravitational-wave multipoles for two simulations. We see that it is not possible to conclude that a particular
set of multipoles will always be reliable. In the \texttt{CF\_7} simulation the $(3,3)$ and $(5,5)$ multipoles cannot be trusted
before merger; we would not expect these to be useful, for example, to calibrate a model of the signal amplitude.
On the other hand, in the \texttt{CF\_79} simulation we see that, despite a low level of noise at early times, all of the 
multipoles in the figure could well be used to model the amplitude. Rather than choose a set of ``trustworthy'' multipoles,
we instead suggest that for most applications one should use only the parts of a post-relaxation-time $\psi_{4,\ell m}$ multipole with an 
amplitude above $10^{-5}$. Depending on the application, of course, one may wish to apply a more or less stringent requirement.

\subsection{Amplitude and Phase accuracy}
\label{sec: cp waveform errors}

In order to estimate the numerical error in the waveform quantities due to the finite resolution of the simulation and the finite radius at which the data 
were extracted, we performed Richardson extrapolation; see Appendix~\ref{sec: mismatch-convergence-relationship}. This requires first estimating 
the convergence order of the code with respect to these quantities. We first processed the data, removing the junk radiation from the waveform in the 
inertial frame in which the simulation was performed. We aligned the waveforms at merger, where merger is defined to be the time at which the quantity 
\begin{align} 
   \mathcal{A}^2 = {}& \sum^2_{m=-2} \left| A_{2m}\left(t\right) \right|^2,
\end{align}
where $A_{2m}$ are the amplitudes of the $\ell=2$ multipoles,
is maximised and resampled using a constant time step of 0.1M. We then rotated the waveform into the co-precessing frame and aligned the 
co-precessing phases at merger. 
The co-precessing frame is one which precesses along with the binary and is advantageous here as it means we can focus on the error in a single multipole (the $\left(2,2\right)$ multipole, which is dominant in this frame, rather than considering the error in each of the $\ell=2$ multipoles (which all have appreciable power in the inertial frame) independently.
The quantities for which we are interested in quantifying the numerical error are the amplitude and phase of 
the (2,2) multipole
in the co-precessing frame as well as two of the Euler angles $\alpha$ and $\beta$ required to rotate the waveform from the 
inertial frame into the co-precessing frame. The Euler angles were calculated using the method detailed in Ref.~\cite{OShaughnessy:2011pmr, Boyle:2011gg}. 

The standard way to perform a convergence test with respect to the resolution is to perform a set of three simulations where the resolution 
improves by a factor of two between each of the simulations. This is computationally prohibitive --- the high resolution run would use $2^6$ 
times as much memory as the low resolution run. Similarly, with {\tt BAM}'s box-based mesh refinement we cannot extract a waveform at three 
different radii on the same level a reasonable distance from the source if each of the extraction radii is twice as far away from the source as 
the previous one.

\begin{figure}[t]
   \centering
   \includegraphics[width=\columnwidth]{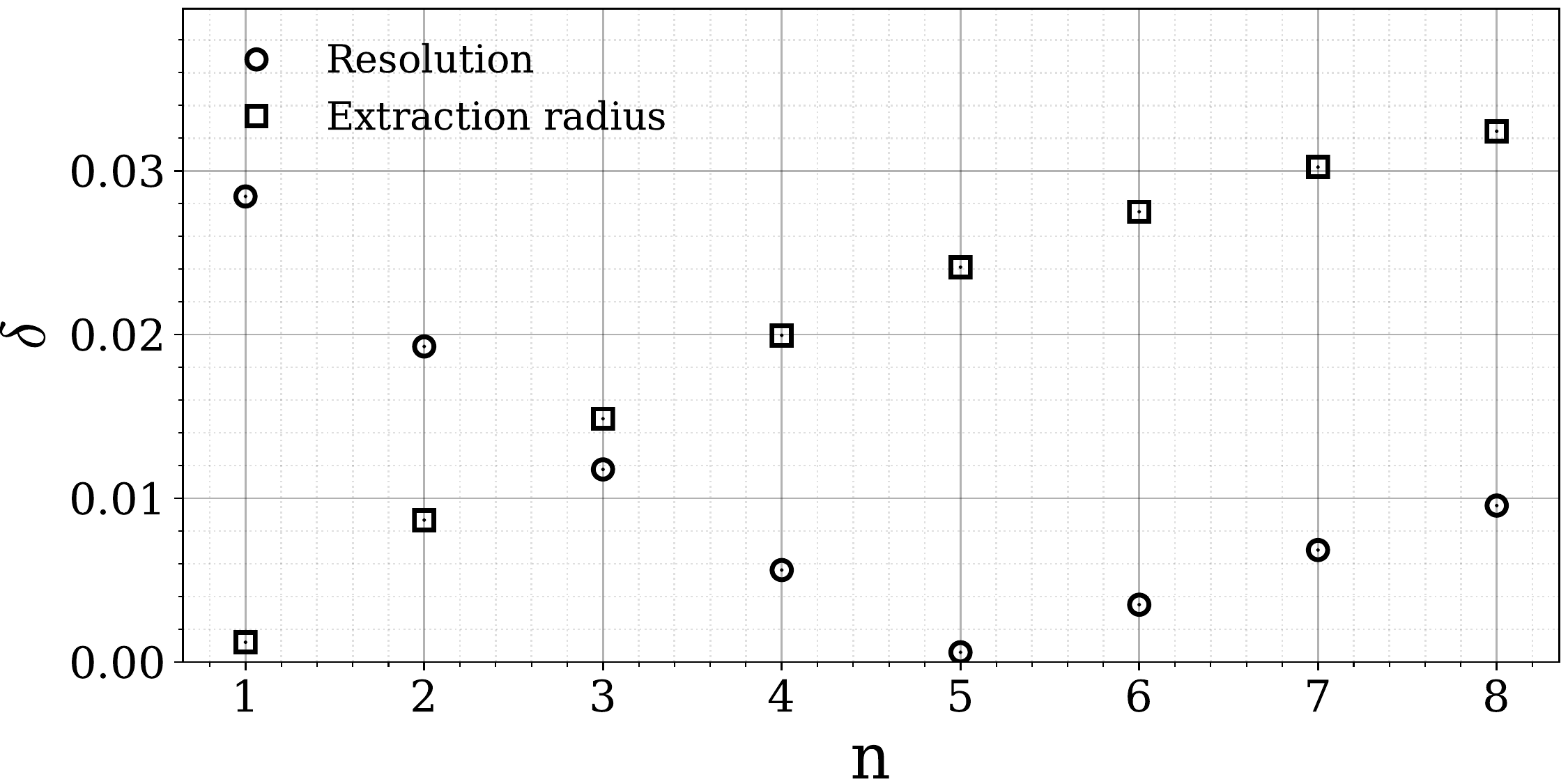}
   \caption{The value $\delta$ as given by equation \ref{eqn: convergence order} as a function of convergence order. The circle markers represent waveforms of differing resolution. The square markers represent waveforms at differing extraction radii. The configuration $q=8$, $\chi=0.8$, $\theta_\text{LS}=150^\circ$ was used in this analysis.}
   \label{fig: RE convergence order}
\end{figure}

We nominally expect the error due to extraction radius $R_{\rm ext}$ to fall off as $1/R_{\rm ext}$, although we will confirm that in our analysis. The
 numerical-resolution convergence order is less clear. The spatial finite-differencing in the bulk is sixth-order, but the time-evolution is fourth-order; either
 may dominate the error budget, depending on the resolution choices and length of the simulation~\cite{Brugmann:2008zz,Husa:2007hp}. 
 For both the extraction-radius and numerical-resolution,
 we determine the appropriate convergence order by studying the convergence behaviour of the phase of the $(\ell=2,m=2)$ multipole in the co-precessing
 frame. We then identify the value of $n$ for which the quantity,
\begin{align} \label{eqn: convergence order}
   \delta = {}& \left(\phi\left(\Delta_1\right) - \phi\left(\Delta_2\right) \right) - 
   \mathcal{C} \left( \phi\left(\Delta_2\right) - \phi\left(\Delta_3\right) \right),
\end{align}
is minimised, where where $\phi$ is the phase of the $(2,2)$ multipole in the co-precessing frame, $\mathcal{C} = \frac{\Delta_1^n - \Delta_2^n}{\Delta_2^n - \Delta_3^n}$ as in Eq.~(\ref{eqn: convergence factor}), and $\Delta_i$ is the 
variable in the error expansion, i.e., numerical resolution or the inverse extraction radius. 
The quantity $\delta(t)$ was calculated over the length of the waveform up to merger and the mean value $\delta=\langle\delta(t)\rangle$ is shown in Fig.~\ref{fig: RE convergence order}. 
This was done for both waveforms of varying resolution and extraction radius for the $q=8$, $\chi=0.8$, $\theta_\text{LS}=150^\circ$ configuration. 
In calculating the convergence order with respect to varying resolution we used waveforms with $\Delta_{\{1,2,3\}} = \{1/144,1/120,1/96\}$ while when considering the convergence order with respect to extraction radius we used $\Delta_{\{1,2,3\}} = \{1/90,1/70,1/60\}$.
From the results shown in Fig.~\ref{fig: RE convergence order}, we make the conservative conclusion that the code is consistent with fourth order finite-differencing, which implies that the time-stepping
dominates the error budget.
From inspecting time-dependent $\delta(t)$ calculated over the length of the waveform, we also find that the spatial differencing (with sixth-order accuracy) 
dominates the error over the first $\sim1000M$ of the waveforms, but the fourth-order-accurate time stepping dominates 
in the last $\sim500M$ before merger, and dominates overall. 
As expected, we see that the radiation extraction errors fall off as $1/R_\text{ext}$.

\begin{table}[t]
   \centering
   \begin{tabular}{@{} lccc @{}} 
      \toprule
      & \multicolumn{3}{c}{ \% Error} \\
      & Resolution & Extraction Radius & Total \\
      \midrule
      $\phi$ & 0.08 & 0.4 & 0.4 \\
      $A$ & 2.5 & 0.9 & 2.7 \\
      $\alpha$ & 0.4 & 0.05 & 0.4 \\
      $\beta$ & 0.2 & 0.06 & 0.2 \\
      \bottomrule
   \end{tabular}
   \caption{Relative error in the waveform quantities compared with the Richardson extrapolated quantities for the $q=8$, $\chi=0.8$, $\theta_\text{LS}=150^\circ$ configuration.}
   \label{tab: relative errors}
\end{table}

\begin{figure}[t]
   \centering
   \includegraphics[width=\columnwidth]{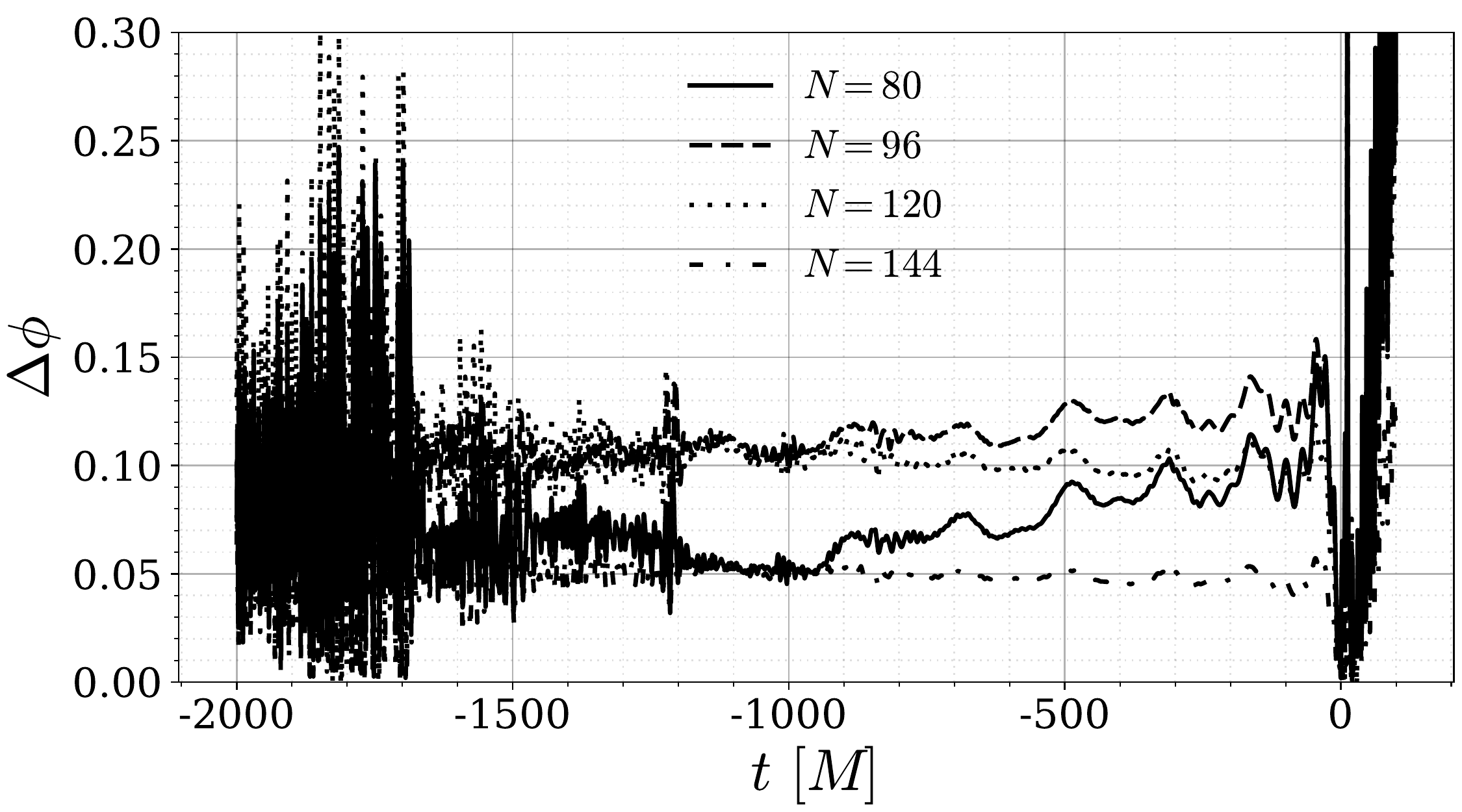}
   \caption{Resolution dependence of the absolute error in the time domain co-precessing phase, relative to the Richardson-extrapolated phase. 
   The phases have been aligned at merger.}
   \label{fig: phase error resolution}
\end{figure}

\begin{figure}[t]
   \centering
   \includegraphics[width=\columnwidth]{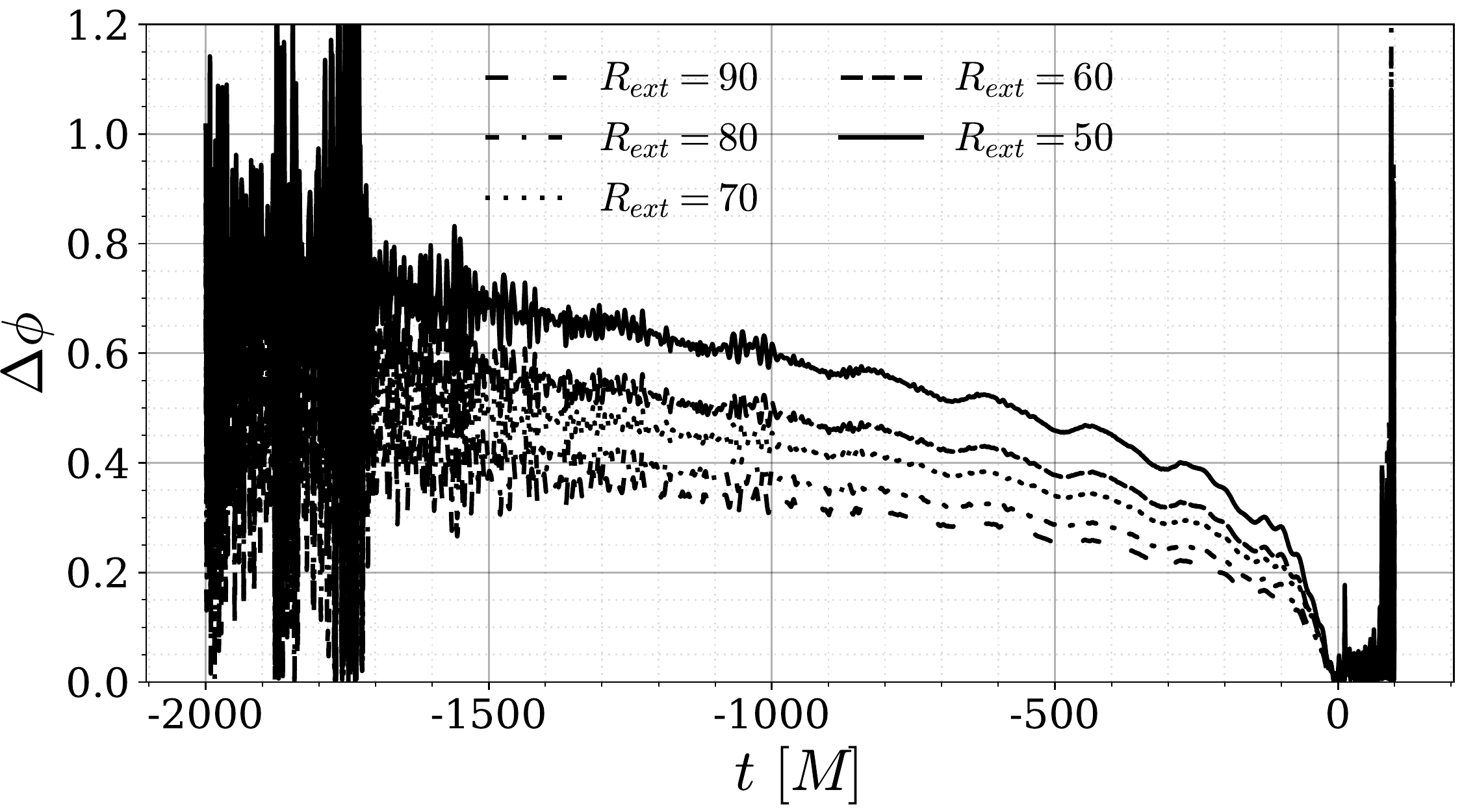}
   \caption{Extraction radius dependence of the absolute error in the time domain co-precessing phase, relative to the Richardson-extrapolated
   phase. The phases have been aligned at merger.}
   \label{fig: phase error extraction radius}
\end{figure}

Assuming these convergence orders, we then calculate the Richardson-extrapolated values of the amplitude, phase, $\alpha$ and $\beta$, as functions 
of time, using Eq.~(\ref{eqn: Richardson extrapolation}) in App.~\ref{sec: mismatch-convergence-relationship}. 
We used the resolutions $N=\{120,144\}$ and the extraction radii $R_\text{ext}=\{80,90\}$ in calculating these Richardson-extrapolated values, which were used to estimate the error in each of these quantities. The error in the waveform quantities does not monotonically increase with 
decreasing resolution since not all of the resolutions lie in the convergence regime. The error in the low and medium resolution waveforms is 
therefore estimated from the combination of the error between these waveforms and the very high resolution waveform and between the very high 
resolution waveform and the ``true'' waveform. 

Since a time shift was performed to align the waveforms at merger, where the phases were then aligned,
rather than aligning the frequencies at merger,
the phase difference does not show a quadratic fall off to zero but rather tends to a constant value and then falls rapidly at merger.
As can be seen from Fig.~\ref{fig: phase error resolution}, the dephasing of the waveform due to the finite resolution is $\sim0.1$ radians for the
 medium resolution ($N = 96$) simulation. Similarly, from Fig.~\ref{fig: phase error extraction radius}, the dephasing due to the finite 
extraction radius is $\sim0.4$ radians for the waveform extracted at $90M$. 
Since the simulations comprising the bulk of the catalogue were performed with medium resolution and we recommend using the waveform 
extracted at $90M$, these are the key values to focus on.
The total phase error in the waveform is therefore estimated to be about 0.4 radians by combining the errors in quadrature.

The relative error for each of the quantities we are interested in are given in Tab.~\ref{tab: relative errors}. 
The quantities presented in this table are calculated as follows; the relative error is taken to be the maximum error, found from Richardson extrapolation, as described above, divided by the maximum value of the quantity, over the length of the waveform. Since we aligned the phases at merger, both the error in the phase and the phase itself are maximum at the start of the waveform. In contrast, both the amplitude and the error in the amplitude peak at merger. We therefore report here the relative error in the peak of the (2,2) multipole in the co-precessing frame. This gives an error around an order of magnitude larger than during the inspiral, where we see a total relative error of 0.1\%. The order of the relative error in the precession angles is fairly consistent over the length of the waveform.

From the values given in Tab.~\ref{tab: relative errors}, we can see that the maximum relative error in the amplitude of the co-precessing waveform is of the order of a few percent, while the relative error in the phase and in the precession angles is around half a percent. This is relevant for the production of a tuned precessing model using data from these simulations since it implies that the model for the precession angles cannot be accurate to more than 0.5\%. Similar results were seen for the other simulations for which we have multiple resolutions.

The errors in the amplitude and the precession angles are affected by the dephasing in the waveform. Therefore, although these results are a good diagnostic for the reliability of the code and a good way to compare accuracy between different simulations performed with the same code, they are difficult to translate into meaningful measures of the accuracy for waveform modelling or other \gw{} applications. In order to get a more meaningful estimate of the accuracy of the waveform we performed the mismatch analysis presented in the following section.

\subsection{Matches}
\label{sec: waveform accuracy mismatches}

The waveform quantities examined in the previous section are the standard quantities used when estimating the convergence order and 
accuracy of a \nr{} code. While useful when comparing the accuracy between simulations and codes, these accuracy measures are difficult 
to interpret in gravitational-wave astronomy applications --- the sensitivity of a search, or the accuracy of a measurement of the properties of a 
binary system. When assessing the accuracy of a waveform it is usually more useful to consider an estimate of the mismatch error. 

The match between two waveforms is defined to be the standard inner product weighted by the power spectral density of the detector $S_n\left(f\right)$ optimised 
over various sets of parameters $\Theta$~\cite{Cutler:1994ys},
\begin{align} \label{eqn: match def}
   M\left(h_1,h_2\right) = {}& \max_{\Theta} \left[ 4 \text{Re} \int^{f_\text{max}}_{f_\text{min}}
   \frac{\tilde{h}_1 \left(f\right) \tilde{h}^*_2 \left(f\right)}{S_n\left(f\right)} \text{d}f \right],
\end{align} where the individual waveforms have been normalised so that $M(h,h) = 1$. 
We also define the mismatch:
\begin{align}\label{eqn: mismatch} 
   \mathcal{M} = {}& 1 - M\left(h_1,h_2\right).
\end{align}

Since these are precessing configurations, we calculate precessing matches as described in Appendix B of Ref.~\cite{Schmidt:2014iyl}. 
In order to see how the match varies over a range of total masses that might be observed by current ground-based detectors, we further 
calculate the power-weighted match as described in Appendix~\ref{match-expressions}, based on the work in Ref.~\cite{Ohme:2011zm}, 
using PhenomPv3~\cite{PhysRevD.100.024059} as the model for the low frequency part of the waveform. We then calculate the mismatch 
as given by Eq.~(\ref{eqn: mismatch}).

We first extrapolate the mismatch due to finite resolution and extraction radius separately, assuming a particular fall-off in the respective errors. In order to then find the overall mismatch due to \emph{both} finite resolution and extraction radius-- i.e. the mismatch with the infinitely far away, infinitely well resolved ``true" waveform-- we need to correctly combine these calculations. The motivation for the correct way to combine such errors is sketched out in the following section and given in more detail in Appendix~\ref{sec: mismatch addition}. We then perform an independent calculation to confirm the mismatches follow the same convergence relation as the waveform quantities discussed in the previous section and finally calculate the extrapolated mismatch.

\subsubsection{Dependence on expansion parameter and addition of mismatch}
\label{sec: brief derivation}

In the following we look at how the mismatch behaves with respect to an expansion parameter, e.g., the numerical resolution 
or the radius at which the gravitational-wave signal is extracted. We then consider the addition of mismatch errors. The
calculation below, where the mismatch is expanded in terms of either the amplitude or phase, is a standard calculation, 
but we discuss it in detail here to help motivate the final result, which is somewhat surprising: although contributions to the 
error in the amplitude or phase of the signal combine in quadrature, as one might expect (see, for example, Sec.~II.A of
Ref.~\cite{Lindblom:2008cm}), separate mismatches should added according to Eq.~(\ref{eqn: adding matches}) below. 

To find how the ratio of two matches between waveforms of differing expansion parameter depends on the expansion parameter, we can examine how the match depends on the amplitude and phase of the waveform. 
From Eq.~(\ref{eqn: match def}), the match goes as
\begin{align}\label{eqn: match approx}
   M \sim {}& \text{Re} \left[ \frac{1}{N_1 N_2} \int h_1\left(f\right) h_2^{*}\left(f\right) \text{d}f \right],
\end{align}
where $N_i$ are the normalisation of each of the waveforms respectively. We take $h_1$ to be the waveform containing either the amplitude or phase error and $h_2$ to be the ``true'' waveform-- i.e.,
\begin{align} 
   h_1\left(f\right) = {}& h_2\left(f\right) + \Delta h\left(f\right),
\end{align}
where $h_2\left(f\right)=A\left(f\right)e^{i\phi\left(f\right)}$, and $A(f)$ is the real amplitude and $\phi(f)$ is the phase. 
We assume the true waveform to be normalised, so
\begin{align} 
   N_2 = {}& \left[ \int \left| h_2\left(f\right) \right|^2 \text{d}f \right]^\frac{1}{2}
   = \left[ \int A^2\left(f\right) \text{d}f \right]^\frac{1}{2} = 1.
\end{align}

A waveform containing some amplitude error $\Delta A$ is given by 
\begin{align}\label{eqn: amp error}
   h_1\left(f\right) = {}& \left( A\left(f\right) + \Delta A\left(f\right) \right) e^{i\phi\left(f\right)}.
\end{align}
Substituting Eq.~(\ref{eqn: amp error}) into (\ref{eqn: match approx}) we find
\begin{align} 
   M \sim {}& \left[ \int \left(A + \Delta A\right)^2 \text{d} f\right]^{-\frac{1}{2}}
   \text{Re} \int A \left( A + \Delta A \right) \text{d} f \nonumber \\
   = {}& \left(1 + 2 b + c \right)^{-\frac{1}{2}} \left( 1 + b \right) \nonumber \\
   \simeq {}& 1 + \frac{1}{2} \left( b^2 - c \right),
\end{align}
where $b = \int A \Delta A \text{d}f$, $c = \int \left( \Delta A\right)^2 \text{d}f$ and we have assumed that $\Delta A$ is small in order to make the approximation in the final step. The mismatch, as defined in Eq.~(\ref{eqn: mismatch}), therefore goes as $b^2 + c$ and so is dominated by the square of the amplitude error.

Similarly for a normalised waveform that contains some phase error $\Delta\phi$,
\begin{align}\label{eqn: phase error}
   h_1\left(f\right) = {}& A\left(f\right)e^{i\left(\phi\left(f\right)+\Delta\phi\left(f\right)\right)},
\end{align}
where $\phi$ is the ``true'' phase. Substituting this expression into Eq.~(\ref{eqn: match approx}) we find
\begin{align} 
   M \sim {}& \text{Re} \left[ \int A^2 e^{i \Delta \phi} \text{d}f \right] \nonumber \\
   \simeq {}& \text{Re} \int A^2 
   \left( 1 + i\Delta\phi - \frac{1}{2} \left(\Delta\phi\right)^2 \right) \text{d}f \nonumber \\
   = {}& \int A^2 \left( 1 - \frac{1}{2} \left(\Delta\phi\right)^2 \right) \text{d}f \nonumber \\
   = {}& 1 - \frac{1}{2} \int A^2 \left(\Delta\phi\right)^2 \text{d}f,
\end{align}
where again we have assumed that the error in the phase is small in order to perform the expansion in the middle step. The mismatch is therefore dominated by the square of the phase error. 

\begin{figure*}[htbp]
   \centering
   \includegraphics[width=\textwidth]{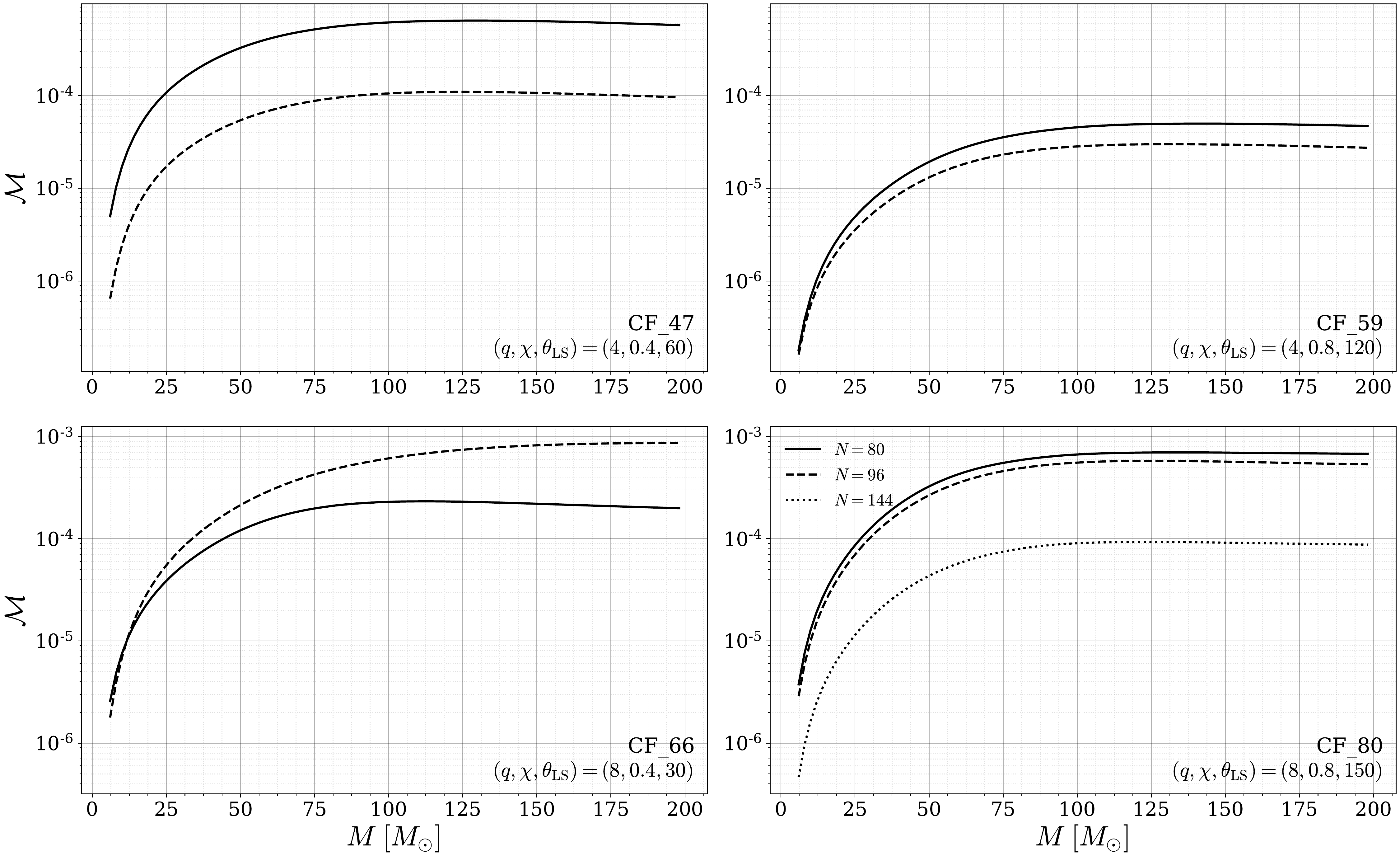}
   \caption{Mismatch between waveforms at varying resolution against the high resolution ($N=120$) waveform as a function of total mass.
   }
   \label{fig: resolution matches}
\end{figure*}

\begin{figure*}[htbp]
   \centering
   \includegraphics[width=\textwidth]{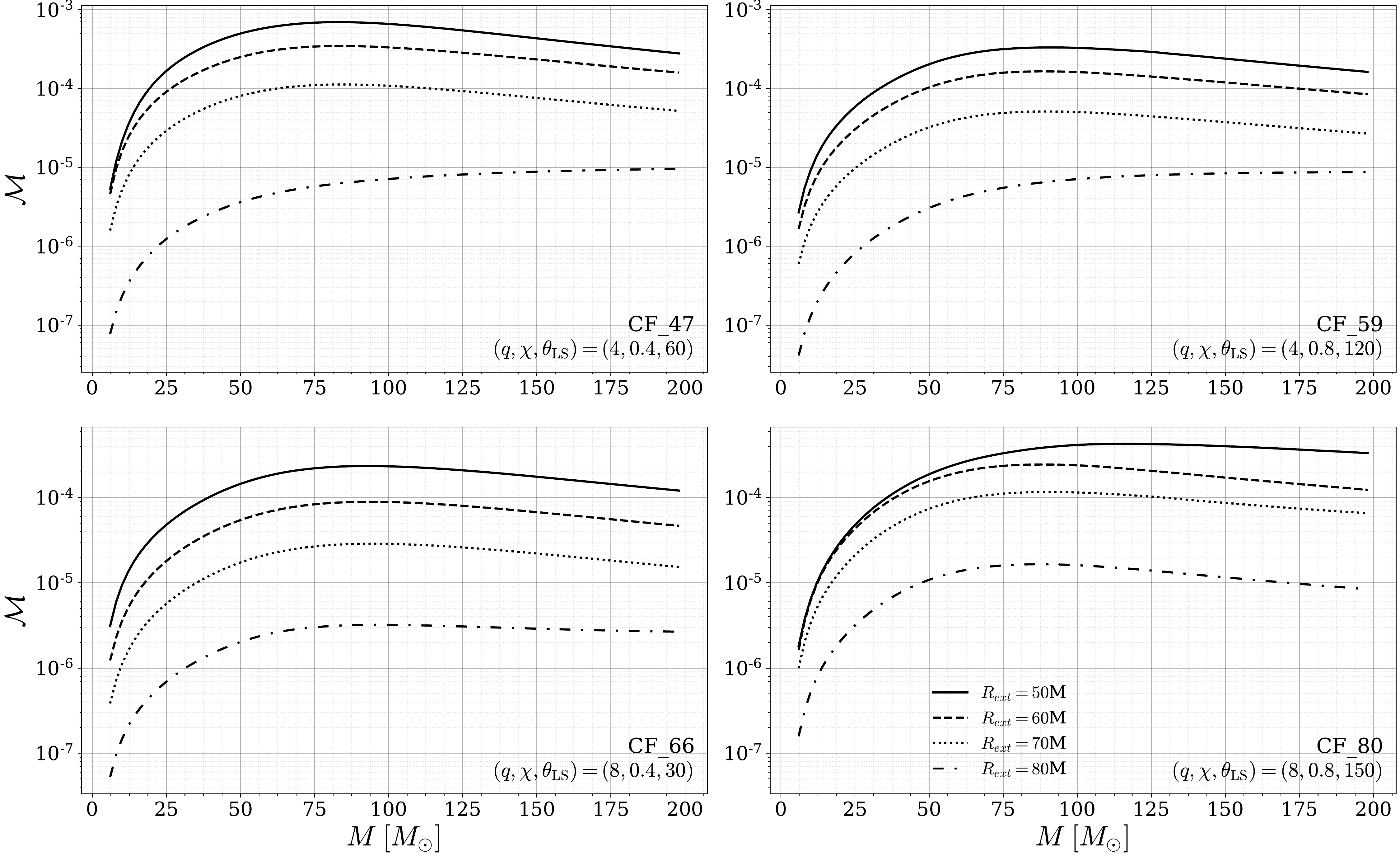}
   \caption{Mismatch between waveforms at varying extraction radii and the waveform extracted at $R_\text{ext} = 90$M as a function of total mass.}
   \label{fig: extraction radius matches}
\end{figure*}

The waveform quantities at finite resolution or extraction radius can be expressed as a Richardson extrapolation of the appropriate expansion parameter (see Appendix~\ref{sec: Richardson Extrapolation}). The difference in the phase and amplitude between two waveforms (labelled $A$ and $B$) is therefore equal to the difference between the leading order error term (i.e. $\Delta q \sim \Delta_B^n - \Delta_A^n$, see Appendix~\ref{sec: convergence} for more detail). Since the mismatch is proportional to the \emph{square} of the error in these waveform quantities, we find that the convergence relation for the mismatch takes the form
\begin{align} \label{eqn: convergence relation}
   \mathcal{M}\left(\Delta_A:\Delta_B\right) = {}& \kappa \left( \Delta_A^n - \Delta_B^n \right) ^2,
\end{align}
where $\Delta_i$ is the value of the expansion parameter for the $i$th waveform and $\kappa$ is a co-efficient to be found. If one of the waveforms being considered is the ``true'' waveform and thus contains no numerical error, then the mismatch between any reference waveform and this true waveform will be given by $\mathcal{M}\left(\Delta_\text{ref}:\Delta_\text{true}=0\right) = \kappa \Delta_\text{ref}^{2n}$. A similar derivation to the one discussed here is also presented in Ref.~\cite{Ferguson:2020xnm}.
From this we can see that the ratio of the mismatch $\mathcal{M}$ between two pairs of waveforms is given by
\begin{align}\label{eqn: mismatch ratio}
   \frac{ \mathcal{M}\left(A:B\right) }{ \mathcal{M}\left(B:C\right) } = {}& 
   \frac{\left(\Delta_A^n - \Delta_B^n\right)^2}{\left(\Delta_B^n - \Delta_C^n\right)^2}.
\end{align} This result will be used in the following sections to study the convergence properties of our numerical-relativity waveforms
via their mismatch error. 

We can also see that the correct way to combine the mismatches between two sets of waveforms $\mathcal{M}\left(A:B\right)$ and $\mathcal{M}\left(B:C\right)$ in order to get the mismatch between the final pair $\mathcal{M}\left(A:C\right)$ is given by
\begin{align}\label{eqn: adding matches} 
   \mathcal{M}\left(A:C\right) = {}& \left(
   \sqrt{\mathcal{M}\left(A:B\right)} + \sqrt{\mathcal{M}\left(B:C\right)}
   \right)^2.
\end{align}
A more rigorous proof of this result is presented in Appendix~\ref{sec: mismatch addition}. 

As stated above, our main use for this result is to combine the mismatch due to different sources of error in our numerical waveforms. We estimate the mismatch between our waveforms at finite extraction radius and finite resolution and the true waveform using 
\begin{align} \label{eqn: complete mismatch}
   \mathcal{M} = {}& \left( \sqrt{\mathcal{M}_\text{resolution}} + 
   \sqrt{\mathcal{M}_\text{extraction radius}} \right)^2,
\end{align}
where $\mathcal{M}_\text{resolution}$ is the mismatch due to the finite resolution of the numerical simulation and $\mathcal{M}_\text{extraction radius}$ is the mismatch due to the finite distance from the source at which the waveforms were extracted.

\subsubsection{Convergence order}

We performed matches between waveforms extracted at $R_\text{ext}=90M$ for the high resolution simulations against all other resolutions available for a 
given configuration. These results are shown in Fig.~\ref{fig: resolution matches}. We also performed matches between waveforms extracted at 
$R_\text{ext}=90$M and all other available extraction radii for the medium resolution simulations for each configuration. These results are shown in Fig.~\ref{fig: extraction radius matches}. 
In both of these comparisons, we have calculated the match against a single resolution or extraction radius. We therefore expect that the matches will improve for cases where the values of the resolution or extraction radius are closer to each other. 
From Fig.~\ref{fig: extraction radius matches} we can see that the matches generally follow this trend, implying that it is reasonable to assume the waveform is being extracted sufficiently far from the source that we may be in the convergence regime. 
This is not true for the mismatches with respect to resolution shown in Fig.~\ref{fig: resolution matches}. The matches between (i) the low and high resolutions and (ii) the medium and high resolutions clearly do not follow any trend for most of the configurations. From this we can see that it is not reasonable to treat the low and medium resolutions as if they lie within the convergence regime.

\begin{figure}[t]
   \centering
   \includegraphics[width=\columnwidth]{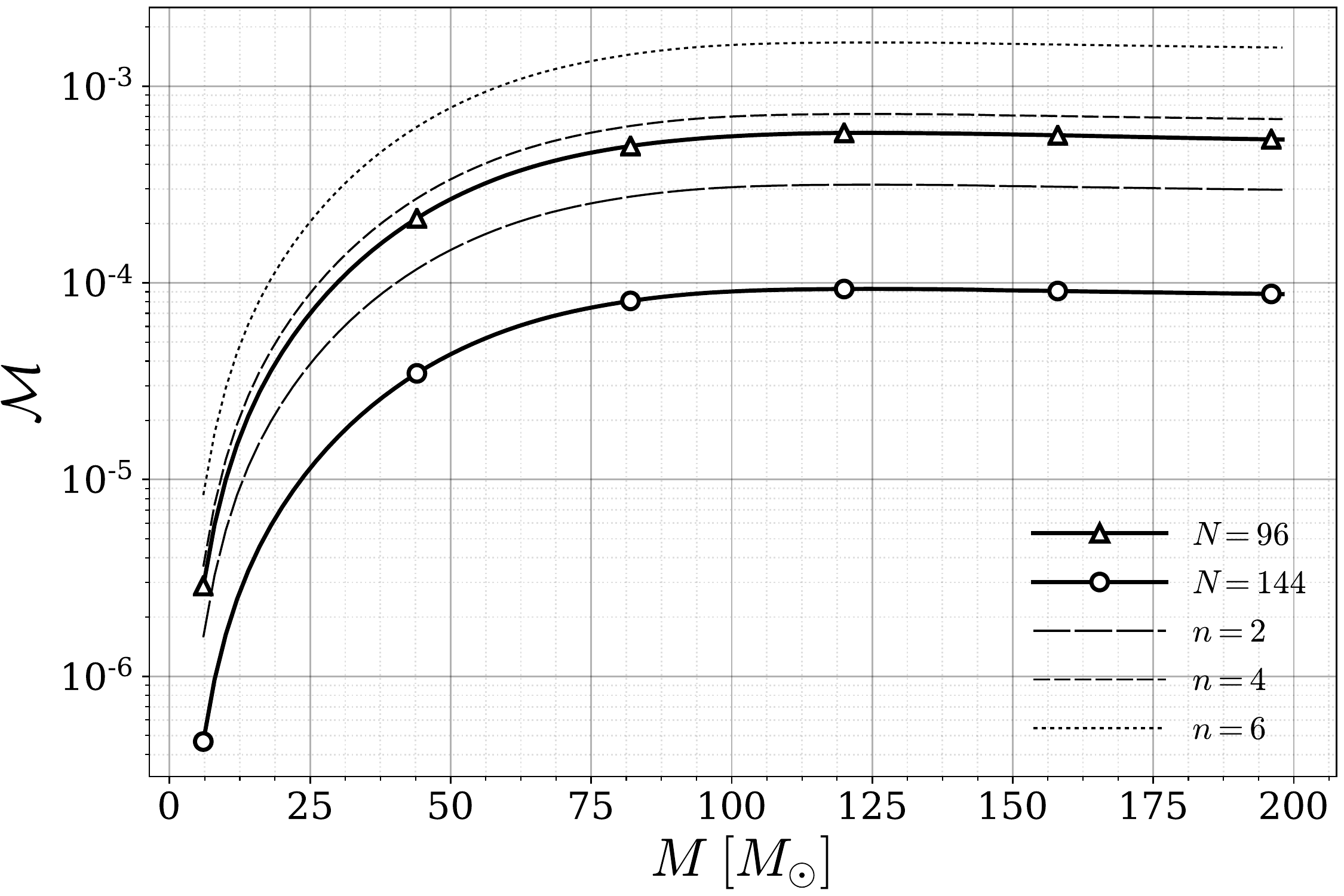}
   \caption{Mismatches demonstrating fourth order convergence of the BAM code with respect to resolution. The solid lines show the calculated mismatch while the dashed lines show the predicted mismatch for varying convergence order. The mismatch was calculated with respect to the $N=120$ resolution run. This is for the $q=8$, $\chi=0.8$, $\theta_\text{LS}=150^\circ$ configuration.}
   \label{fig: resolution convergence}
\end{figure}

That the medium resolution does not lie quite within the convergence regime is demonstrated clearly in Fig.~\ref{fig: resolution convergence}, 
where we show the mismatch between the medium and high and the high and very high resolutions using 
Eq.~(\ref{eqn: mismatch ratio}) for varying convergence order. 
From this analysis it is clear that the mismatch is closest to being fourth-order 
convergent. This analysis could only be done for the case ($q=8$, $\chi=0.8$, $\theta_\text{LS}=150^\circ$) since this is the only case for 
which we have the very high resolution run. 

\begin{figure*}[htbp]
   \centering
   \includegraphics[width=\textwidth]{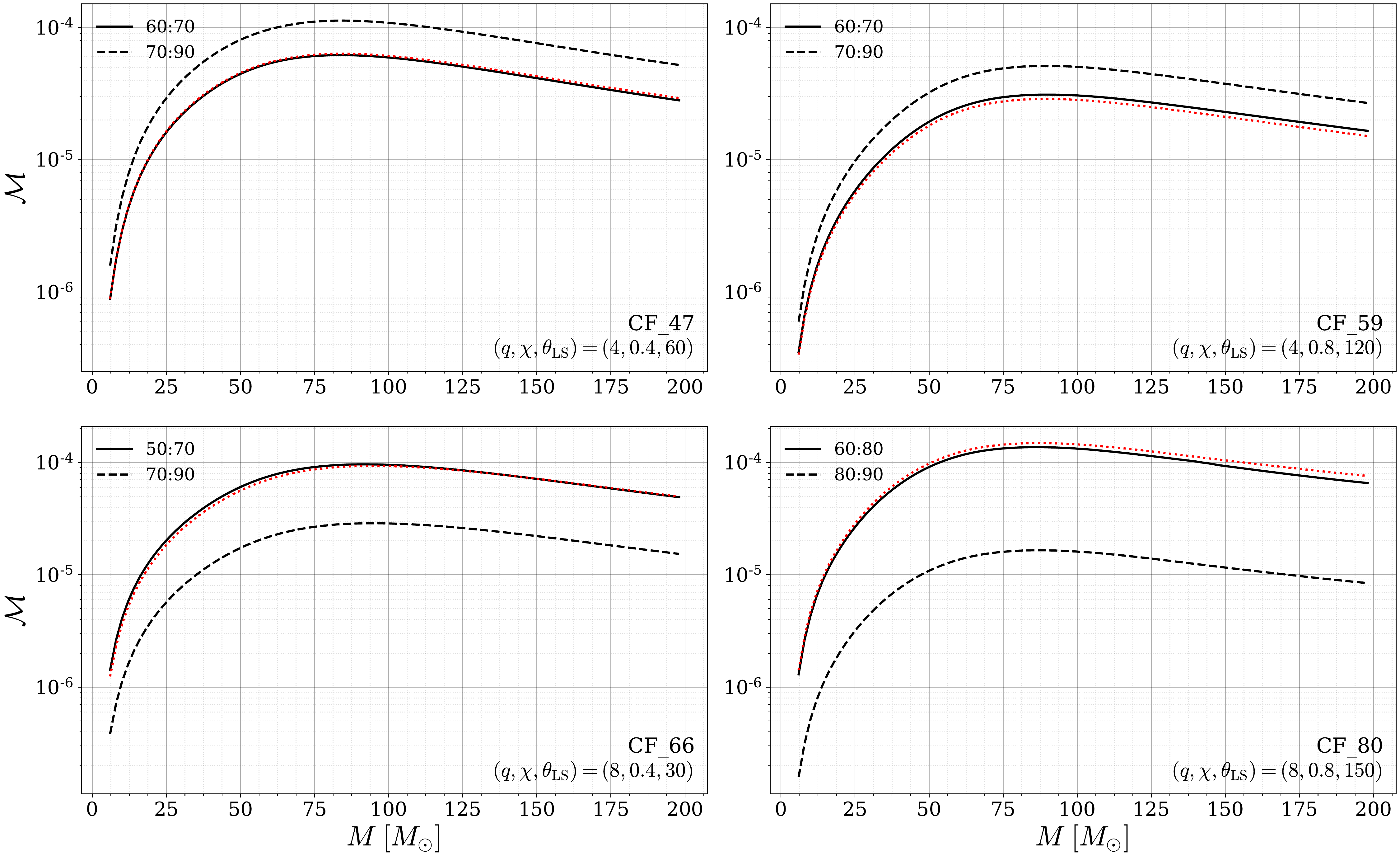}
   \caption{Mismatches demonstrating first order convergence of the BAM code with respect to extraction radius. The black lines indicated in the legend show the calculated mismatch, while the dotted red line shows the predicted mismatch for the pair of waveforms indicated by the solid line based on the mismatch indicated by the dashed line, assuming first order convergence.
   }
   \label{fig: extraction radius convergence}
\end{figure*}

Conversely, since it seems reasonable to assume the waveforms extracted at varying extraction radii mostly lie within the convergence regime, we calculated the ratio of the mismatch between each of the pairs of waveforms from different extraction radii using Eq.~(\ref{eqn: mismatch ratio}) for varying convergence order. For each of the four configurations we investigated it was found that the results were most consistent with first order convergence. This is demonstrated in Fig.~\ref{fig: extraction radius convergence}, where the solid lines show the calculated mismatch between two waveforms of different extraction radii and the dotted red line shows the expected value of the match for first order convergence. 

Not all the waveforms from the different extraction radii show perfect convergence for every configuration. The mismatch between $R_\text{ext}=80$ and $R_\text{ext}=90$ often does not follow the trend-- we expect this is because the mismatch between these waveforms is so small $\left(\mathcal{O}\left(10^{-6}\right)\right)$ that it is very sensitive to any data processing performed in the course of calculating the match. The mismatch between $R_\text{ext}=50$ and $R_\text{ext}=90$ also often does not follow the trend and we do not expect it to hold for small extraction radii.

The convergence order calculated using this method agrees with the estimate calculated in the previous section; the code is approximately fourth order convergent with respect to resolution and first order convergent with respect to the extraction radius. 

\subsubsection{Extrapolation}

\begin{figure*}[htbp]
   \centering
   \includegraphics[width=\textwidth]{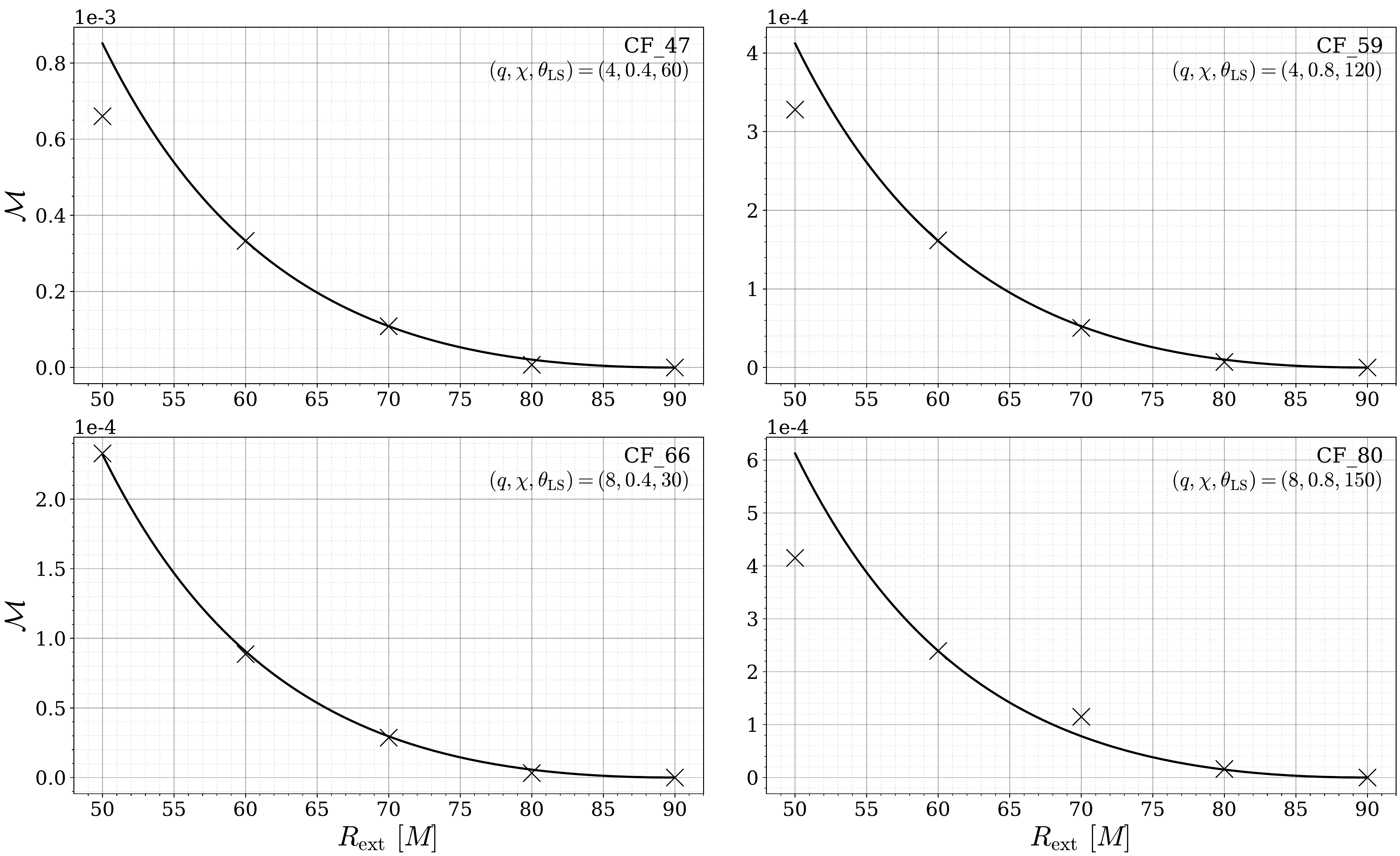}
   \caption{Variation of the mismatch with extraction radius for a system with a total mass of 100M. The mismatches shown are with respect to 
   the waveform extracted at $R_\text{ext}=90$M. The line shows the fit based on Eq.~(\ref{eqn: convergence order}) assuming first order 
   convergence. It is consistent with all extraction radii except at $R_\text{ext}=50$M.}
   \label{fig: fit at 100M}
\end{figure*}

\begin{figure}[htbp]
   \centering
   \includegraphics[width=\columnwidth]{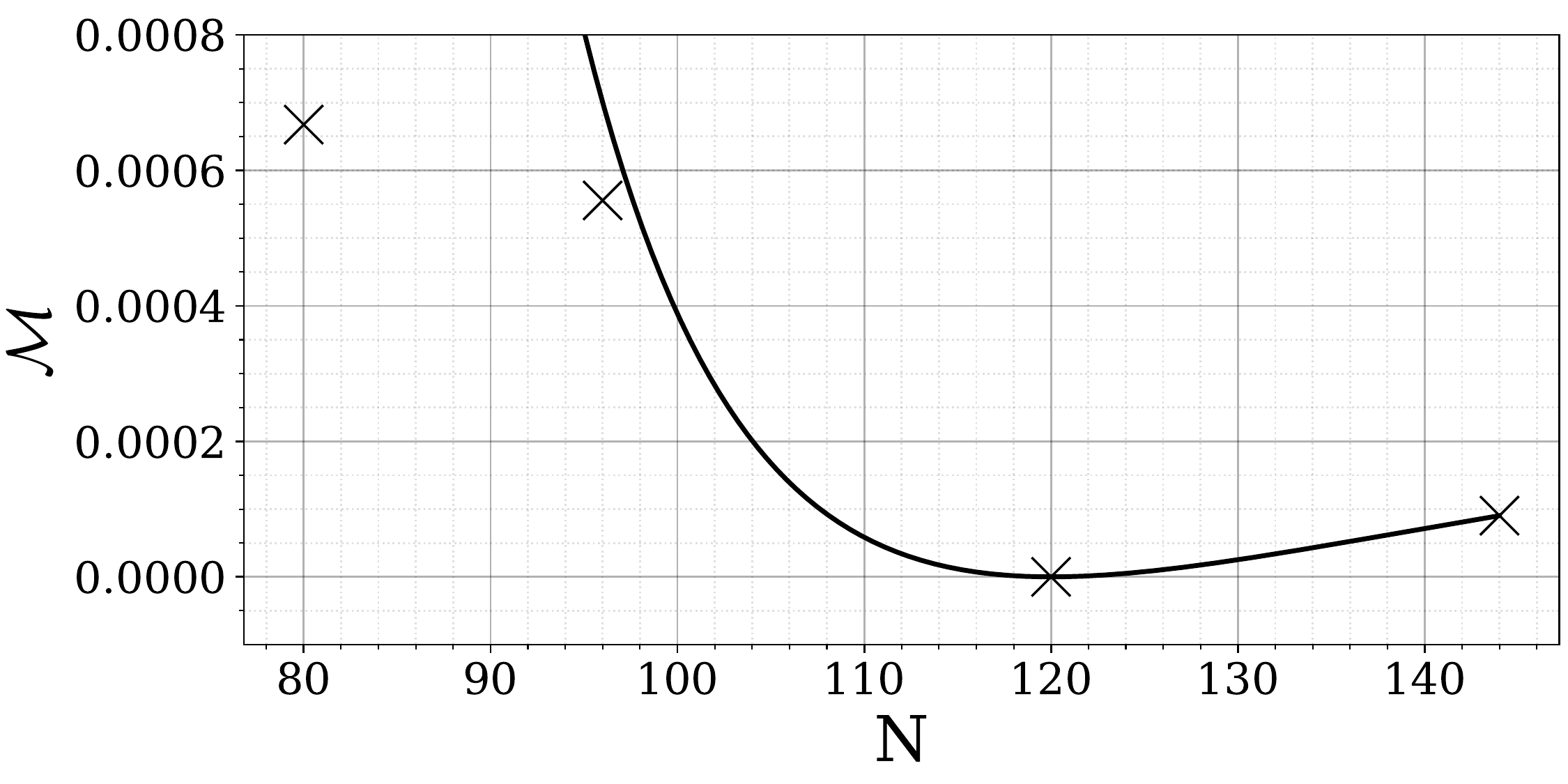}
   \caption{Variation of the mismatch with resolution for a $q=8$, $\chi=0.8$, $\theta_\text{LS}=150^\circ$ system with a total mass of 100M. 
   The mismatches shown are with respect to the medium resolution run and so the mismatch is zero at $N=120$. The line shows the relation 
   in Eq.~(\ref{eqn: convergence relation}) with $\kappa_\text{res}$ assuming fourth order convergence calculated using 
   Eq.~(\ref{eqn: kappa_res}) and is consistent with all resolutions except the lowest one at $N=80$. }
   \label{fig: resolution fit at 100M}
\end{figure}

\begin{figure}[htbp]
   \centering
   \includegraphics[width=\columnwidth]{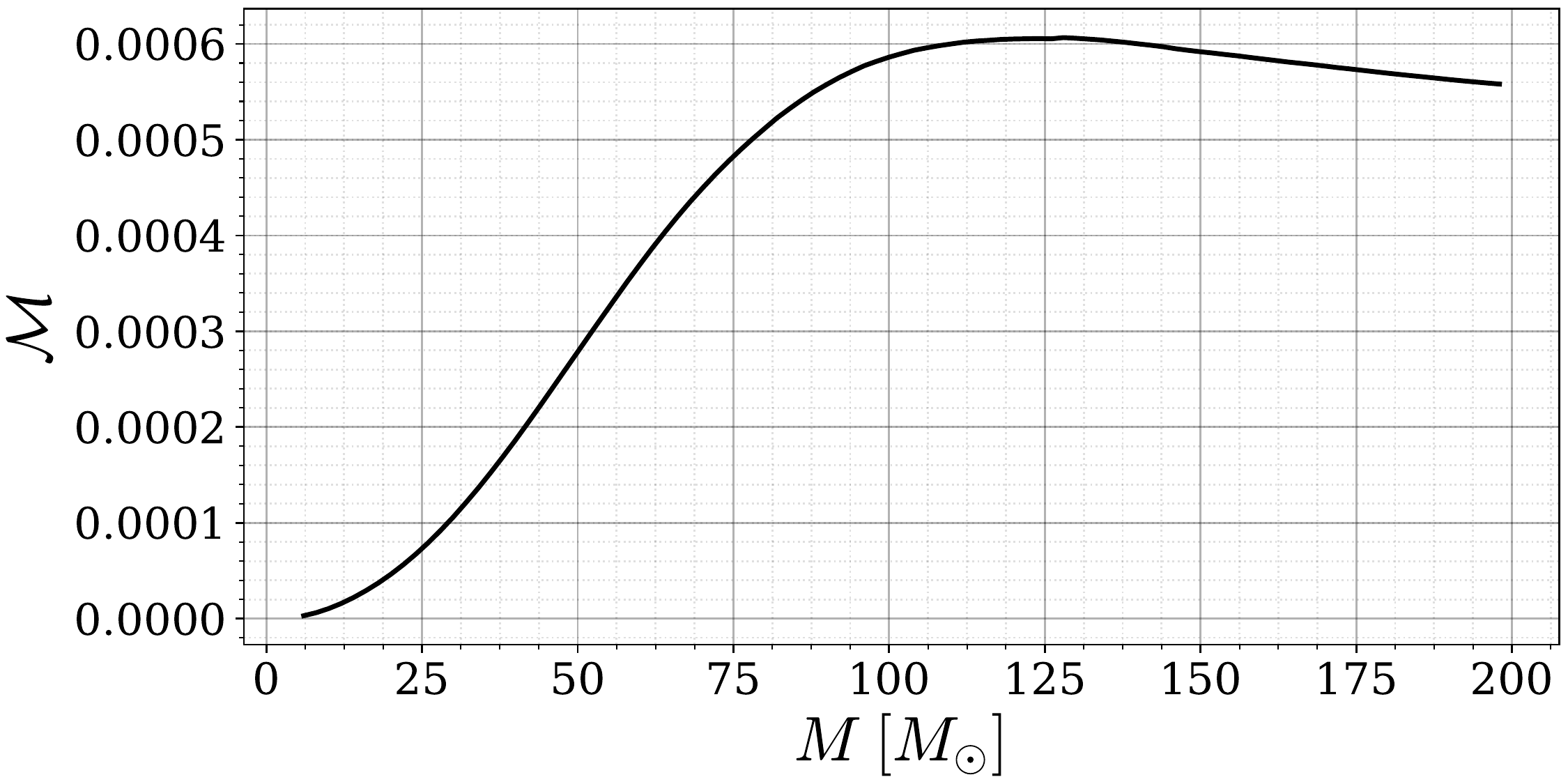}
   \caption{Projected mismatch between a waveform extracted at a resolution of $N = 96$ and one that is infinitely well resolved.}
   \label{fig: infinite resolution}
\end{figure}

Having verified the convergence order of the code $n$, we can calculate the convergence relation of the mismatches shown in 
Figs.~\ref{fig: resolution matches} and \ref{fig: extraction radius matches}. 
We first look at the mismatch due to the finite resolution of the simulation. We assume that the two highest resolution simulations we performed ($N=120$ and $N=144$) lie in the convergence regime but we know the two lower resolution simulations do not. Assuming fourth order convergence, we use Eq.~(\ref{eqn: convergence relation}) to calculate $\kappa_\text{res}$ using
\begin{align} \label{eqn: kappa_res}
   \kappa_\text{res} = {}& 
   \frac{\mathcal{M}\left(\Delta_{144}:\Delta_{120}\right)}{\left(144^4 - 120^4\right)^2}.
\end{align}
From Fig.~\ref{fig: resolution fit at 100M}, which shows the convergence relation for the mismatches calculated for a system with total mass 
100M$_\odot$, we can see that this appears to be a reasonable assumption. From $\kappa_\text{res}$ we can estimate the mismatch 
between the high or very high resolution waveforms with an infinitely well resolved waveform. However, we want to know the mismatch for 
the medium resolution runs since this is the resolution that was used to perform the simulations for the catalogue of waveforms presented 
in Tabs.~\ref{table:metadata} and~\ref{table:metadata-2}. Since this resolution does not lie in the convergence regime (and the phase error 
does not improve monotonically from the medium resolution to the high and very high resolution waveforms) we cannot simply use the 
calculated convergence relation in order to estimate the mismatch between a waveform at this resolution and the ``true'' waveform. 
Instead we use the formula given in Eq.~(\ref{eqn: adding matches}) 
to add the mismatch between the medium resolution and the very high resolution waveforms to the mismatch between the very high resolution waveform and the ``true'' waveform:
\begin{align} 
   \mathcal{M}\left(\Delta_{96}:\Delta_\infty\right) = {}& 
   \left( \sqrt{\mathcal{M}\left(\Delta_{96}:\Delta_{144}\right)} + \right. \nonumber \\
   {}& \left. \qquad \sqrt{\mathcal{M}\left(\Delta_{144}:\Delta_\infty\right)} \right)^2 \nonumber \\
   = {}& \left( \sqrt{\mathcal{M}\left(\Delta_{96}:\Delta_{144}\right)} + 
   \sqrt{\frac{\kappa_\text{res}}{144^4}} \right)^2.
\end{align}
The result of this extrapolation procedure is shown in Fig.~\ref{fig: infinite resolution}. We could only perform this calculation for the case 
($q=8$, $\chi=0.8$, $\theta_\text{LS}=150^\circ$) since this is the only case for which we have the very high resolution run. However, from 
Fig.~\ref{fig: resolution matches} we can see that the mismatch between the medium and high resolution runs is the worst for this case, 
so this estimate should give an upper bound for the mismatch between the medium resolution run and the ``true'' waveform. 
Fig.~\ref{fig: infinite resolution} shows the projected mismatch between a medium resolution waveform and one that is infinitely well 
resolved for a range of total masses. The maximum mismatch between a medium resolution waveform and an infinitely well resolved 
one is $6.0\times10^{-4}$.

\begin{figure}[t]
   \centering
   \includegraphics[width=\columnwidth]{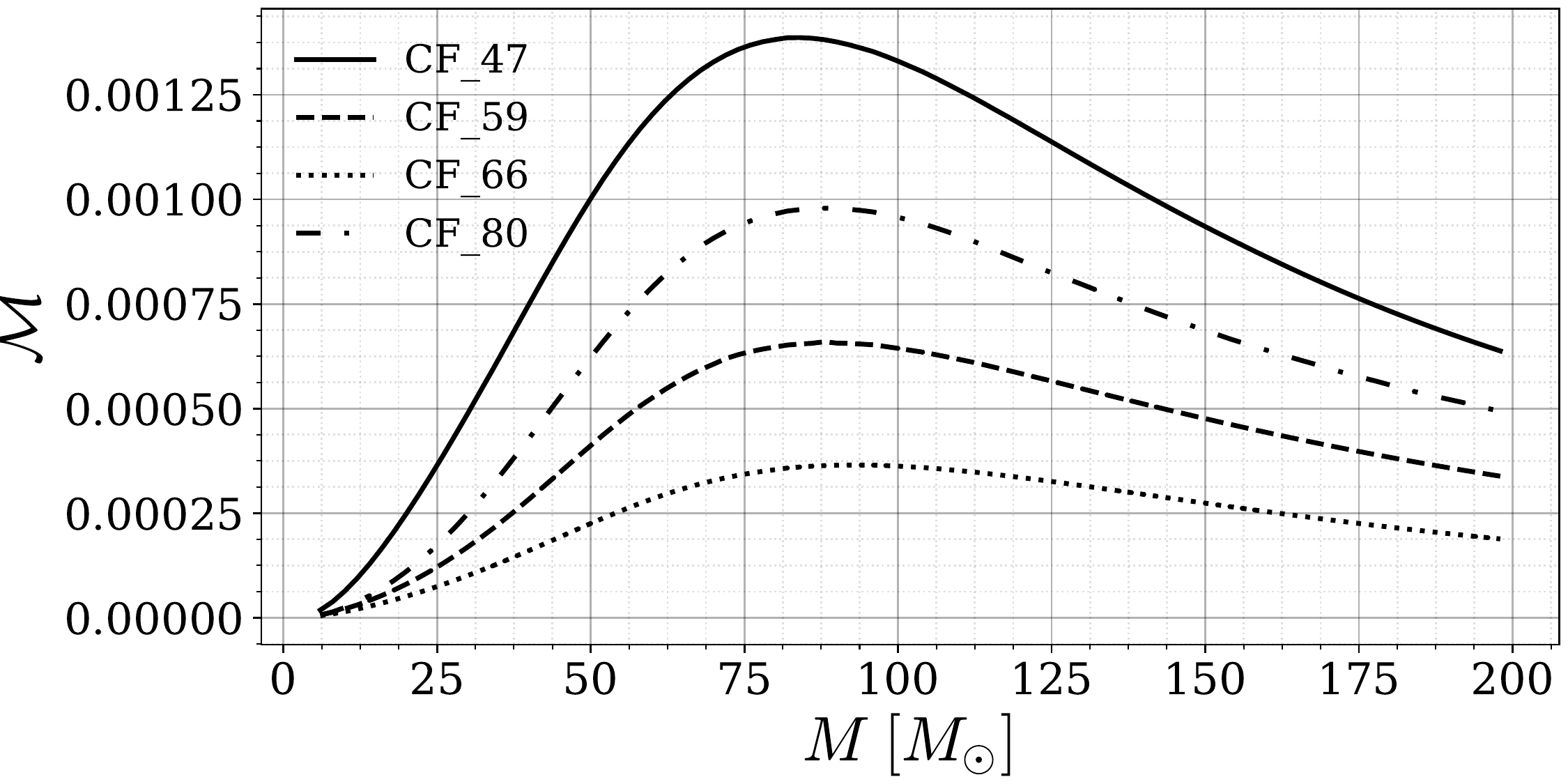}
   \caption{Projected mismatch between a waveform extracted at a radius of $R_\text{ext} = 90$M and one extracted infinitely far away.}
   \label{fig: infinitely far away}
\end{figure}

We next examine the mismatch due to the finite distance from the source at which the waveform is extracted. To calculate the first order 
convergence relation with respect to the extraction radius, we performed a fit through each of the mismatches which were found to follow 
the convergence relation. This is demonstrated for mismatches between waveforms of different extraction radii and the waveform at 
$R_\text{ext}=90$M in Fig.~\ref{fig: fit at 100M}, for a system with total mass 100M$_\odot$. This fit gives the value of $\kappa_\text{ext}$ 
for every value of the total mass of the system. From this we can calculate the mismatch between the waveform at $R\text{ext}=90$M 
and the ``true'' waveform from $\mathcal{M}\left(\Delta_{90}:\Delta_\infty\right) = \frac{\kappa_\text{ext}}{90^2}$. 

The mismatch between the waveform extracted at $R_\text{ext}=90$M and the ``true'' waveform is shown in Fig.~\ref{fig: infinitely far away}. 
The configuration that gives the greatest mismatch is $q=4$, $\chi=0.4$, $\theta_\text{LS}=60$ since, as noted above, this simulation 
was much longer than the others and so has greater opportunity to accumulate phase error between the two waveforms. The maximum 
mismatch between a waveform at $R_\text{ext}=90$ and at $R_\text{ext}\rightarrow\infty$  is taken to be $1.4\times10^{-3}$.

\begin{figure*}[t]
	\includegraphics[width=0.9\textwidth]{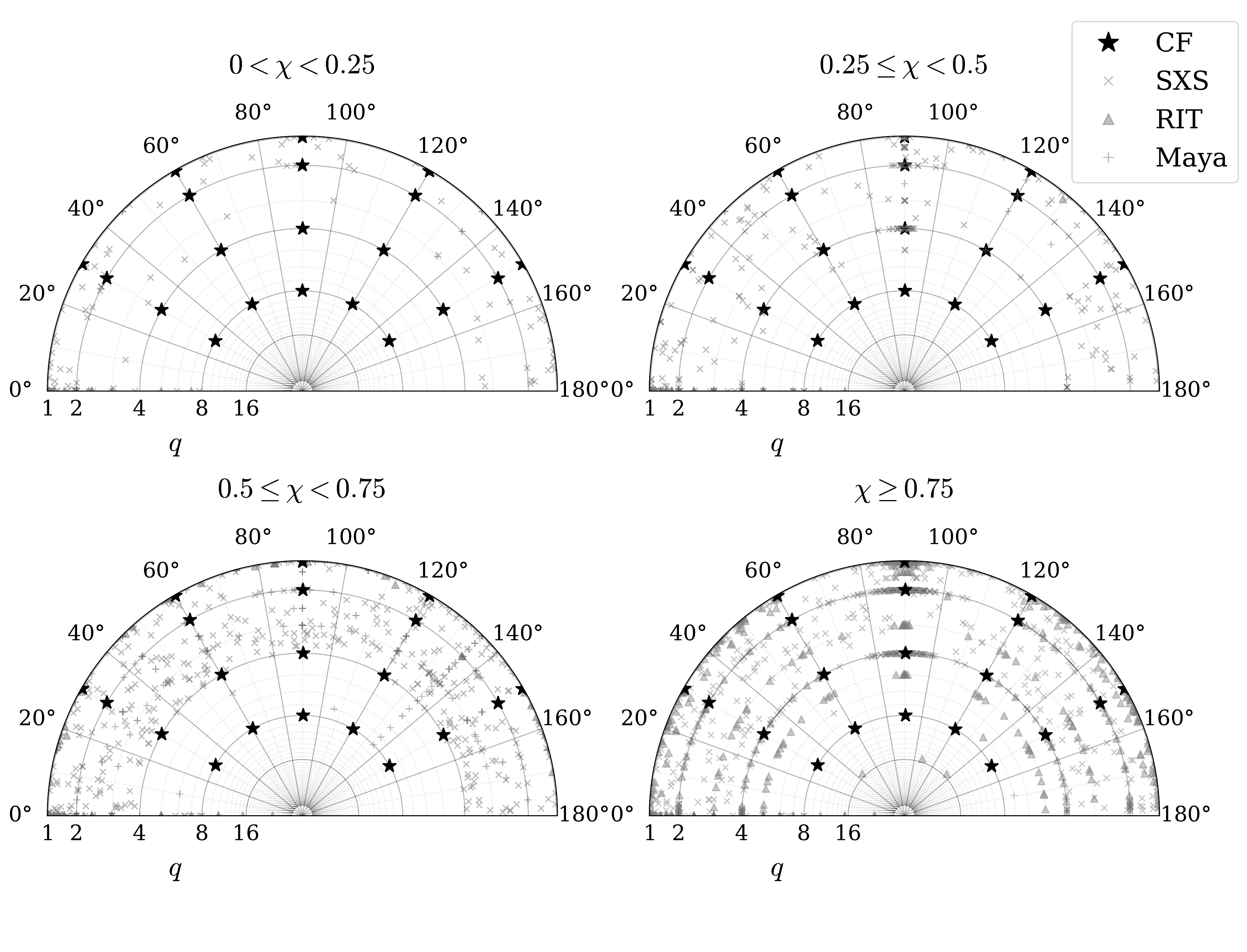}
	\vspace{-3.5em}
	\caption{Comparison between the parameters of the new BBH simulations
	presented here (CF) and the existing BBH simulations in the the SXS,
	RIT and Maya catalogues. The \emph{Top left} spin disk shows
	simulations with spin on the larger black hole $0 < \chi < 0.25$,
	\emph{Top right} $0.25 \leq \chi < 0.5$, \emph{Bottom left} $0.5 \leq \chi < 0.75$
        and \emph{Bottom right} $\chi \geq 0.75$. The radius of each disk shows the mass ratio of the binary
	and the orientation shows the spin tilt angle of the larger black hole.
        Spin tilt angles of $90^{\circ}$ means that the spin vector lies in the binary's orbital plane.}
	\label{figure:bam_spec_comparison}
\end{figure*}

We estimate the mismatch between our medium resolution waveform extracted at $R_\text{ext} = 90$M and the true waveform using 
Eq.~(\ref{eqn: complete mismatch}).
A conservative estimate of the mismatch between a waveform extracted at a finite distance of 90M from the source for a simulation performed with a grid spacing $d=0.0104$ and the theoretical `analytical' solution is therefore $3.83\times10^{-3}$. This provides a limit on the mismatch error of the waveforms presented in this catalogue of 0.004, or 0.4\%.

\section{Catalogue comparison}
\label{sec:comparison}

A number of numerical relativity groups have started building larger and more comprehensive catalogues
that span a growing region of the parameter space. At time of publication, there are a total of 4,352 publicly available BBH simulations in a combination of the Simulating eXtreme Spacetimes (SXS) Collaboration~\cite{Mroue:2013xna, Boyle:2019kee} and the Rochester Institute of Technology (RIT)~\cite{Healy:2017psd, Healy:2019jyf, Healy:2020vre, Healy:2022wdn} and Maya catalogues~\cite{Jani:2016wkt}. 

The SXS collaboration has produced the largest catalogue to date with 2,019 BBH simulations spanning $1 \leq q \leq 10$ and $0 \leq \chi \leq 1$. The RIT catalogue contains 1,881 BBH simulations covering $1 \leq q \leq 128$ and $0 \leq \chi \leq 0.99$ and the Maya catalogue contains 452 unique BBH waveforms from more than 600 BBH simulations ranging between $1 \leq q \leq 15$ and $0 \leq \chi \leq 0.8$. Unlike the simulations presented here, the SXS, RIT and Maya catalogues all contain simulations 
where the individual black hole spins can be zero or perfectly aligned/anti-aligned with the orbital angular momentum. Considering only the precessing parameter space, the SXS collaboration has produced 1,429 simulations spanning $1 \leq q \leq 6$ and $0 \leq \chi \leq 0.99$, the RIT catalogue contains $561$ simulations covering $1 \leq q \leq 15$ and $0 \leq \chi \leq 0.99$ and the Maya catalogue contains 315 waveforms ranging between $1 \leq q \leq 8$ and $0.1 \leq \chi \leq 0.8$.

Fig.~\ref{figure:bam_spec_comparison} compares the parameter space coverage of
the simulations presented here with the existing non-zero-spin simulations included in the SXS, RIT and Maya catalogues over the mass ratio and larger black hole spin tilt angle and spin magnitude parameter space. We see that although the existing catalogues provide good coverage for high black hole spins $\chi \geq 0.5$ and near equal mass ratios, there is a dearth of existing precessing simulations for low black hole spins $\chi < 0.25$ and unequal mass ratios $q \gtrsim 4$.

Recent gravitational wave observations~\cite[see e.g.][]{LIGOScientific:2020stg, LIGOScientific:2020zkf} have shown a need for 
BBH simulations in this low black hole spin and unequal mass ratio region of the parameter space in order to build reliable and 
accurate waveform models for use in Bayesian inference. The uniform coverage of the single-spin space up the $q=8$ has made it 
possible to construct an accurate generic precessing-binary model for future observations~\cite{Hamilton:2021pkf}. 
Most astrophysical models suggest
that BBH at larger mass ratios will be rare (e.g., Ref.~\cite{LIGOScientific:2021psn}), but given that there has been one observation 
to date at $q\sim10$~\cite{LIGOScientific:2020zkf} extension of this parameter-space coverage to yet higher mass ratios will be necessary in the future.

\section{Discussion}
\label{sec:discussion}

We have produced a catalogue containing 80 waveforms from single-spin precessing systems with mass ratios up to $q=8$, dimensionless spin magnitudes up to $\chi=0.8$ and a range of spin inclination angles. In all cases the spin was placed on the larger black hole. We estimate our uncertainty in the masses of the initial black holes to be $\mathcal{O}(0.05\%)$. We estimate the uncertainty in the initial spin magnitude to be $\mathcal{O}(10^{-3})$ while the uncertainty in the initial spin inclination is $\mathcal{O}(1^\circ)$. Similarly, we obtain estimates of the uncertainty of the remnant properties reported in this paper. We find the final mass has an uncertainty of $5\times10^{-4}$, while the final spin magnitude is accurate to within $5\times10^{-3}$.

The starting frequency of the simulations was chosen such that the simulations were all around a similar length ($\sim2000M$) in order to limit the dephasing in the waveform and thus ensure sufficient accuracy throughout the evolution of the binary. We performed a careful analysis of the errors due to the finite resolution of the simulations and due to the finite distance from the source at which the waveforms were extracted. From this we were able to provide a conservative estimate of the mismatch uncertainty of our waveforms of 0.4\%.

The catalogue presented here is sufficient to capture a wide range of single-spin precession effects. Most notably, the systems contained within it have a non-zero opening angle of the precession cone ranging from $\sim1^\circ$ to $\sim115^\circ$ radians. The cases with the largest opening angles display the most dominant precession effects. In particular, for initial configurations with mass ratios $q\sim8$, high spins and large spin inclination angles, the final spin will be in the opposite direction to the binary's angular momentum prior to merger, thus producing a ``negative'' final spin. 
The majority of precessing simulations in other catalogues do not extend beyond $q=5$ so consequently, this region of parameter space is poorly covered by \nr{} simulations. Indeed, in this catalogue, despite having 20 precessing simulations at $q=8$ we see only two cases where we have a negative final spin (\texttt{CF\_75} and \texttt{CF\_80}). The phenomenology of this region has therefore not yet been thoroughly explored and a more detailed study is planned for future work.
Finally, we also see a wide range of recoil velocities for the configurations included in this catalogue, with the highest values seen for equal mass systems. We also see the greatest range of values for equal mass systems, depending on the in-plane spin angle. For $q=8$ systems we see much lower values in general across all cases.

While this catalogue was sufficient to produce the first \imr{} model of precessing systems tuned to \nr{}, \texttt{PhenomPNR}, it will need to be greatly expanded in order to meet modelling requirements of future gravitational wave observations. 
Existing catalogues (such as the SXS, Maya and RIT catalogues) provide a comprehensive coverage of the two-spin precessing parameter space up to $q=4$. This catalogue provides a systematic coverage of the single-spin precessing parameter space up to $q=8$. However, while it uses a consistent in-plane spin direction at the starting frequency (the initial configurations all have the in-plane spin component along the binary's separation vector), these will translate into quite different spin directions at merger. Consequently, for any modelling that includes effects due to the in-plane spin direction, this catalogue contains an incomplete and possibly random sampling of points.

There are many directions in which this catalogue can be expanded to: include higher mass ratios, comprehensively cover rotations of the in-plane spin component, include two-spin systems, produce longer and more accurate waveforms and include binaries on eccentric orbits. Since the production of these simulations are expensive (the catalogue presented here is estimated to have required around 25 million CPU hours in total) and generic modelling at higher mass ratios and for longer waveforms is not a completely solved problem, it is an open question as to which direction in parameter space is most urgent. 

Assuming the mass ratio distribution reported in~\cite{LIGOScientific:2021psn}, we estimate that only 1.3\% of observed binaries will have $q>8$. 
This is supported by gravitational wave detections so far since, out of the 90 binaries reported by the LVK 
collaborations~\cite{LIGOScientific:2021djp}, only one has been found to have a mass ratio clearly greater than 8~\cite{LIGOScientific:2020zkf}. 
Similarly, from the production of \texttt{PhenomPNR}, we know that we will require longer waveforms for binaries with higher mass ratios and spins, particularly those with a spin inclination angle of $\theta_\text{LS} > 90^\circ$, since inaccuracies in \pn{} estimates of precession effects become more appreciable towards lower frequencies in this region of the parameter space. For the existing catalogue, the cases \texttt{CF\_79} and \texttt{CF\_80} (at $q=8$ and $\chi=0.8$) are already sufficiently short to limit model accuracy and we anticipate this will continue for decreasing spin magnitude as we go to higher mass ratios~\cite{Hamilton:2021pkf}. 
This is therefore a smaller fraction of binaries than the simple requirement to extend to high mass ratios.
Similarly, we expect to see two-spin effects in signals with SNRs greater than 100~\cite{Purrer:2015nkh,Khan:2019kot}. 
Taking the detection threshold to be SNR 10, then we expect to be able to identify two-spin effects in ~0.1\% of detections. 

We have not considered eccentricity here, but eccentric modelling and \nr{} simulations have been discussed in other works~\cite{Boyle:2019kee, Ramos-Buades:2018azo, Ramos-Buades:2019uvh, Huerta:2017kez, Hinder:2017sxy, Nagar:2021gss, Islam:2021mha, Liu:2021pkr}.

From this, we conclude that the most urgent extension is required to systems with higher mass ratios. This is closely followed by a systematic sampling that will explore the most dominant physical effects of two-spin systems (such as those that will impact the opening angle of the precession cone at merger) or the in-plane direction (such as the recoil velocity).

\section{Acknowledgements}

We would like to thank Steve Fairhurst, Frank Ohme, Vivien Raymond for many useful discussions; Kieran Philips for extensive 
optimisation of the BAM code that resulted in a greater than 30\% increase in speed and a much lower memory footprint, and also
for his work in testing the code on the Oracle Cloud Infrastructure (OCI); Paul Hopkins for wide-ranging cluster support and also
assistance in setting up OCI runs; Phil Bates at Oracle for initiating the OCI work and to Phil and his team for extensive technical
support and guidance. 

The authors were supported in part by Science and Technology Facilities Council (STFC) grant ST/V00154X/1 and European Research Council (ERC)
Consolidator Grant 647839.
E. Hamilton was supported in part by Swiss National Science Foundation (SNSF) grant IZCOZ0-189876. 
L. London was supported at Massachusetts Institute of Technology~(MIT) by National Science Foundation Grant No. PHY-1707549 
as well as support from MIT’s School of Science and Department of Physics.
A. Vano-Vinuales thanks FCT for financial support through Project No. UIDB/00099/2020.

%\mdh{Those grant acknowledgements are copied from the PNR paper, so some updates are needed!}

This work used the DiRAC@Durham facility managed by the Institute for Computational Cosmology on behalf of the STFC DiRAC HPC Facility (www.dirac.ac.uk). The equipment was funded by BEIS capital funding via STFC capital grants ST/P002293/1, ST/R002371/1 and ST/S002502/1, Durham University and STFC operations grant ST/R000832/1. DiRAC is part of the National e-Infrastructure.
Additionally, this research was undertaken using the supercomputing facilities at Cardiff University operated by Advanced Research Computing at Cardiff (ARCCA) on behalf of the Cardiff Supercomputing Facility and the HPC Wales and Supercomputing Wales (SCW) projects. We acknowledge the support of the latter, which is part-funded by the European Regional Development Fund (ERDF) via the Welsh Government.
This work was also supported in part by Oracle Cloud credits and related resources provided by the Oracle for Research program.

\begin{appendices}

\renewcommand{\thesubsection}{\arabic{subsection}}

\section{Iterative initial data construction for single spin precessing systems}
\label{app:initial-data-method}

The initial data construction method of~\cite{Husa:2015iqa} for aligned-spin
systems is extended to precessing systems using the follow iterative brute-force 
algorithm. A sequence of input parameter sets
\begin{equation}
\bm{\theta}_i \equiv
\left(q_i^D, \bm{S}_{1,i}^D, \bm{S}_{2,i}^D\right) =
\left(q, \bm{0}, S_2\hat{\bm{t}}_i\right),
\end{equation}
will be iteratively refined, defined at a user-specified
separation of $D_{\text{start}}$ and used as initial conditions for the EOB
solver. The single-spin orienation unit vector $\hat{\bm{t}}$ can be expressed in terms of angles
$\theta$ and $\phi$
\begin{equation}
\hat{\bm{t}_i} = \big(\cos(\phi_i)\cos(\theta_i),\sin(\phi_i)\cos(\theta_i),\cos(\theta_i)\big)
\end{equation}
The EOB simulations are started with both component black holes placed on the
x-axis and the orbital angular momentum parallel to the $z$-axis. For this work
the black holes are placed $\sim 40M$ apart. Each successive set of parameters
are chosen based on the EOB evolution of previous parameter sets in the
sequence. This is repeated until the EOB evolution results in the required
parameters $(q, \bm{S}_1, \bm{S}_2)$ at $M \Omega_\mathrm{orb}$ using the
following algorithm

\begin{enumerate}

\item[\textbf{(1)} ] \textbf{Initial candidate parameters at $D_{\text{start}}$ ($n=0$)} \\
$\phi_0$ is chosen to be the target azimuthal spin angle $\phi$. The EOB solver
is then run until $M \Omega_\mathrm{orb}$ is reached and the azimuthal spin
angle at that time is recorded as $\phi_{\omega,0}$. The EOB spin dynamics are
explored in the region around $M \Omega_\mathrm{orb}$ to find the closest time
when the spin angle is equal to $\phi$. The difference in frequency $\Delta
M\omega_0$ between this time and $M \Omega_\mathrm{orb}$ is recorded. If
$|\Delta M\omega_0| \leqslant M\omega_\mathrm{tol}$ where
$M\omega_\mathrm{tol}$ is a user specified tolerance, then the algorithm stops.
For the initial data generated in this work the percentage error tolerance of
the orbital frequency is specified to be $M\omega_\mathrm{tol} = 1 \%$. If
$|\Delta M\omega_0| > M\omega_\mathrm{tol}$ then proceed to the next step.

\item[\textbf{(2)} ] \textbf{Second candidate parameters at $D_{\text{start}}$ ($n=1$)} \\
$\phi_1$ is chosen to be the difference between the target azimuthal spin angle
and the azimuthal spin angle of the EOB spin dynamics at $M
\omega_\mathrm{orb}$, given by $\phi_c \equiv \phi - \phi_{\omega,0}$. The EOB
spin dynamics are explored again, recording $\phi_{\omega,1}$ at $M
\omega_\mathrm{orb}$, and calculating $|\Delta M\omega_1|$ as in the same way
as $|\Delta M\omega_0|$ in the previous step. If $|\Delta M\omega_1| \leqslant
M\omega_\mathrm{tol}$ the algorithm stops otherwise proceed to the next step.

\item[\textbf{(3)} ] \textbf{Third candidate parameters at $D_{\text{start}}$ ($n=2$)} \\
Set $\Delta\phi$ to be $10^{\circ}$ if $||\Delta M\omega_1| -
M\omega_\mathrm{tol}| > M\omega_\mathrm{tol}/2$ otherwise set $\Delta\phi$ to be
$5^{\circ}$. $\phi_2$ is chosen to be the target azimuthal spin angle $\phi_c +
\Delta\phi$. $\phi_{\omega,2}$ and $|\Delta M\omega_2|$ are calculated in the
same way as previous steps. If $|\Delta M\omega_2| \leqslant
M\omega_\mathrm{tol}$ the algorithm stops otherwise proceed to the next step.

\item[\textbf{(4)} ] \textbf{Further candidate parameters at $D_{\text{start}}$ ($n>2$)} \\
Set $\Delta\phi$ to be $10^{\circ}$ if $||\Delta M\omega_{n-1}| -
M\omega_\mathrm{tol}| > M\omega_\mathrm{tol}/2$ otherwise set $\Delta\phi$ to be
$5^{\circ}$. If $\phi_{\omega,2} > \phi_{\omega,1}$ this indicates that the
azimuthal spin angle is being rotated in the wrong direction. As such, if
$\phi_{\omega,2} > \phi_{\omega,1}$ then $\phi_n = \phi_c +
(2-n)\Delta\phi$, otherwise set $\phi_n = \phi_c + (n-1)\Delta\phi$. $|\Delta
M\omega_n|$ is calculated in the same way as previous steps. If $|\Delta
M\omega_n| \leqslant M\omega_\mathrm{tol}$ the algorithm stops otherwise repeat
this step until this inequality is satisfied.

\end{enumerate}

Once the required tolerance is met and the algorithm stops, the position,
linear momentum and spin of each black hole are taken from the EOB dynamics at
$M \Omega_\mathrm{orb}$ and used as input for the Bowen-York initial data
solver.

\section{NR simulations with cloud computing} 
\label{sec:cloud}

For this work we have run individual {\tt BAM} simulations on up to 512 processors, and these require fast inter-processor
connections to ensure that inter-processor communication is not the dominant limitation on the calculation speed.
This is typical for NR codes, and as such these are usually run on clusters that have been constructed primarily for
highly parallelised high-performance computing (HPC), such as the DiRAC Cosma clusters that were used for the majority 
of the runs presented here. An increasingly popular source of computing resources are cloud services. These have
historically been set up with large numbers of independent (high-throughput computing) applications in mind. 
However, recently some services have improved the speed of inter-processor communication, with the goal of 
making cloud computing services competitive for HPC.

As part of the NR work presented here, we investigated the performance of {\tt BAM} on the Oracle Cloud Infrastructure (OCI). 
We performed a series of experiments to determine the optimal performance we could achieve with the hardware 
available at the time (2018). These tests used a bespoke ``bare metal'' setup and ran at about 60\% of the speed on the DiRAC 
cosma5 cluster. 
(Note that since these tests were done, both the cosma clusters and the cores used at OCI have been superseded by newer
hardware.) 

We also completed a set of production simulations: these were the five NR simulations at mass-ratio $q=2$ and black-hole spin 
$S_2/m_2 = 0.6$ (\texttt{CF\_31}--\texttt{CF\_35}). Each run was performed on 128 cores and required approximately 
140,000 CPU hours. 

The production simulations used the ``cluster-in-the-cloud'' infrastructure~\cite{citc} to create container-based cluster instances
using OCI Terraform on AMD64 128-core BM.Standard.E2.64 nodes, and ran at an average 1.8 $M$/hr. Similar runs on 
cosma6 ran at about 3.7 $M$/hr on double the number of cores. These suggest that in terms of computational cost and
efficiency, cloud-based resources have the potential to be competitive to standard clusters.

\section{Convergence estimates and Richardson extrapolation}
\label{sec: mismatch-convergence-relationship}

\subsection{Richardson Extrapolation}
\label{sec: Richardson Extrapolation}

A quantity $q$ calculated at finite resolution or extraction radius can be given by 
\begin{align}
   q^{*} = {}& q\left(\Delta\right) + e_i \Delta^i,
\end{align}
where $\Delta$ is the expansion parameter ($\frac{1}{N}$ for resolution or $\frac{1}{R_\text{ext}}$ for extraction radius), $e$ is the finite order error and $i$ is the order at which the error contributes. In this paper the quantity $q$ we are considering is the waveform extracted from the numerical simulation. We therefore have that
\begin{align}\label{eqn: error expansion}
   q^{*} = {}& q\left(\Delta\right) + e_n \Delta^n + \mathcal{O}\left( \Delta^{n'} \right),
\end{align}
where $e_n \Delta^n$ is the leading order error contribution, $n$ is the convergence order of the simulation and $n'>n$.

Considering two waveforms computed using different expansion parameters $\Delta_1$ and $\Delta_2$, we can solve the two simultaneous equations that arise from Eq.~(\ref{eqn: error expansion}) to give
\begin{align}\label{eqn: accurate q*}
   q^{*} = {}& R\left(\Delta_1,\Delta_2\right) 
   + \mathcal{O}\left(\left(\frac{\Delta_1}{\Delta_2}\right)^{n'}\right),
\end{align}
where
\begin{align}\label{eqn: Richardson extrapolation}
   R\left(\Delta_1,\Delta_2\right) = {}&\frac{\left(
   \frac{\Delta_1}{\Delta_2}\right)^n q\left( \Delta_2 \right) - q\left( \Delta_1 \right)}
   {\left(\frac{\Delta_1}{\Delta_2}\right)^n - 1},
\end{align}
is the \emph{Richardson extrapolation}~\cite{doi:10.1098/rsta.1911.0009,Alcubierre:1138167} of $q\left(\Delta_1\right)$. $R\left(\Delta_1,\Delta_2\right)$ has a higher order error due to the truncation of the expansion in $\Delta$ than $q\left(\Delta_1\right)$. $R\left(\Delta_1,\Delta_2\right) - q\left(\Delta_1\right)$ gives the truncation error of the quantity $q$.

\subsection{Convergence}
\label{sec: convergence}

Considering now three waveforms computed with expansion parameters $\Delta_1 > \Delta_2 > \Delta_3$ we can eliminate $q^{*}$ in 
Eq.~(\ref{eqn: accurate q*}). Neglecting higher order error terms, the ratio of the difference between two sets of numerical waveforms with expansion parameter $\Delta_1 > \Delta_2 > \Delta_3$ is then given by
\begin{align}\label{eqn: convergence factor}
   \mathcal{C} = {}& 
   \frac{q\left(\Delta_1\right) - q\left(\Delta_2\right)}{q\left(\Delta_2\right) - q\left(\Delta_3\right)}
   = \frac{\Delta_1^n - \Delta_2^n}{\Delta_2^n - \Delta_3^n}.
\end{align}
This relation holds for features of a waveform, such as its amplitude and phase, but not for derived quantities such as the match.

To understand how the match between a set of waveforms in a convergence series varies, 
consider a detector response derived from gravitational wave strain solutions
of a finite difference approximation numerical relativity code. This can be
represented by a \emph{Richardson
expansion}~\cite{doi:10.1098/rsta.1911.0009,Alcubierre:1138167} as a
power series in an expansion parameter $\Delta$. Consider two detector
responses, $h_1$ and $h_2$, at two resolutions with expansion parameters
$\Delta_1$ and $\Delta_2$ respectively. As seen in Eq.~(\ref{eqn: error expansion}), 
for an $i^\mathrm{th}$ order accurate
finite difference method these can be represented by their truncated Richardson
expansions
\newcommand{\diffdel}{\Delta_2^n - \Delta_1^n}
\begin{align}
h_1(f) & = h(f) + e_n(f) \Delta_1^n, \label{eqn: h1}\\
h_2(f) & = h(f) + e_n(f) \Delta_2^n, \\
& = h_1(f) + e_n(f) (\diffdel),
\end{align}
where $h(f)$ is the detector response of the exact solution, and $e_i(f)$ are
the leading order error functions.

The match between these two detector responses can be expanded in the
expansion parameter $(\diffdel)$. Utilising linearity in the inner product, we have
\begin{widetext}
\begin{align}
M(h_1,h_2) & =
    \max_{\Theta}\left[ \frac{\left\langle h_1 \middle| h_2 \right\rangle}{\|h_1\|\|h_2\|} \right], \nonumber \\
& =
    \max_{\Theta}\left[ \frac{
        \|h_1\|^2 + (\diffdel)\left\langle h_1 \middle| e_n \right\rangle
    }{
        \|h_1\|\left(\|h_1\|^2 + 2(\diffdel)\left\langle h_1 \middle| e_n \right\rangle + (\diffdel)^2\|e_n\|^2 \right)^{1/2}
    } \right], \nonumber \\
& \approx
    \max_{\Theta}\left[
        1 - \frac{1}{2}\left( \frac{\|e_n\|^2}{\|h_1\|^2} - \left( \frac{\left\langle h_1 \middle| e_n \right\rangle}{\|h_1\|^2} \right)^2 \right) (\diffdel)^2
    \right], \nonumber \\
& \approx 1 - \kappa(\diffdel)^2, \label{eq:match-convergence}
\end{align}
\end{widetext}
where between the second and third lines we have performed a binomial expansion of the denominator and terms of higher order in the expansion parameter are dropped between steps. 
The constant coefficient $\kappa$ is defined as
\begin{equation}
\kappa \equiv \min_{\Theta} \left[ \frac{1}{2} \left(
    \frac{\|e_n\|^2}{\|h\|^2} - \frac{\left\langle h \middle| e_n \right\rangle^2  }{\|h\|^4}
\right)\right],
\end{equation}
where we have used Eq.~(\ref{eqn: h1}) to re-write $h_1$ in terms of $h$ and again neglected terms of higher order in the expansion parameter. $\kappa$ can be seen to be bounded below by zero from the Cauchy-Schwarz inequality.
The mismatch may then be approximated as
\begin{equation}
\label{eq:mismatch-convergence}
\mathcal{M}(h_1,h_2) =  \kappa(\diffdel)^2.
\end{equation}

It is important to note that the leading order expansion parameter terms in the
approximation Eq.~(\ref{eq:mismatch-convergence}) are quadratic in the expansion
parameters. In addition, while it is likely that the leading order coefficient
$\kappa$ cannot be calculated directly, it is independent of any resolution-specific 
expressions. As such $\kappa$ is constant within any convergence
series. This leads to the following two results, describing ratios of
mismatches in convergence series and the combination of mismatches in
convergence series,
\begin{align}
\frac{\mathcal{M}(h_1,h_2)}{\mathcal{M}(h_2,h_3)}
    & = \frac{(\Delta_1^n - \Delta_2^n)^2}{(\Delta_2^n - \Delta_3^n)^2}, \\ 
\mathcal{M}(h_1,h_3)
    & = \left( \sqrt{\mathcal{M}(h_1,h_2)} + \sqrt{\mathcal{M}(h_2,h_3)} \right)^2. \label{eqn: mismatch combo} 
\end{align}
Eq.~(\ref{eqn: mismatch combo}) holds generally, not just for the case of mismatches between waveforms in a convergence series. 
This is shown in  Appendix~\ref{sec: mismatch addition}.

\section{Addition of mismatches}\label{sec: mismatch addition}

Consider three waveforms, $h_1$, $h_2$, $h_3$, which are all normalised, $|h_1| = |h_2| = |h_3| = 1$. If
$\langle h_1 | h_2 \rangle = A$, and $\langle h_2 | h_3 \rangle = B$, we would like to estimate an upper bound on
$C = \langle h_1 | h_3 \rangle$.

We write each waveform with reference to one of the others. Choose $h_2$, since that is our ``middle'' waveform. We can
write, \begin{eqnarray}
h_1 & = & A h_2 + \sqrt{1-A^2} h_{2\perp}, \\
h_3 & = & B h_2 + \sqrt{1-B^2} h'_{2\perp}.
\end{eqnarray} Both $h_{2\perp}$ and $h'_{2\perp}$ are orthogonal to $h_2$, but are not necessarily the same waveform,
and the weights are chosen to ensure that all waveforms are normalised. We can also write this as $A = \cos(\theta_{12})$ and
$B = \cos(\theta_{23})$, and we therefore have,
\begin{eqnarray}
h_1 & = & \cos(\theta_{12}) h_2 + \sin(\theta_{12}) h_{2\perp}, \\
h_3 & = &\cos(\theta_{23}) h_2 + \sin(\theta_{23}) h'_{2\perp},
\end{eqnarray} and so,
\begin{equation}
\langle h_1 | h_3 \rangle =  \cos(\theta_{12}) \cos(\theta_{23}) +  \sin(\theta_{12}) \sin(\theta_{23}) \langle h_{2\perp} | h'_{2\perp} \rangle.
\end{equation} If the two orthogonal contributions are the same, then the combined match will be $\cos(\theta_{12} - \theta_{23})$, which
is the best match we can have; if $A = B$ then $C = 1$, i.e., $h_1 = h_3$. Alternatively, if $\langle h_{2\perp} | h'_{2\perp} \rangle = 0$,
then $C = AB$. In general we are interested in cases where all matches are close to unity, and so if $A = 1 - \mathcal{M}_{12}$ and
$B = 1 - \mathcal{M}_{23}$, where $\mathcal{M}_{12}$ and $\mathcal{M}_{23}$ are the mismatches, then we will have
$C \approx 1 - \mathcal{M}_{12} - \mathcal{M}_{23}$, i.e., the mismatches add linearly.

The worst combined match, and therefore the
upper bound on the combined mismatch, occurs when $\langle h_{2\perp} | h'_{2\perp} \rangle = -1$, which can be thought of
geometrically as $h_1$ and $h_3$ differing from $h_2$ in opposite directions, and so we must add their differences to calculate the
combined difference. In this case, we have $C = \cos(\theta_{12} + \theta_{23})$. This allows the extreme case where
$h_1 = ( h_2 + h_{2\perp})/\sqrt{2}$ and $h_3 = ( h_2 - h_{2\perp})/\sqrt{2}$, and in this case we
have $A = B = 1/\sqrt{2}$, so the two waveforms are ``equally far apart'', but the combined match is $C = 0$, and so $h_1$ and $h_3$
are orthogonal to each other.

For the situations we are interested in, where the mismatches are small, we recall that $\cos(\theta) \approx 1 - \theta^2/2$, and so
the mismatches can be approximated as \begin{eqnarray}
\mathcal{M}_{12} & = & \frac{1}{2} \theta_{12}^2, \\
\mathcal{M}_{23} & = & \frac{1}{2} \theta_{23}^2, \\
 \mathcal{M}_{13} & = & \frac{1}{2} (\theta_{12} + \theta_{23})^2,
 \end{eqnarray} and therefore \begin{equation}
 \mathcal{M}_{13} \approx \left( \sqrt{\mathcal{M}_{12}} +  \sqrt{\mathcal{M}_{23}} \right)^{2}. \label{eq:matchsum}
 \end{equation}

 This is not strictly an upper bound on the match that we calculate, because we also optimise over time and phase shifts, and the
 optimisation is not captured in our calculation, and will have a different effect on each individual match. However, we have found
 in toy examples that Eq.~(\ref{eq:matchsum}) provides an excellent estimate of the combined mismatch.

\section{Power-weighted precessing mismatch}
\label{match-expressions}

The \emph{match} between two real valued detector response waveforms $h_1(t)$
and $h_2(t)$ is defined to be the standard inner product weighted by the power
spectral density of the detector $S_n\left(f\right)$ maximised over various
sets of parameters $\Theta$~\cite{Cutler:1994ys}, as given by Eq.~(\ref{eqn: match def}). 
The \emph{mismatch} may then by defined by Eq.~(\ref{eqn: mismatch}).
For precessing waveforms, the set of parameters $\Theta$ that are maximised
over are a relative time shift $t_0$ between the waveforms, a relative phase
shift $\phi_0$, and the detector response polarisation angle
$\psi_0$~\cite{Schmidt:2014iyl}. The precessing matches performed in this work
are calculated as described in Appendix B of Ref.~\cite{Schmidt:2014iyl}.

In order to see how the match varied over a range of total masses that might be
observed by current ground based detectors, we further calculated the
power-weighted match as is described in Ref.~\cite{Ohme:2011zm}

To perform matches over a frequency range $f \in [f_\mathrm{min},
f_\mathrm{max}]$ that extends below the minimum frequency $f_\mathrm{NR}$ of 
one of the waveforms in this catalogue scaled to a specified total mass, the full
integral from $f_\mathrm{min}$ to the maximum NR frequency can be approximated
using a \emph{power-weighted mismatch} using the method described
in Ref.~\cite{Ohme:2011zm}. This method takes into account the missing inspiral part
of the waveform between $f_\mathrm{min}$ and the start of the NR waveform.

To perform a power-weighted mismatch the constituent waveforms are first split
up into contributions from NR defined over the frequency range $f \in
[f_\mathrm{NR}, f_\mathrm{max}]$ and the contributions from the inspiral
below the lowest NR frequency defined over the frequency range $f \in
[f_\mathrm{min}, f_\mathrm{NR})$,
\begin{equation}
h(f) =
\begin{cases}
h_\mathrm{ins}, & f \in [f_\mathrm{min}, f_\mathrm{NR}), \\
h_\mathrm{NR}, & f \in [f_\mathrm{NR}, f_\mathrm{max}].
\end{cases}
\end{equation}
The power-weighted mismatch is then the mismatch in each region weighted
by the fraction of power in each region,
\begin{align}
\mathcal{M}_\mathrm{pow} & \equiv
    \frac{\|h\|^2_{(f_\mathrm{min}, f_\mathrm{NR})}}{\|h\|^2}\mathcal{M}_\mathrm{ins} +
    \frac{\|h\|^2_{(f_\mathrm{NR},f_\mathrm{max})}}{\|h\|^2}\mathcal{M}_\text{NR}, \\
\mathcal{M}_\mathrm{ins} & \equiv
   \mathcal{M}_{(f_\mathrm{min}, f_\mathrm{NR})}(h_{1,\mathrm{ins}}, h_{2,\mathrm{ins}}), \\
\mathcal{M}_\mathrm{NR} & \equiv
   \mathcal{M}_{(f_\mathrm{NR},f_\mathrm{max})}(h_{1,\mathrm{NR}}, h_{2,\mathrm{NR}}),
\end{align}
where the subscript ranges $(f_1, f_2)$ denote the frequency ranges over which
the inner product Eq.~(\ref{eqn: mismatch}) is evaluated for that
expression.

The inspiral parts are assumed to perfectly agree which means that
$\mathcal{M}_\mathrm{ins}$ can be set to 0. This reduces the power-weighted
mismatch to,
\begin{equation}
\mathcal{M}_\mathrm{pow} =
    \frac{\|h\|^2_{(f_\mathrm{NR},f_\mathrm{min})}}{\|h\|^2}\mathcal{M}_\text{NR}.
\end{equation}
The inspiral contribution $\|h\|^2_{(f_\mathrm{min}, f_\mathrm{NR})}$ to
$\|h\|$ can be calculated using any appropriate inspiral waveform. For this
work the precessing waveform model PhenomPv3~\cite{PhysRevD.100.024059} was
used as the inspiral waveform.

It is important to make clear as described in~\cite{Ohme:2011zm} that
$\mathcal{M}_\mathrm{pow}$ will be a lower bound to the mismatch $\mathcal{M}
\geqslant \mathcal{M}_\mathrm{pow}$, however it is a sufficiently accurate
 approximation for NR accuracy assessment in this context.

%%%%%%%%%%%%%%%%%%%%

\end{appendices}

%\newpage
\bibliographystyle{apsrev}
\bibliography{paper.bib}

\begin{thebibliography}{135}
\expandafter\ifx\csname natexlab\endcsname\relax\def\natexlab#1{#1}\fi
\expandafter\ifx\csname bibnamefont\endcsname\relax
  \def\bibnamefont#1{#1}\fi
\expandafter\ifx\csname bibfnamefont\endcsname\relax
  \def\bibfnamefont#1{#1}\fi
\expandafter\ifx\csname citenamefont\endcsname\relax
  \def\citenamefont#1{#1}\fi
\expandafter\ifx\csname url\endcsname\relax
  \def\url#1{\texttt{#1}}\fi
\expandafter\ifx\csname urlprefix\endcsname\relax\def\urlprefix{URL }\fi
\providecommand{\bibinfo}[2]{#2}
\providecommand{\eprint}[2][]{\url{#2}}

\bibitem[{\citenamefont{Pretorius}(2005)}]{Pretorius:2005gq}
\bibinfo{author}{\bibfnamefont{F.}~\bibnamefont{Pretorius}},
  \bibinfo{journal}{Phys. Rev. Lett.} \textbf{\bibinfo{volume}{95}},
  \bibinfo{pages}{121101} (\bibinfo{year}{2005}), \eprint{gr-qc/0507014}.

\bibitem[{\citenamefont{Campanelli et~al.}(2006)\citenamefont{Campanelli,
  Lousto, Marronetti, and Zlochower}}]{Campanelli:2005dd}
\bibinfo{author}{\bibfnamefont{M.}~\bibnamefont{Campanelli}},
  \bibinfo{author}{\bibfnamefont{C.~O.} \bibnamefont{Lousto}},
  \bibinfo{author}{\bibfnamefont{P.}~\bibnamefont{Marronetti}},
  \bibnamefont{and}
  \bibinfo{author}{\bibfnamefont{Y.}~\bibnamefont{Zlochower}},
  \bibinfo{journal}{Phys. Rev. Lett.} \textbf{\bibinfo{volume}{96}},
  \bibinfo{pages}{111101} (\bibinfo{year}{2006}), \eprint{gr-qc/0511048}.

\bibitem[{\citenamefont{Baker et~al.}(2006)\citenamefont{Baker, Centrella,
  Choi, Koppitz, and van Meter}}]{Baker:2005vv}
\bibinfo{author}{\bibfnamefont{J.~G.} \bibnamefont{Baker}},
  \bibinfo{author}{\bibfnamefont{J.}~\bibnamefont{Centrella}},
  \bibinfo{author}{\bibfnamefont{D.-I.} \bibnamefont{Choi}},
  \bibinfo{author}{\bibfnamefont{M.}~\bibnamefont{Koppitz}}, \bibnamefont{and}
  \bibinfo{author}{\bibfnamefont{J.}~\bibnamefont{van Meter}},
  \bibinfo{journal}{Phys. Rev. Lett.} \textbf{\bibinfo{volume}{96}},
  \bibinfo{pages}{111102} (\bibinfo{year}{2006}), \eprint{gr-qc/0511103}.

\bibitem[{\citenamefont{Bruegmann et~al.}(2008)\citenamefont{Bruegmann,
  Gonzalez, Hannam, Husa, Sperhake, and Tichy}}]{Brugmann:2008zz}
\bibinfo{author}{\bibfnamefont{B.}~\bibnamefont{Bruegmann}},
  \bibinfo{author}{\bibfnamefont{J.~A.} \bibnamefont{Gonzalez}},
  \bibinfo{author}{\bibfnamefont{M.}~\bibnamefont{Hannam}},
  \bibinfo{author}{\bibfnamefont{S.}~\bibnamefont{Husa}},
  \bibinfo{author}{\bibfnamefont{U.}~\bibnamefont{Sperhake}}, \bibnamefont{and}
  \bibinfo{author}{\bibfnamefont{W.}~\bibnamefont{Tichy}},
  \bibinfo{journal}{Phys. Rev. D} \textbf{\bibinfo{volume}{77}},
  \bibinfo{pages}{024027} (\bibinfo{year}{2008}), \eprint{gr-qc/0610128}.

\bibitem[{\citenamefont{Husa et~al.}(2008{\natexlab{a}})\citenamefont{Husa,
  Gonzalez, Hannam, Bruegmann, and Sperhake}}]{Husa:2007hp}
\bibinfo{author}{\bibfnamefont{S.}~\bibnamefont{Husa}},
  \bibinfo{author}{\bibfnamefont{J.~A.} \bibnamefont{Gonzalez}},
  \bibinfo{author}{\bibfnamefont{M.}~\bibnamefont{Hannam}},
  \bibinfo{author}{\bibfnamefont{B.}~\bibnamefont{Bruegmann}},
  \bibnamefont{and} \bibinfo{author}{\bibfnamefont{U.}~\bibnamefont{Sperhake}},
  \bibinfo{journal}{Class. Quant. Grav.} \textbf{\bibinfo{volume}{25}},
  \bibinfo{pages}{105006} (\bibinfo{year}{2008}{\natexlab{a}}),
  \eprint{0706.0740}.

\bibitem[{\citenamefont{Scheel et~al.}(2006)\citenamefont{Scheel, Pfeiffer,
  Lindblom, Kidder, Rinne, and Teukolsky}}]{Scheel:2006gg}
\bibinfo{author}{\bibfnamefont{M.~A.} \bibnamefont{Scheel}},
  \bibinfo{author}{\bibfnamefont{H.~P.} \bibnamefont{Pfeiffer}},
  \bibinfo{author}{\bibfnamefont{L.}~\bibnamefont{Lindblom}},
  \bibinfo{author}{\bibfnamefont{L.~E.} \bibnamefont{Kidder}},
  \bibinfo{author}{\bibfnamefont{O.}~\bibnamefont{Rinne}}, \bibnamefont{and}
  \bibinfo{author}{\bibfnamefont{S.~A.} \bibnamefont{Teukolsky}},
  \bibinfo{journal}{Phys. Rev. D} \textbf{\bibinfo{volume}{74}},
  \bibinfo{pages}{104006} (\bibinfo{year}{2006}), \eprint{gr-qc/0607056}.

\bibitem[{\citenamefont{Hemberger et~al.}(2013)\citenamefont{Hemberger, Scheel,
  Kidder, Szil\'agyi, Lovelace, Taylor, and Teukolsky}}]{Hemberger:2012jz}
\bibinfo{author}{\bibfnamefont{D.~A.} \bibnamefont{Hemberger}},
  \bibinfo{author}{\bibfnamefont{M.~A.} \bibnamefont{Scheel}},
  \bibinfo{author}{\bibfnamefont{L.~E.} \bibnamefont{Kidder}},
  \bibinfo{author}{\bibfnamefont{B.}~\bibnamefont{Szil\'agyi}},
  \bibinfo{author}{\bibfnamefont{G.}~\bibnamefont{Lovelace}},
  \bibinfo{author}{\bibfnamefont{N.~W.} \bibnamefont{Taylor}},
  \bibnamefont{and} \bibinfo{author}{\bibfnamefont{S.~A.}
  \bibnamefont{Teukolsky}}, \bibinfo{journal}{Class. Quant. Grav.}
  \textbf{\bibinfo{volume}{30}}, \bibinfo{pages}{115001}
  (\bibinfo{year}{2013}), \eprint{1211.6079}.

\bibitem[{\citenamefont{Herrmann
  et~al.}(2007{\natexlab{a}})\citenamefont{Herrmann, Hinder, Shoemaker, and
  Laguna}}]{Herrmann:2007cwl}
\bibinfo{author}{\bibfnamefont{F.}~\bibnamefont{Herrmann}},
  \bibinfo{author}{\bibfnamefont{I.}~\bibnamefont{Hinder}},
  \bibinfo{author}{\bibfnamefont{D.}~\bibnamefont{Shoemaker}},
  \bibnamefont{and} \bibinfo{author}{\bibfnamefont{P.}~\bibnamefont{Laguna}},
  \bibinfo{journal}{Class. Quant. Grav.} \textbf{\bibinfo{volume}{24}},
  \bibinfo{pages}{S33} (\bibinfo{year}{2007}{\natexlab{a}}).

\bibitem[{\citenamefont{Zlochower et~al.}(2005)\citenamefont{Zlochower, Baker,
  Campanelli, and Lousto}}]{Zlochower:2005bj}
\bibinfo{author}{\bibfnamefont{Y.}~\bibnamefont{Zlochower}},
  \bibinfo{author}{\bibfnamefont{J.~G.} \bibnamefont{Baker}},
  \bibinfo{author}{\bibfnamefont{M.}~\bibnamefont{Campanelli}},
  \bibnamefont{and} \bibinfo{author}{\bibfnamefont{C.~O.}
  \bibnamefont{Lousto}}, \bibinfo{journal}{Phys. Rev. D}
  \textbf{\bibinfo{volume}{72}}, \bibinfo{pages}{024021}
  (\bibinfo{year}{2005}), \eprint{gr-qc/0505055}.

\bibitem[{\citenamefont{Sperhake}(2007)}]{Sperhake:2006cy}
\bibinfo{author}{\bibfnamefont{U.}~\bibnamefont{Sperhake}},
  \bibinfo{journal}{Phys. Rev. D} \textbf{\bibinfo{volume}{76}},
  \bibinfo{pages}{104015} (\bibinfo{year}{2007}), \eprint{gr-qc/0606079}.

\bibitem[{\citenamefont{Loffler et~al.}(2012)}]{Loffler:2011ay}
\bibinfo{author}{\bibfnamefont{F.}~\bibnamefont{Loffler}} \bibnamefont{et~al.},
  \bibinfo{journal}{Class. Quant. Grav.} \textbf{\bibinfo{volume}{29}},
  \bibinfo{pages}{115001} (\bibinfo{year}{2012}), \eprint{1111.3344}.

\bibitem[{\citenamefont{Lousto and Zlochower}(2011)}]{Lousto:2010ut}
\bibinfo{author}{\bibfnamefont{C.~O.} \bibnamefont{Lousto}} \bibnamefont{and}
  \bibinfo{author}{\bibfnamefont{Y.}~\bibnamefont{Zlochower}},
  \bibinfo{journal}{Phys. Rev. Lett.} \textbf{\bibinfo{volume}{106}},
  \bibinfo{pages}{041101} (\bibinfo{year}{2011}), \eprint{1009.0292}.

\bibitem[{\citenamefont{Sperhake et~al.}(2011)\citenamefont{Sperhake, Cardoso,
  Ott, Schnetter, and Witek}}]{Sperhake:2011ik}
\bibinfo{author}{\bibfnamefont{U.}~\bibnamefont{Sperhake}},
  \bibinfo{author}{\bibfnamefont{V.}~\bibnamefont{Cardoso}},
  \bibinfo{author}{\bibfnamefont{C.~D.} \bibnamefont{Ott}},
  \bibinfo{author}{\bibfnamefont{E.}~\bibnamefont{Schnetter}},
  \bibnamefont{and} \bibinfo{author}{\bibfnamefont{H.}~\bibnamefont{Witek}},
  \bibinfo{journal}{Phys. Rev. D} \textbf{\bibinfo{volume}{84}},
  \bibinfo{pages}{084038} (\bibinfo{year}{2011}), \eprint{1105.5391}.

\bibitem[{\citenamefont{Scheel et~al.}(2015)\citenamefont{Scheel, Giesler,
  Hemberger, Lovelace, Kuper, Boyle, Szil\'agyi, and Kidder}}]{Scheel:2014ina}
\bibinfo{author}{\bibfnamefont{M.~A.} \bibnamefont{Scheel}},
  \bibinfo{author}{\bibfnamefont{M.}~\bibnamefont{Giesler}},
  \bibinfo{author}{\bibfnamefont{D.~A.} \bibnamefont{Hemberger}},
  \bibinfo{author}{\bibfnamefont{G.}~\bibnamefont{Lovelace}},
  \bibinfo{author}{\bibfnamefont{K.}~\bibnamefont{Kuper}},
  \bibinfo{author}{\bibfnamefont{M.}~\bibnamefont{Boyle}},
  \bibinfo{author}{\bibfnamefont{B.}~\bibnamefont{Szil\'agyi}},
  \bibnamefont{and} \bibinfo{author}{\bibfnamefont{L.~E.}
  \bibnamefont{Kidder}}, \bibinfo{journal}{Class. Quant. Grav.}
  \textbf{\bibinfo{volume}{32}}, \bibinfo{pages}{105009}
  (\bibinfo{year}{2015}), \eprint{1412.1803}.

\bibitem[{\citenamefont{Husa et~al.}(2016)\citenamefont{Husa, Khan, Hannam,
  Pürrer, Ohme, Jiménez~Forteza, and Bohé}}]{Husa:2015iqa}
\bibinfo{author}{\bibfnamefont{S.}~\bibnamefont{Husa}},
  \bibinfo{author}{\bibfnamefont{S.}~\bibnamefont{Khan}},
  \bibinfo{author}{\bibfnamefont{M.}~\bibnamefont{Hannam}},
  \bibinfo{author}{\bibfnamefont{M.}~\bibnamefont{Pürrer}},
  \bibinfo{author}{\bibfnamefont{F.}~\bibnamefont{Ohme}},
  \bibinfo{author}{\bibfnamefont{X.}~\bibnamefont{Jiménez~Forteza}},
  \bibnamefont{and} \bibinfo{author}{\bibfnamefont{A.}~\bibnamefont{Bohé}},
  \bibinfo{journal}{Phys. Rev.} \textbf{\bibinfo{volume}{D93}},
  \bibinfo{pages}{044006} (\bibinfo{year}{2016}), \eprint{1508.07250}.

\bibitem[{\citenamefont{Khan et~al.}(2016)\citenamefont{Khan, Husa, Hannam,
  Ohme, Pürrer, Jiménez~Forteza, and Bohé}}]{Khan:2015jqa}
\bibinfo{author}{\bibfnamefont{S.}~\bibnamefont{Khan}},
  \bibinfo{author}{\bibfnamefont{S.}~\bibnamefont{Husa}},
  \bibinfo{author}{\bibfnamefont{M.}~\bibnamefont{Hannam}},
  \bibinfo{author}{\bibfnamefont{F.}~\bibnamefont{Ohme}},
  \bibinfo{author}{\bibfnamefont{M.}~\bibnamefont{Pürrer}},
  \bibinfo{author}{\bibfnamefont{X.}~\bibnamefont{Jiménez~Forteza}},
  \bibnamefont{and} \bibinfo{author}{\bibfnamefont{A.}~\bibnamefont{Bohé}},
  \bibinfo{journal}{Phys. Rev.} \textbf{\bibinfo{volume}{D93}},
  \bibinfo{pages}{044007} (\bibinfo{year}{2016}), \eprint{1508.07253}.

\bibitem[{\citenamefont{Pratten et~al.}(2020)\citenamefont{Pratten, Husa,
  Garcia-Quiros, Colleoni, Ramos-Buades, Estelles, and
  Jaume}}]{Pratten:2020fqn}
\bibinfo{author}{\bibfnamefont{G.}~\bibnamefont{Pratten}},
  \bibinfo{author}{\bibfnamefont{S.}~\bibnamefont{Husa}},
  \bibinfo{author}{\bibfnamefont{C.}~\bibnamefont{Garcia-Quiros}},
  \bibinfo{author}{\bibfnamefont{M.}~\bibnamefont{Colleoni}},
  \bibinfo{author}{\bibfnamefont{A.}~\bibnamefont{Ramos-Buades}},
  \bibinfo{author}{\bibfnamefont{H.}~\bibnamefont{Estelles}}, \bibnamefont{and}
  \bibinfo{author}{\bibfnamefont{R.}~\bibnamefont{Jaume}},
  \bibinfo{journal}{Phys. Rev. D} \textbf{\bibinfo{volume}{102}},
  \bibinfo{pages}{064001} (\bibinfo{year}{2020}), \eprint{2001.11412}.

\bibitem[{\citenamefont{Garc\'\i{}a-Quir\'os
  et~al.}(2020)\citenamefont{Garc\'\i{}a-Quir\'os, Colleoni, Husa, Estell\'es,
  Pratten, Ramos-Buades, Mateu-Lucena, and Jaume}}]{Garcia-Quiros:2020qpx}
\bibinfo{author}{\bibfnamefont{C.}~\bibnamefont{Garc\'\i{}a-Quir\'os}},
  \bibinfo{author}{\bibfnamefont{M.}~\bibnamefont{Colleoni}},
  \bibinfo{author}{\bibfnamefont{S.}~\bibnamefont{Husa}},
  \bibinfo{author}{\bibfnamefont{H.}~\bibnamefont{Estell\'es}},
  \bibinfo{author}{\bibfnamefont{G.}~\bibnamefont{Pratten}},
  \bibinfo{author}{\bibfnamefont{A.}~\bibnamefont{Ramos-Buades}},
  \bibinfo{author}{\bibfnamefont{M.}~\bibnamefont{Mateu-Lucena}},
  \bibnamefont{and} \bibinfo{author}{\bibfnamefont{R.}~\bibnamefont{Jaume}},
  \bibinfo{journal}{Phys. Rev. D} \textbf{\bibinfo{volume}{102}},
  \bibinfo{pages}{064002} (\bibinfo{year}{2020}), \eprint{2001.10914}.

\bibitem[{\citenamefont{Estell\'es et~al.}(2020)\citenamefont{Estell\'es, Husa,
  Colleoni, Keitel, Mateu-Lucena, Garc\'\i{}a-Quir\'os, Ramos-Buades, and
  Borchers}}]{Estelles:2020twz}
\bibinfo{author}{\bibfnamefont{H.}~\bibnamefont{Estell\'es}},
  \bibinfo{author}{\bibfnamefont{S.}~\bibnamefont{Husa}},
  \bibinfo{author}{\bibfnamefont{M.}~\bibnamefont{Colleoni}},
  \bibinfo{author}{\bibfnamefont{D.}~\bibnamefont{Keitel}},
  \bibinfo{author}{\bibfnamefont{M.}~\bibnamefont{Mateu-Lucena}},
  \bibinfo{author}{\bibfnamefont{C.}~\bibnamefont{Garc\'\i{}a-Quir\'os}},
  \bibinfo{author}{\bibfnamefont{A.}~\bibnamefont{Ramos-Buades}},
  \bibnamefont{and} \bibinfo{author}{\bibfnamefont{A.}~\bibnamefont{Borchers}}
  (\bibinfo{year}{2020}), \eprint{2012.11923}.

\bibitem[{\citenamefont{Hamilton et~al.}(2021)\citenamefont{Hamilton, London,
  Thompson, Fauchon-Jones, Hannam, Kalaghatgi, Khan, Pannarale, and
  Vano-Vinuales}}]{Hamilton:2021pkf}
\bibinfo{author}{\bibfnamefont{E.}~\bibnamefont{Hamilton}},
  \bibinfo{author}{\bibfnamefont{L.}~\bibnamefont{London}},
  \bibinfo{author}{\bibfnamefont{J.~E.} \bibnamefont{Thompson}},
  \bibinfo{author}{\bibfnamefont{E.}~\bibnamefont{Fauchon-Jones}},
  \bibinfo{author}{\bibfnamefont{M.}~\bibnamefont{Hannam}},
  \bibinfo{author}{\bibfnamefont{C.}~\bibnamefont{Kalaghatgi}},
  \bibinfo{author}{\bibfnamefont{S.}~\bibnamefont{Khan}},
  \bibinfo{author}{\bibfnamefont{F.}~\bibnamefont{Pannarale}},
  \bibnamefont{and}
  \bibinfo{author}{\bibfnamefont{A.}~\bibnamefont{Vano-Vinuales}},
  \bibinfo{journal}{Phys. Rev. D} \textbf{\bibinfo{volume}{104}},
  \bibinfo{pages}{124027} (\bibinfo{year}{2021}), \eprint{2107.08876}.

\bibitem[{\citenamefont{Buonanno et~al.}(2009)\citenamefont{Buonanno, Pan,
  Pfeiffer, Scheel, Buchman, and Kidder}}]{Buonanno:2009qa}
\bibinfo{author}{\bibfnamefont{A.}~\bibnamefont{Buonanno}},
  \bibinfo{author}{\bibfnamefont{Y.}~\bibnamefont{Pan}},
  \bibinfo{author}{\bibfnamefont{H.~P.} \bibnamefont{Pfeiffer}},
  \bibinfo{author}{\bibfnamefont{M.~A.} \bibnamefont{Scheel}},
  \bibinfo{author}{\bibfnamefont{L.~T.} \bibnamefont{Buchman}},
  \bibnamefont{and} \bibinfo{author}{\bibfnamefont{L.~E.}
  \bibnamefont{Kidder}}, \bibinfo{journal}{Phys. Rev. D}
  \textbf{\bibinfo{volume}{79}}, \bibinfo{pages}{124028}
  (\bibinfo{year}{2009}), \eprint{0902.0790}.

\bibitem[{\citenamefont{Taracchini et~al.}(2014)}]{Taracchini:2013rva}
\bibinfo{author}{\bibfnamefont{A.}~\bibnamefont{Taracchini}}
  \bibnamefont{et~al.}, \bibinfo{journal}{Phys. Rev. D}
  \textbf{\bibinfo{volume}{89}}, \bibinfo{pages}{061502}
  (\bibinfo{year}{2014}), \eprint{1311.2544}.

\bibitem[{\citenamefont{Pan et~al.}(2014)\citenamefont{Pan, Buonanno,
  Taracchini, Kidder, Mroué, Pfeiffer, Scheel, and Szilágyi}}]{Pan:2013rra}
\bibinfo{author}{\bibfnamefont{Y.}~\bibnamefont{Pan}},
  \bibinfo{author}{\bibfnamefont{A.}~\bibnamefont{Buonanno}},
  \bibinfo{author}{\bibfnamefont{A.}~\bibnamefont{Taracchini}},
  \bibinfo{author}{\bibfnamefont{L.~E.} \bibnamefont{Kidder}},
  \bibinfo{author}{\bibfnamefont{A.~H.} \bibnamefont{Mroué}},
  \bibinfo{author}{\bibfnamefont{H.~P.} \bibnamefont{Pfeiffer}},
  \bibinfo{author}{\bibfnamefont{M.~A.} \bibnamefont{Scheel}},
  \bibnamefont{and}
  \bibinfo{author}{\bibfnamefont{B.}~\bibnamefont{Szilágyi}},
  \bibinfo{journal}{Phys. Rev.} \textbf{\bibinfo{volume}{D89}},
  \bibinfo{pages}{084006} (\bibinfo{year}{2014}), \eprint{1307.6232}.

\bibitem[{\citenamefont{Boh\'e et~al.}(2017)}]{Bohe:2016gbl}
\bibinfo{author}{\bibfnamefont{A.}~\bibnamefont{Boh\'e}} \bibnamefont{et~al.},
  \bibinfo{journal}{Phys. Rev. D} \textbf{\bibinfo{volume}{95}},
  \bibinfo{pages}{044028} (\bibinfo{year}{2017}), \eprint{1611.03703}.

\bibitem[{\citenamefont{Babak et~al.}(2017)\citenamefont{Babak, Taracchini, and
  Buonanno}}]{Babak:2016tgq}
\bibinfo{author}{\bibfnamefont{S.}~\bibnamefont{Babak}},
  \bibinfo{author}{\bibfnamefont{A.}~\bibnamefont{Taracchini}},
  \bibnamefont{and} \bibinfo{author}{\bibfnamefont{A.}~\bibnamefont{Buonanno}},
  \bibinfo{journal}{Phys. Rev. D} \textbf{\bibinfo{volume}{95}},
  \bibinfo{pages}{024010} (\bibinfo{year}{2017}), \eprint{1607.05661}.

\bibitem[{\citenamefont{Cotesta et~al.}(2018)\citenamefont{Cotesta, Buonanno,
  Bohé, Taracchini, Hinder, and Ossokine}}]{Cotesta:2018fcv}
\bibinfo{author}{\bibfnamefont{R.}~\bibnamefont{Cotesta}},
  \bibinfo{author}{\bibfnamefont{A.}~\bibnamefont{Buonanno}},
  \bibinfo{author}{\bibfnamefont{A.}~\bibnamefont{Bohé}},
  \bibinfo{author}{\bibfnamefont{A.}~\bibnamefont{Taracchini}},
  \bibinfo{author}{\bibfnamefont{I.}~\bibnamefont{Hinder}}, \bibnamefont{and}
  \bibinfo{author}{\bibfnamefont{S.}~\bibnamefont{Ossokine}},
  \bibinfo{journal}{Phys. Rev.} \textbf{\bibinfo{volume}{D98}},
  \bibinfo{pages}{084028} (\bibinfo{year}{2018}), \eprint{1803.10701}.

\bibitem[{\citenamefont{Blackman
  et~al.}(2017{\natexlab{a}})\citenamefont{Blackman, Field, Scheel, Galley,
  Hemberger, Schmidt, and Smith}}]{Blackman:2017dfb}
\bibinfo{author}{\bibfnamefont{J.}~\bibnamefont{Blackman}},
  \bibinfo{author}{\bibfnamefont{S.~E.} \bibnamefont{Field}},
  \bibinfo{author}{\bibfnamefont{M.~A.} \bibnamefont{Scheel}},
  \bibinfo{author}{\bibfnamefont{C.~R.} \bibnamefont{Galley}},
  \bibinfo{author}{\bibfnamefont{D.~A.} \bibnamefont{Hemberger}},
  \bibinfo{author}{\bibfnamefont{P.}~\bibnamefont{Schmidt}}, \bibnamefont{and}
  \bibinfo{author}{\bibfnamefont{R.}~\bibnamefont{Smith}},
  \bibinfo{journal}{Phys. Rev. D} \textbf{\bibinfo{volume}{95}},
  \bibinfo{pages}{104023} (\bibinfo{year}{2017}{\natexlab{a}}),
  \eprint{1701.00550}.

\bibitem[{\citenamefont{Blackman
  et~al.}(2017{\natexlab{b}})\citenamefont{Blackman, Field, Scheel, Galley,
  Ott, Boyle, Kidder, Pfeiffer, and Szil\'agyi}}]{Blackman:2017pcm}
\bibinfo{author}{\bibfnamefont{J.}~\bibnamefont{Blackman}},
  \bibinfo{author}{\bibfnamefont{S.~E.} \bibnamefont{Field}},
  \bibinfo{author}{\bibfnamefont{M.~A.} \bibnamefont{Scheel}},
  \bibinfo{author}{\bibfnamefont{C.~R.} \bibnamefont{Galley}},
  \bibinfo{author}{\bibfnamefont{C.~D.} \bibnamefont{Ott}},
  \bibinfo{author}{\bibfnamefont{M.}~\bibnamefont{Boyle}},
  \bibinfo{author}{\bibfnamefont{L.~E.} \bibnamefont{Kidder}},
  \bibinfo{author}{\bibfnamefont{H.~P.} \bibnamefont{Pfeiffer}},
  \bibnamefont{and}
  \bibinfo{author}{\bibfnamefont{B.}~\bibnamefont{Szil\'agyi}},
  \bibinfo{journal}{Phys. Rev. D} \textbf{\bibinfo{volume}{96}},
  \bibinfo{pages}{024058} (\bibinfo{year}{2017}{\natexlab{b}}),
  \eprint{1705.07089}.

\bibitem[{\citenamefont{Varma et~al.}(2019{\natexlab{a}})\citenamefont{Varma,
  Field, Scheel, Blackman, Gerosa, Stein, Kidder, and
  Pfeiffer}}]{Varma:2019csw}
\bibinfo{author}{\bibfnamefont{V.}~\bibnamefont{Varma}},
  \bibinfo{author}{\bibfnamefont{S.~E.} \bibnamefont{Field}},
  \bibinfo{author}{\bibfnamefont{M.~A.} \bibnamefont{Scheel}},
  \bibinfo{author}{\bibfnamefont{J.}~\bibnamefont{Blackman}},
  \bibinfo{author}{\bibfnamefont{D.}~\bibnamefont{Gerosa}},
  \bibinfo{author}{\bibfnamefont{L.~C.} \bibnamefont{Stein}},
  \bibinfo{author}{\bibfnamefont{L.~E.} \bibnamefont{Kidder}},
  \bibnamefont{and} \bibinfo{author}{\bibfnamefont{H.~P.}
  \bibnamefont{Pfeiffer}} (\bibinfo{year}{2019}{\natexlab{a}}),
  \eprint{1905.09300}.

\bibitem[{\citenamefont{Gonzalez
  et~al.}(2007{\natexlab{a}})\citenamefont{Gonzalez, Sperhake, Bruegmann,
  Hannam, and Husa}}]{Gonzalez:2006md}
\bibinfo{author}{\bibfnamefont{J.~A.} \bibnamefont{Gonzalez}},
  \bibinfo{author}{\bibfnamefont{U.}~\bibnamefont{Sperhake}},
  \bibinfo{author}{\bibfnamefont{B.}~\bibnamefont{Bruegmann}},
  \bibinfo{author}{\bibfnamefont{M.}~\bibnamefont{Hannam}}, \bibnamefont{and}
  \bibinfo{author}{\bibfnamefont{S.}~\bibnamefont{Husa}},
  \bibinfo{journal}{Phys. Rev. Lett.} \textbf{\bibinfo{volume}{98}},
  \bibinfo{pages}{091101} (\bibinfo{year}{2007}{\natexlab{a}}),
  \eprint{gr-qc/0610154}.

\bibitem[{\citenamefont{Gonzalez
  et~al.}(2007{\natexlab{b}})\citenamefont{Gonzalez, Hannam, Sperhake,
  Bruegmann, and Husa}}]{Gonzalez:2007hi}
\bibinfo{author}{\bibfnamefont{J.~A.} \bibnamefont{Gonzalez}},
  \bibinfo{author}{\bibfnamefont{M.~D.} \bibnamefont{Hannam}},
  \bibinfo{author}{\bibfnamefont{U.}~\bibnamefont{Sperhake}},
  \bibinfo{author}{\bibfnamefont{B.}~\bibnamefont{Bruegmann}},
  \bibnamefont{and} \bibinfo{author}{\bibfnamefont{S.}~\bibnamefont{Husa}},
  \bibinfo{journal}{Phys. Rev. Lett.} \textbf{\bibinfo{volume}{98}},
  \bibinfo{pages}{231101} (\bibinfo{year}{2007}{\natexlab{b}}),
  \eprint{gr-qc/0702052}.

\bibitem[{\citenamefont{Herrmann
  et~al.}(2007{\natexlab{b}})\citenamefont{Herrmann, Hinder, Shoemaker, Laguna,
  and Matzner}}]{Herrmann:2007ex}
\bibinfo{author}{\bibfnamefont{F.}~\bibnamefont{Herrmann}},
  \bibinfo{author}{\bibfnamefont{I.}~\bibnamefont{Hinder}},
  \bibinfo{author}{\bibfnamefont{D.~M.} \bibnamefont{Shoemaker}},
  \bibinfo{author}{\bibfnamefont{P.}~\bibnamefont{Laguna}}, \bibnamefont{and}
  \bibinfo{author}{\bibfnamefont{R.~A.} \bibnamefont{Matzner}},
  \bibinfo{journal}{Phys. Rev. D} \textbf{\bibinfo{volume}{76}},
  \bibinfo{pages}{084032} (\bibinfo{year}{2007}{\natexlab{b}}),
  \eprint{0706.2541}.

\bibitem[{\citenamefont{Campanelli
  et~al.}(2007{\natexlab{a}})\citenamefont{Campanelli, Lousto, Zlochower, and
  Merritt}}]{Campanelli:2007cga}
\bibinfo{author}{\bibfnamefont{M.}~\bibnamefont{Campanelli}},
  \bibinfo{author}{\bibfnamefont{C.~O.} \bibnamefont{Lousto}},
  \bibinfo{author}{\bibfnamefont{Y.}~\bibnamefont{Zlochower}},
  \bibnamefont{and} \bibinfo{author}{\bibfnamefont{D.}~\bibnamefont{Merritt}},
  \bibinfo{journal}{Phys. Rev. Lett.} \textbf{\bibinfo{volume}{98}},
  \bibinfo{pages}{231102} (\bibinfo{year}{2007}{\natexlab{a}}),
  \eprint{gr-qc/0702133}.

\bibitem[{\citenamefont{Campanelli
  et~al.}(2007{\natexlab{b}})\citenamefont{Campanelli, Lousto, Zlochower, and
  Merritt}}]{Campanelli:2007ew}
\bibinfo{author}{\bibfnamefont{M.}~\bibnamefont{Campanelli}},
  \bibinfo{author}{\bibfnamefont{C.~O.} \bibnamefont{Lousto}},
  \bibinfo{author}{\bibfnamefont{Y.}~\bibnamefont{Zlochower}},
  \bibnamefont{and} \bibinfo{author}{\bibfnamefont{D.}~\bibnamefont{Merritt}},
  \bibinfo{journal}{Astrophys. J. Lett.} \textbf{\bibinfo{volume}{659}},
  \bibinfo{pages}{L5} (\bibinfo{year}{2007}{\natexlab{b}}),
  \eprint{gr-qc/0701164}.

\bibitem[{\citenamefont{Kesden}(2008)}]{Kesden:2008ga}
\bibinfo{author}{\bibfnamefont{M.}~\bibnamefont{Kesden}},
  \bibinfo{journal}{Phys. Rev. D} \textbf{\bibinfo{volume}{78}},
  \bibinfo{pages}{084030} (\bibinfo{year}{2008}), \eprint{0807.3043}.

\bibitem[{\citenamefont{Lousto and Zlochower}(2013)}]{Lousto:2012gt}
\bibinfo{author}{\bibfnamefont{C.~O.} \bibnamefont{Lousto}} \bibnamefont{and}
  \bibinfo{author}{\bibfnamefont{Y.}~\bibnamefont{Zlochower}},
  \bibinfo{journal}{Phys. Rev. D} \textbf{\bibinfo{volume}{87}},
  \bibinfo{pages}{084027} (\bibinfo{year}{2013}), \eprint{1211.7099}.

\bibitem[{\citenamefont{Healy et~al.}(2014)\citenamefont{Healy, Lousto, and
  Zlochower}}]{Healy:2014yta}
\bibinfo{author}{\bibfnamefont{J.}~\bibnamefont{Healy}},
  \bibinfo{author}{\bibfnamefont{C.~O.} \bibnamefont{Lousto}},
  \bibnamefont{and}
  \bibinfo{author}{\bibfnamefont{Y.}~\bibnamefont{Zlochower}},
  \bibinfo{journal}{Phys. Rev. D} \textbf{\bibinfo{volume}{90}},
  \bibinfo{pages}{104004} (\bibinfo{year}{2014}), \eprint{1406.7295}.

\bibitem[{\citenamefont{Healy and Lousto}(2017)}]{Healy:2016lce}
\bibinfo{author}{\bibfnamefont{J.}~\bibnamefont{Healy}} \bibnamefont{and}
  \bibinfo{author}{\bibfnamefont{C.~O.} \bibnamefont{Lousto}},
  \bibinfo{journal}{Phys. Rev. D} \textbf{\bibinfo{volume}{95}},
  \bibinfo{pages}{024037} (\bibinfo{year}{2017}), \eprint{1610.09713}.

\bibitem[{\citenamefont{Healy and Lousto}(2018)}]{Healy:2018swt}
\bibinfo{author}{\bibfnamefont{J.}~\bibnamefont{Healy}} \bibnamefont{and}
  \bibinfo{author}{\bibfnamefont{C.~O.} \bibnamefont{Lousto}},
  \bibinfo{journal}{Phys. Rev. D} \textbf{\bibinfo{volume}{97}},
  \bibinfo{pages}{084002} (\bibinfo{year}{2018}), \eprint{1801.08162}.

\bibitem[{\citenamefont{Aylott et~al.}(2009)}]{Aylott:2009ya}
\bibinfo{author}{\bibfnamefont{B.}~\bibnamefont{Aylott}} \bibnamefont{et~al.},
  \bibinfo{journal}{Class. Quant. Grav.} \textbf{\bibinfo{volume}{26}},
  \bibinfo{pages}{165008} (\bibinfo{year}{2009}), \eprint{0901.4399}.

\bibitem[{\citenamefont{Aasi et~al.}(2014)}]{LIGOScientific:2014oec}
\bibinfo{author}{\bibfnamefont{J.}~\bibnamefont{Aasi}} \bibnamefont{et~al.}
  (\bibinfo{collaboration}{LIGO Scientific, VIRGO, NINJA-2}),
  \bibinfo{journal}{Class. Quant. Grav.} \textbf{\bibinfo{volume}{31}},
  \bibinfo{pages}{115004} (\bibinfo{year}{2014}), \eprint{1401.0939}.

\bibitem[{\citenamefont{Abbott
  et~al.}(2019{\natexlab{a}})}]{LIGOScientific:2018mvr}
\bibinfo{author}{\bibfnamefont{B.~P.} \bibnamefont{Abbott}}
  \bibnamefont{et~al.} (\bibinfo{collaboration}{LIGO Scientific, Virgo}),
  \bibinfo{journal}{Phys. Rev. X} \textbf{\bibinfo{volume}{9}},
  \bibinfo{pages}{031040} (\bibinfo{year}{2019}{\natexlab{a}}),
  \eprint{1811.12907}.

\bibitem[{\citenamefont{Abbott
  et~al.}(2021{\natexlab{a}})}]{LIGOScientific:2020ibl}
\bibinfo{author}{\bibfnamefont{R.}~\bibnamefont{Abbott}} \bibnamefont{et~al.}
  (\bibinfo{collaboration}{LIGO Scientific, Virgo}), \bibinfo{journal}{Phys.
  Rev. X} \textbf{\bibinfo{volume}{11}}, \bibinfo{pages}{021053}
  (\bibinfo{year}{2021}{\natexlab{a}}), \eprint{2010.14527}.

\bibitem[{\citenamefont{Abbott
  et~al.}(2021{\natexlab{b}})}]{LIGOScientific:2021djp}
\bibinfo{author}{\bibfnamefont{R.}~\bibnamefont{Abbott}} \bibnamefont{et~al.}
  (\bibinfo{collaboration}{LIGO Scientific, VIRGO, KAGRA})
  (\bibinfo{year}{2021}{\natexlab{b}}), \eprint{2111.03606}.

\bibitem[{\citenamefont{Abbott
  et~al.}(2016{\natexlab{a}})}]{LIGOScientific:2016aoc}
\bibinfo{author}{\bibfnamefont{B.~P.} \bibnamefont{Abbott}}
  \bibnamefont{et~al.} (\bibinfo{collaboration}{LIGO Scientific, Virgo}),
  \bibinfo{journal}{Phys. Rev. Lett.} \textbf{\bibinfo{volume}{116}},
  \bibinfo{pages}{061102} (\bibinfo{year}{2016}{\natexlab{a}}),
  \eprint{1602.03837}.

\bibitem[{\citenamefont{Abbott
  et~al.}(2016{\natexlab{b}})}]{LIGOScientific:2016dsl}
\bibinfo{author}{\bibfnamefont{B.~P.} \bibnamefont{Abbott}}
  \bibnamefont{et~al.} (\bibinfo{collaboration}{LIGO Scientific, Virgo}),
  \bibinfo{journal}{Phys. Rev. X} \textbf{\bibinfo{volume}{6}},
  \bibinfo{pages}{041015} (\bibinfo{year}{2016}{\natexlab{b}}),
  \bibinfo{note}{[Erratum: Phys.Rev.X 8, 039903 (2018)]}, \eprint{1606.04856}.

\bibitem[{\citenamefont{Abbott
  et~al.}(2016{\natexlab{c}})}]{LIGOScientific:2016sjg}
\bibinfo{author}{\bibfnamefont{B.~P.} \bibnamefont{Abbott}}
  \bibnamefont{et~al.} (\bibinfo{collaboration}{LIGO Scientific, Virgo}),
  \bibinfo{journal}{Phys. Rev. Lett.} \textbf{\bibinfo{volume}{116}},
  \bibinfo{pages}{241103} (\bibinfo{year}{2016}{\natexlab{c}}),
  \eprint{1606.04855}.

\bibitem[{\citenamefont{Abbott
  et~al.}(2017{\natexlab{a}})}]{LIGOScientific:2017bnn}
\bibinfo{author}{\bibfnamefont{B.~P.} \bibnamefont{Abbott}}
  \bibnamefont{et~al.} (\bibinfo{collaboration}{LIGO Scientific, VIRGO}),
  \bibinfo{journal}{Phys. Rev. Lett.} \textbf{\bibinfo{volume}{118}},
  \bibinfo{pages}{221101} (\bibinfo{year}{2017}{\natexlab{a}}),
  \bibinfo{note}{[Erratum: Phys.Rev.Lett. 121, 129901 (2018)]},
  \eprint{1706.01812}.

\bibitem[{\citenamefont{Abbott
  et~al.}(2017{\natexlab{b}})}]{LIGOScientific:2017ycc}
\bibinfo{author}{\bibfnamefont{B.~P.} \bibnamefont{Abbott}}
  \bibnamefont{et~al.} (\bibinfo{collaboration}{LIGO Scientific, Virgo}),
  \bibinfo{journal}{Phys. Rev. Lett.} \textbf{\bibinfo{volume}{119}},
  \bibinfo{pages}{141101} (\bibinfo{year}{2017}{\natexlab{b}}),
  \eprint{1709.09660}.

\bibitem[{\citenamefont{Abbott
  et~al.}(2017{\natexlab{c}})}]{LIGOScientific:2017vox}
\bibinfo{author}{\bibfnamefont{B.~. P.~.} \bibnamefont{Abbott}}
  \bibnamefont{et~al.} (\bibinfo{collaboration}{LIGO Scientific, Virgo}),
  \bibinfo{journal}{Astrophys. J. Lett.} \textbf{\bibinfo{volume}{851}},
  \bibinfo{pages}{L35} (\bibinfo{year}{2017}{\natexlab{c}}),
  \eprint{1711.05578}.

\bibitem[{\citenamefont{Abbott
  et~al.}(2020{\natexlab{a}})}]{LIGOScientific:2020aai}
\bibinfo{author}{\bibfnamefont{B.~P.} \bibnamefont{Abbott}}
  \bibnamefont{et~al.} (\bibinfo{collaboration}{LIGO Scientific, Virgo}),
  \bibinfo{journal}{Astrophys. J. Lett.} \textbf{\bibinfo{volume}{892}},
  \bibinfo{pages}{L3} (\bibinfo{year}{2020}{\natexlab{a}}),
  \eprint{2001.01761}.

\bibitem[{\citenamefont{Abbott
  et~al.}(2020{\natexlab{b}})}]{LIGOScientific:2020stg}
\bibinfo{author}{\bibfnamefont{R.}~\bibnamefont{Abbott}} \bibnamefont{et~al.}
  (\bibinfo{collaboration}{LIGO Scientific, Virgo}), \bibinfo{journal}{Phys.
  Rev. D} \textbf{\bibinfo{volume}{102}}, \bibinfo{pages}{043015}
  (\bibinfo{year}{2020}{\natexlab{b}}), \eprint{2004.08342}.

\bibitem[{\citenamefont{Abbott
  et~al.}(2020{\natexlab{c}})}]{LIGOScientific:2020zkf}
\bibinfo{author}{\bibfnamefont{R.}~\bibnamefont{Abbott}} \bibnamefont{et~al.}
  (\bibinfo{collaboration}{LIGO Scientific, Virgo}),
  \bibinfo{journal}{Astrophys. J. Lett.} \textbf{\bibinfo{volume}{896}},
  \bibinfo{pages}{L44} (\bibinfo{year}{2020}{\natexlab{c}}),
  \eprint{2006.12611}.

\bibitem[{\citenamefont{Abbott
  et~al.}(2021{\natexlab{c}})}]{LIGOScientific:2021qlt}
\bibinfo{author}{\bibfnamefont{R.}~\bibnamefont{Abbott}} \bibnamefont{et~al.}
  (\bibinfo{collaboration}{LIGO Scientific, KAGRA, VIRGO}),
  \bibinfo{journal}{Astrophys. J. Lett.} \textbf{\bibinfo{volume}{915}},
  \bibinfo{pages}{L5} (\bibinfo{year}{2021}{\natexlab{c}}),
  \eprint{2106.15163}.

\bibitem[{\citenamefont{Abbott
  et~al.}(2021{\natexlab{d}})}]{LIGOScientific:2021usb}
\bibinfo{author}{\bibfnamefont{R.}~\bibnamefont{Abbott}} \bibnamefont{et~al.}
  (\bibinfo{collaboration}{LIGO Scientific, VIRGO})
  (\bibinfo{year}{2021}{\natexlab{d}}), \eprint{2108.01045}.

\bibitem[{\citenamefont{London et~al.}(2018)\citenamefont{London, Khan,
  Fauchon-Jones, Garc\'{i}a, Hannam, Husa, Jim\'enez-Forteza, Kalaghatgi, Ohme,
  and Pannarale}}]{London:2017bcn}
\bibinfo{author}{\bibfnamefont{L.}~\bibnamefont{London}},
  \bibinfo{author}{\bibfnamefont{S.}~\bibnamefont{Khan}},
  \bibinfo{author}{\bibfnamefont{E.}~\bibnamefont{Fauchon-Jones}},
  \bibinfo{author}{\bibfnamefont{C.}~\bibnamefont{Garc\'{i}a}},
  \bibinfo{author}{\bibfnamefont{M.}~\bibnamefont{Hannam}},
  \bibinfo{author}{\bibfnamefont{S.}~\bibnamefont{Husa}},
  \bibinfo{author}{\bibfnamefont{X.}~\bibnamefont{Jim\'enez-Forteza}},
  \bibinfo{author}{\bibfnamefont{C.}~\bibnamefont{Kalaghatgi}},
  \bibinfo{author}{\bibfnamefont{F.}~\bibnamefont{Ohme}}, \bibnamefont{and}
  \bibinfo{author}{\bibfnamefont{F.}~\bibnamefont{Pannarale}},
  \bibinfo{journal}{Phys. Rev. Lett.} \textbf{\bibinfo{volume}{120}},
  \bibinfo{pages}{161102} (\bibinfo{year}{2018}), \eprint{1708.00404}.

\bibitem[{\citenamefont{Khan et~al.}(2019)\citenamefont{Khan, Chatziioannou,
  Hannam, and Ohme}}]{PhysRevD.100.024059}
\bibinfo{author}{\bibfnamefont{S.}~\bibnamefont{Khan}},
  \bibinfo{author}{\bibfnamefont{K.}~\bibnamefont{Chatziioannou}},
  \bibinfo{author}{\bibfnamefont{M.}~\bibnamefont{Hannam}}, \bibnamefont{and}
  \bibinfo{author}{\bibfnamefont{F.}~\bibnamefont{Ohme}},
  \bibinfo{journal}{Phys. Rev. D} \textbf{\bibinfo{volume}{100}},
  \bibinfo{pages}{024059} (\bibinfo{year}{2019}),
  \urlprefix\url{https://link.aps.org/doi/10.1103/PhysRevD.100.024059}.

\bibitem[{\citenamefont{Khan et~al.}(2020)\citenamefont{Khan, Ohme,
  Chatziioannou, and Hannam}}]{Khan:2019kot}
\bibinfo{author}{\bibfnamefont{S.}~\bibnamefont{Khan}},
  \bibinfo{author}{\bibfnamefont{F.}~\bibnamefont{Ohme}},
  \bibinfo{author}{\bibfnamefont{K.}~\bibnamefont{Chatziioannou}},
  \bibnamefont{and} \bibinfo{author}{\bibfnamefont{M.}~\bibnamefont{Hannam}},
  \bibinfo{journal}{Phys. Rev.} \textbf{\bibinfo{volume}{D101}},
  \bibinfo{pages}{024056} (\bibinfo{year}{2020}), \eprint{1911.06050}.

\bibitem[{\citenamefont{Thompson et~al.}(2020)\citenamefont{Thompson,
  Fauchon-Jones, Khan, Nitoglia, Pannarale, Dietrich, and
  Hannam}}]{Thompson:2020nei}
\bibinfo{author}{\bibfnamefont{J.~E.} \bibnamefont{Thompson}},
  \bibinfo{author}{\bibfnamefont{E.}~\bibnamefont{Fauchon-Jones}},
  \bibinfo{author}{\bibfnamefont{S.}~\bibnamefont{Khan}},
  \bibinfo{author}{\bibfnamefont{E.}~\bibnamefont{Nitoglia}},
  \bibinfo{author}{\bibfnamefont{F.}~\bibnamefont{Pannarale}},
  \bibinfo{author}{\bibfnamefont{T.}~\bibnamefont{Dietrich}}, \bibnamefont{and}
  \bibinfo{author}{\bibfnamefont{M.}~\bibnamefont{Hannam}},
  \bibinfo{journal}{Phys. Rev. D} \textbf{\bibinfo{volume}{101}},
  \bibinfo{pages}{124059} (\bibinfo{year}{2020}), \eprint{2002.08383}.

\bibitem[{\citenamefont{Pratten et~al.}(2021)}]{Pratten:2020ceb}
\bibinfo{author}{\bibfnamefont{G.}~\bibnamefont{Pratten}} \bibnamefont{et~al.},
  \bibinfo{journal}{Phys. Rev. D} \textbf{\bibinfo{volume}{103}},
  \bibinfo{pages}{104056} (\bibinfo{year}{2021}), \eprint{2004.06503}.

\bibitem[{\citenamefont{Estell\'es et~al.}(2021)\citenamefont{Estell\'es,
  Colleoni, Garc\'\i{}a-Quir\'os, Husa, Keitel, Mateu-Lucena, Planas, and
  Ramos-Buades}}]{Estelles:2021gvs}
\bibinfo{author}{\bibfnamefont{H.}~\bibnamefont{Estell\'es}},
  \bibinfo{author}{\bibfnamefont{M.}~\bibnamefont{Colleoni}},
  \bibinfo{author}{\bibfnamefont{C.}~\bibnamefont{Garc\'\i{}a-Quir\'os}},
  \bibinfo{author}{\bibfnamefont{S.}~\bibnamefont{Husa}},
  \bibinfo{author}{\bibfnamefont{D.}~\bibnamefont{Keitel}},
  \bibinfo{author}{\bibfnamefont{M.}~\bibnamefont{Mateu-Lucena}},
  \bibinfo{author}{\bibfnamefont{M.~d.~L.} \bibnamefont{Planas}},
  \bibnamefont{and}
  \bibinfo{author}{\bibfnamefont{A.}~\bibnamefont{Ramos-Buades}}
  (\bibinfo{year}{2021}), \eprint{2105.05872}.

\bibitem[{\citenamefont{Ossokine et~al.}(2020)}]{Ossokine:2020kjp}
\bibinfo{author}{\bibfnamefont{S.}~\bibnamefont{Ossokine}}
  \bibnamefont{et~al.}, \bibinfo{journal}{Phys. Rev. D}
  \textbf{\bibinfo{volume}{102}}, \bibinfo{pages}{044055}
  (\bibinfo{year}{2020}), \eprint{2004.09442}.

\bibitem[{\citenamefont{Matas et~al.}(2020)}]{Matas:2020wab}
\bibinfo{author}{\bibfnamefont{A.}~\bibnamefont{Matas}} \bibnamefont{et~al.},
  \bibinfo{journal}{Phys. Rev. D} \textbf{\bibinfo{volume}{102}},
  \bibinfo{pages}{043023} (\bibinfo{year}{2020}), \eprint{2004.10001}.

\bibitem[{\citenamefont{Abbott et~al.}(2017{\natexlab{d}})}]{Abbott:2016wiq}
\bibinfo{author}{\bibfnamefont{B.~P.} \bibnamefont{Abbott}}
  \bibnamefont{et~al.} (\bibinfo{collaboration}{LIGO Scientific, Virgo}),
  \bibinfo{journal}{Class. Quant. Grav.} \textbf{\bibinfo{volume}{34}},
  \bibinfo{pages}{104002} (\bibinfo{year}{2017}{\natexlab{d}}),
  \eprint{1611.07531}.

\bibitem[{\citenamefont{Pürrer and Haster}(2019)}]{Purrer:2019jcp}
\bibinfo{author}{\bibfnamefont{M.}~\bibnamefont{Pürrer}} \bibnamefont{and}
  \bibinfo{author}{\bibfnamefont{C.-J.} \bibnamefont{Haster}}
  (\bibinfo{year}{2019}), \eprint{1912.10055}.

\bibitem[{\citenamefont{Abbott et~al.}(2019{\natexlab{b}})}]{Salemi:2019ovz}
\bibinfo{author}{\bibfnamefont{B.~P.} \bibnamefont{Abbott}}
  \bibnamefont{et~al.} (\bibinfo{collaboration}{LIGO Scientific, Virgo})
  (\bibinfo{year}{2019}{\natexlab{b}}), \eprint{1906.08000}.

\bibitem[{\citenamefont{Abbott et~al.}(2016{\natexlab{d}})}]{Abbott:2016apu}
\bibinfo{author}{\bibfnamefont{B.~P.} \bibnamefont{Abbott}}
  \bibnamefont{et~al.} (\bibinfo{collaboration}{LIGO Scientific, Virgo}),
  \bibinfo{journal}{Phys. Rev.} \textbf{\bibinfo{volume}{D94}},
  \bibinfo{pages}{064035} (\bibinfo{year}{2016}{\natexlab{d}}),
  \eprint{1606.01262}.

\bibitem[{\citenamefont{Lange et~al.}(2017)}]{Lange:2017wki}
\bibinfo{author}{\bibfnamefont{J.}~\bibnamefont{Lange}} \bibnamefont{et~al.},
  \bibinfo{journal}{Phys. Rev.} \textbf{\bibinfo{volume}{D96}},
  \bibinfo{pages}{104041} (\bibinfo{year}{2017}), \eprint{1705.09833}.

\bibitem[{\citenamefont{Rezzolla
  et~al.}(2008{\natexlab{a}})\citenamefont{Rezzolla, Barausse, Dorband,
  Pollney, Reisswig, Seiler, and Husa}}]{Rezzolla:2007rz}
\bibinfo{author}{\bibfnamefont{L.}~\bibnamefont{Rezzolla}},
  \bibinfo{author}{\bibfnamefont{E.}~\bibnamefont{Barausse}},
  \bibinfo{author}{\bibfnamefont{E.~N.} \bibnamefont{Dorband}},
  \bibinfo{author}{\bibfnamefont{D.}~\bibnamefont{Pollney}},
  \bibinfo{author}{\bibfnamefont{C.}~\bibnamefont{Reisswig}},
  \bibinfo{author}{\bibfnamefont{J.}~\bibnamefont{Seiler}}, \bibnamefont{and}
  \bibinfo{author}{\bibfnamefont{S.}~\bibnamefont{Husa}},
  \bibinfo{journal}{Phys. Rev. D} \textbf{\bibinfo{volume}{78}},
  \bibinfo{pages}{044002} (\bibinfo{year}{2008}{\natexlab{a}}),
  \eprint{0712.3541}.

\bibitem[{\citenamefont{Rezzolla
  et~al.}(2008{\natexlab{b}})\citenamefont{Rezzolla, Diener, Dorband, Pollney,
  Reisswig, Schnetter, and Seiler}}]{Rezzolla:2007rd}
\bibinfo{author}{\bibfnamefont{L.}~\bibnamefont{Rezzolla}},
  \bibinfo{author}{\bibfnamefont{P.}~\bibnamefont{Diener}},
  \bibinfo{author}{\bibfnamefont{E.~N.} \bibnamefont{Dorband}},
  \bibinfo{author}{\bibfnamefont{D.}~\bibnamefont{Pollney}},
  \bibinfo{author}{\bibfnamefont{C.}~\bibnamefont{Reisswig}},
  \bibinfo{author}{\bibfnamefont{E.}~\bibnamefont{Schnetter}},
  \bibnamefont{and} \bibinfo{author}{\bibfnamefont{J.}~\bibnamefont{Seiler}},
  \bibinfo{journal}{Astrophys. J. Lett.} \textbf{\bibinfo{volume}{674}},
  \bibinfo{pages}{L29} (\bibinfo{year}{2008}{\natexlab{b}}),
  \eprint{0710.3345}.

\bibitem[{\citenamefont{Tichy and Marronetti}(2007)}]{Tichy:2007hk}
\bibinfo{author}{\bibfnamefont{W.}~\bibnamefont{Tichy}} \bibnamefont{and}
  \bibinfo{author}{\bibfnamefont{P.}~\bibnamefont{Marronetti}},
  \bibinfo{journal}{Phys. Rev. D} \textbf{\bibinfo{volume}{76}},
  \bibinfo{pages}{061502} (\bibinfo{year}{2007}), \eprint{gr-qc/0703075}.

\bibitem[{\citenamefont{Barausse and Rezzolla}(2009)}]{Barausse:2009uz}
\bibinfo{author}{\bibfnamefont{E.}~\bibnamefont{Barausse}} \bibnamefont{and}
  \bibinfo{author}{\bibfnamefont{L.}~\bibnamefont{Rezzolla}},
  \bibinfo{journal}{Astrophys. J. Lett.} \textbf{\bibinfo{volume}{704}},
  \bibinfo{pages}{L40} (\bibinfo{year}{2009}), \eprint{0904.2577}.

\bibitem[{\citenamefont{Barausse et~al.}(2012)\citenamefont{Barausse, Morozova,
  and Rezzolla}}]{Barausse:2012qz}
\bibinfo{author}{\bibfnamefont{E.}~\bibnamefont{Barausse}},
  \bibinfo{author}{\bibfnamefont{V.}~\bibnamefont{Morozova}}, \bibnamefont{and}
  \bibinfo{author}{\bibfnamefont{L.}~\bibnamefont{Rezzolla}},
  \bibinfo{journal}{Astrophys. J.} \textbf{\bibinfo{volume}{758}},
  \bibinfo{pages}{63} (\bibinfo{year}{2012}), \bibinfo{note}{[Erratum:
  Astrophys.J. 786, 76 (2014)]}, \eprint{1206.3803}.

\bibitem[{\citenamefont{Lousto et~al.}(2012)\citenamefont{Lousto, Zlochower,
  Dotti, and Volonteri}}]{Lousto:2012su}
\bibinfo{author}{\bibfnamefont{C.~O.} \bibnamefont{Lousto}},
  \bibinfo{author}{\bibfnamefont{Y.}~\bibnamefont{Zlochower}},
  \bibinfo{author}{\bibfnamefont{M.}~\bibnamefont{Dotti}}, \bibnamefont{and}
  \bibinfo{author}{\bibfnamefont{M.}~\bibnamefont{Volonteri}},
  \bibinfo{journal}{Phys. Rev. D} \textbf{\bibinfo{volume}{85}},
  \bibinfo{pages}{084015} (\bibinfo{year}{2012}), \eprint{1201.1923}.

\bibitem[{\citenamefont{Jiménez-Forteza
  et~al.}(2017)\citenamefont{Jiménez-Forteza, Keitel, Husa, Hannam, Khan, and
  Pürrer}}]{Jimenez-Forteza:2016oae}
\bibinfo{author}{\bibfnamefont{X.}~\bibnamefont{Jiménez-Forteza}},
  \bibinfo{author}{\bibfnamefont{D.}~\bibnamefont{Keitel}},
  \bibinfo{author}{\bibfnamefont{S.}~\bibnamefont{Husa}},
  \bibinfo{author}{\bibfnamefont{M.}~\bibnamefont{Hannam}},
  \bibinfo{author}{\bibfnamefont{S.}~\bibnamefont{Khan}}, \bibnamefont{and}
  \bibinfo{author}{\bibfnamefont{M.}~\bibnamefont{Pürrer}},
  \bibinfo{journal}{Phys. Rev.} \textbf{\bibinfo{volume}{D95}},
  \bibinfo{pages}{064024} (\bibinfo{year}{2017}), \eprint{1611.00332}.

\bibitem[{\citenamefont{Varma et~al.}(2019{\natexlab{b}})\citenamefont{Varma,
  Gerosa, Stein, H\'ebert, and Zhang}}]{Varma:2018aht}
\bibinfo{author}{\bibfnamefont{V.}~\bibnamefont{Varma}},
  \bibinfo{author}{\bibfnamefont{D.}~\bibnamefont{Gerosa}},
  \bibinfo{author}{\bibfnamefont{L.~C.} \bibnamefont{Stein}},
  \bibinfo{author}{\bibfnamefont{F.}~\bibnamefont{H\'ebert}}, \bibnamefont{and}
  \bibinfo{author}{\bibfnamefont{H.}~\bibnamefont{Zhang}},
  \bibinfo{journal}{Phys. Rev. Lett.} \textbf{\bibinfo{volume}{122}},
  \bibinfo{pages}{011101} (\bibinfo{year}{2019}{\natexlab{b}}),
  \eprint{1809.09125}.

\bibitem[{\citenamefont{Zappa et~al.}(2019)\citenamefont{Zappa, Bernuzzi,
  Pannarale, Mapelli, and Giacobbo}}]{Zappa:2019ntl}
\bibinfo{author}{\bibfnamefont{F.}~\bibnamefont{Zappa}},
  \bibinfo{author}{\bibfnamefont{S.}~\bibnamefont{Bernuzzi}},
  \bibinfo{author}{\bibfnamefont{F.}~\bibnamefont{Pannarale}},
  \bibinfo{author}{\bibfnamefont{M.}~\bibnamefont{Mapelli}}, \bibnamefont{and}
  \bibinfo{author}{\bibfnamefont{N.}~\bibnamefont{Giacobbo}},
  \bibinfo{journal}{Phys. Rev. Lett.} \textbf{\bibinfo{volume}{123}},
  \bibinfo{pages}{041102} (\bibinfo{year}{2019}), \eprint{1903.11622}.

\bibitem[{\citenamefont{Ghosh et~al.}(2018)\citenamefont{Ghosh,
  Johnson-Mcdaniel, Ghosh, Mishra, Ajith, Del~Pozzo, Berry, Nielsen, and
  London}}]{Ghosh:2017gfp}
\bibinfo{author}{\bibfnamefont{A.}~\bibnamefont{Ghosh}},
  \bibinfo{author}{\bibfnamefont{N.~K.} \bibnamefont{Johnson-Mcdaniel}},
  \bibinfo{author}{\bibfnamefont{A.}~\bibnamefont{Ghosh}},
  \bibinfo{author}{\bibfnamefont{C.~K.} \bibnamefont{Mishra}},
  \bibinfo{author}{\bibfnamefont{P.}~\bibnamefont{Ajith}},
  \bibinfo{author}{\bibfnamefont{W.}~\bibnamefont{Del~Pozzo}},
  \bibinfo{author}{\bibfnamefont{C.~P.~L.} \bibnamefont{Berry}},
  \bibinfo{author}{\bibfnamefont{A.~B.} \bibnamefont{Nielsen}},
  \bibnamefont{and} \bibinfo{author}{\bibfnamefont{L.}~\bibnamefont{London}},
  \bibinfo{journal}{Class. Quant. Grav.} \textbf{\bibinfo{volume}{35}},
  \bibinfo{pages}{014002} (\bibinfo{year}{2018}), \eprint{1704.06784}.

\bibitem[{\citenamefont{Abbott
  et~al.}(2016{\natexlab{e}})}]{LIGOScientific:2016lio}
\bibinfo{author}{\bibfnamefont{B.~P.} \bibnamefont{Abbott}}
  \bibnamefont{et~al.} (\bibinfo{collaboration}{LIGO Scientific, Virgo}),
  \bibinfo{journal}{Phys. Rev. Lett.} \textbf{\bibinfo{volume}{116}},
  \bibinfo{pages}{221101} (\bibinfo{year}{2016}{\natexlab{e}}),
  \bibinfo{note}{[Erratum: Phys.Rev.Lett. 121, 129902 (2018)]},
  \eprint{1602.03841}.

\bibitem[{\citenamefont{Abbott
  et~al.}(2021{\natexlab{e}})}]{LIGOScientific:2020tif}
\bibinfo{author}{\bibfnamefont{R.}~\bibnamefont{Abbott}} \bibnamefont{et~al.}
  (\bibinfo{collaboration}{LIGO Scientific, Virgo}), \bibinfo{journal}{Phys.
  Rev. D} \textbf{\bibinfo{volume}{103}}, \bibinfo{pages}{122002}
  (\bibinfo{year}{2021}{\natexlab{e}}), \eprint{2010.14529}.

\bibitem[{\citenamefont{Abbott
  et~al.}(2021{\natexlab{f}})}]{LIGOScientific:2021sio}
\bibinfo{author}{\bibfnamefont{R.}~\bibnamefont{Abbott}} \bibnamefont{et~al.}
  (\bibinfo{collaboration}{LIGO Scientific, VIRGO, KAGRA})
  (\bibinfo{year}{2021}{\natexlab{f}}), \eprint{2112.06861}.

\bibitem[{\citenamefont{Mroue et~al.}(2013)}]{Mroue:2013xna}
\bibinfo{author}{\bibfnamefont{A.~H.} \bibnamefont{Mroue}}
  \bibnamefont{et~al.}, \bibinfo{journal}{Phys. Rev. Lett.}
  \textbf{\bibinfo{volume}{111}}, \bibinfo{pages}{241104}
  (\bibinfo{year}{2013}), \eprint{1304.6077}.

\bibitem[{\citenamefont{Boyle et~al.}(2019)}]{Boyle:2019kee}
\bibinfo{author}{\bibfnamefont{M.}~\bibnamefont{Boyle}} \bibnamefont{et~al.},
  \bibinfo{journal}{Class. Quant. Grav.} \textbf{\bibinfo{volume}{36}},
  \bibinfo{pages}{195006} (\bibinfo{year}{2019}), \eprint{1904.04831}.

\bibitem[{\citenamefont{Healy et~al.}(2017)\citenamefont{Healy, Lousto,
  Zlochower, and Campanelli}}]{Healy:2017psd}
\bibinfo{author}{\bibfnamefont{J.}~\bibnamefont{Healy}},
  \bibinfo{author}{\bibfnamefont{C.~O.} \bibnamefont{Lousto}},
  \bibinfo{author}{\bibfnamefont{Y.}~\bibnamefont{Zlochower}},
  \bibnamefont{and}
  \bibinfo{author}{\bibfnamefont{M.}~\bibnamefont{Campanelli}},
  \bibinfo{journal}{Class. Quant. Grav.} \textbf{\bibinfo{volume}{34}},
  \bibinfo{pages}{224001} (\bibinfo{year}{2017}), \eprint{1703.03423}.

\bibitem[{\citenamefont{Healy et~al.}(2019)\citenamefont{Healy, Lousto, Lange,
  O'Shaughnessy, Zlochower, and Campanelli}}]{Healy:2019jyf}
\bibinfo{author}{\bibfnamefont{J.}~\bibnamefont{Healy}},
  \bibinfo{author}{\bibfnamefont{C.~O.} \bibnamefont{Lousto}},
  \bibinfo{author}{\bibfnamefont{J.}~\bibnamefont{Lange}},
  \bibinfo{author}{\bibfnamefont{R.}~\bibnamefont{O'Shaughnessy}},
  \bibinfo{author}{\bibfnamefont{Y.}~\bibnamefont{Zlochower}},
  \bibnamefont{and}
  \bibinfo{author}{\bibfnamefont{M.}~\bibnamefont{Campanelli}},
  \bibinfo{journal}{Phys. Rev.} \textbf{\bibinfo{volume}{D100}},
  \bibinfo{pages}{024021} (\bibinfo{year}{2019}), \eprint{1901.02553}.

\bibitem[{\citenamefont{Healy and Lousto}(2020)}]{Healy:2020vre}
\bibinfo{author}{\bibfnamefont{J.}~\bibnamefont{Healy}} \bibnamefont{and}
  \bibinfo{author}{\bibfnamefont{C.~O.} \bibnamefont{Lousto}},
  \bibinfo{journal}{Phys. Rev. D} \textbf{\bibinfo{volume}{102}},
  \bibinfo{pages}{104018} (\bibinfo{year}{2020}), \eprint{2007.07910}.

\bibitem[{\citenamefont{Healy and Lousto}(2022)}]{Healy:2022wdn}
\bibinfo{author}{\bibfnamefont{J.}~\bibnamefont{Healy}} \bibnamefont{and}
  \bibinfo{author}{\bibfnamefont{C.~O.} \bibnamefont{Lousto}}
  (\bibinfo{year}{2022}), \eprint{2202.00018}.

\bibitem[{\citenamefont{Jani et~al.}(2016)\citenamefont{Jani, Healy, Clark,
  London, Laguna, and Shoemaker}}]{Jani:2016wkt}
\bibinfo{author}{\bibfnamefont{K.}~\bibnamefont{Jani}},
  \bibinfo{author}{\bibfnamefont{J.}~\bibnamefont{Healy}},
  \bibinfo{author}{\bibfnamefont{J.~A.} \bibnamefont{Clark}},
  \bibinfo{author}{\bibfnamefont{L.}~\bibnamefont{London}},
  \bibinfo{author}{\bibfnamefont{P.}~\bibnamefont{Laguna}}, \bibnamefont{and}
  \bibinfo{author}{\bibfnamefont{D.}~\bibnamefont{Shoemaker}},
  \bibinfo{journal}{Class. Quant. Grav.} \textbf{\bibinfo{volume}{33}},
  \bibinfo{pages}{204001} (\bibinfo{year}{2016}), \eprint{1605.03204}.

\bibitem[{\citenamefont{Bowen and York}(1980)}]{PhysRevD.21.2047}
\bibinfo{author}{\bibfnamefont{J.~M.} \bibnamefont{Bowen}} \bibnamefont{and}
  \bibinfo{author}{\bibfnamefont{J.~W.} \bibnamefont{York}},
  \bibinfo{journal}{Phys. Rev. D} \textbf{\bibinfo{volume}{21}},
  \bibinfo{pages}{2047} (\bibinfo{year}{1980}),
  \urlprefix\url{https://link.aps.org/doi/10.1103/PhysRevD.21.2047}.

\bibitem[{\citenamefont{Brandt and Bruegmann}(1997)}]{Brandt:1997tf}
\bibinfo{author}{\bibfnamefont{S.}~\bibnamefont{Brandt}} \bibnamefont{and}
  \bibinfo{author}{\bibfnamefont{B.}~\bibnamefont{Bruegmann}},
  \bibinfo{journal}{Phys. Rev. Lett.} \textbf{\bibinfo{volume}{78}},
  \bibinfo{pages}{3606} (\bibinfo{year}{1997}), \eprint{gr-qc/9703066}.

\bibitem[{\citenamefont{Ansorg et~al.}(2004)\citenamefont{Ansorg, Br\"ugmann,
  and Tichy}}]{PhysRevD.70.064011}
\bibinfo{author}{\bibfnamefont{M.}~\bibnamefont{Ansorg}},
  \bibinfo{author}{\bibfnamefont{B.}~\bibnamefont{Br\"ugmann}},
  \bibnamefont{and} \bibinfo{author}{\bibfnamefont{W.}~\bibnamefont{Tichy}},
  \bibinfo{journal}{Phys. Rev. D} \textbf{\bibinfo{volume}{70}},
  \bibinfo{pages}{064011} (\bibinfo{year}{2004}),
  \urlprefix\url{https://link.aps.org/doi/10.1103/PhysRevD.70.064011}.

\bibitem[{\citenamefont{Baumgarte and Shapiro}(1998)}]{Baumgarte:1998te}
\bibinfo{author}{\bibfnamefont{T.~W.} \bibnamefont{Baumgarte}}
  \bibnamefont{and} \bibinfo{author}{\bibfnamefont{S.~L.}
  \bibnamefont{Shapiro}}, \bibinfo{journal}{Phys. Rev. D}
  \textbf{\bibinfo{volume}{59}}, \bibinfo{pages}{024007}
  (\bibinfo{year}{1998}), \eprint{gr-qc/9810065}.

\bibitem[{\citenamefont{Shibata and Nakamura}(1995)}]{Shibata:1995we}
\bibinfo{author}{\bibfnamefont{M.}~\bibnamefont{Shibata}} \bibnamefont{and}
  \bibinfo{author}{\bibfnamefont{T.}~\bibnamefont{Nakamura}},
  \bibinfo{journal}{Phys. Rev. D} \textbf{\bibinfo{volume}{52}},
  \bibinfo{pages}{5428} (\bibinfo{year}{1995}).

\bibitem[{\citenamefont{Newman and Penrose}(1962)}]{Newman:1961qr}
\bibinfo{author}{\bibfnamefont{E.}~\bibnamefont{Newman}} \bibnamefont{and}
  \bibinfo{author}{\bibfnamefont{R.}~\bibnamefont{Penrose}},
  \bibinfo{journal}{J. Math. Phys.} \textbf{\bibinfo{volume}{3}},
  \bibinfo{pages}{566} (\bibinfo{year}{1962}), \bibinfo{note}{erratum in J.
  Math. Phys. 4, 998 (1963)}.

\bibitem[{\citenamefont{Baker et~al.}(2002)\citenamefont{Baker, Campanelli, and
  Lousto}}]{Baker:2001sf}
\bibinfo{author}{\bibfnamefont{J.~G.} \bibnamefont{Baker}},
  \bibinfo{author}{\bibfnamefont{M.}~\bibnamefont{Campanelli}},
  \bibnamefont{and} \bibinfo{author}{\bibfnamefont{C.~O.}
  \bibnamefont{Lousto}}, \bibinfo{journal}{Phys. Rev. D}
  \textbf{\bibinfo{volume}{65}}, \bibinfo{pages}{044001}
  (\bibinfo{year}{2002}), \eprint{gr-qc/0104063}.

\bibitem[{\citenamefont{Reisswig and Pollney}(2011)}]{Reisswig:2010di}
\bibinfo{author}{\bibfnamefont{C.}~\bibnamefont{Reisswig}} \bibnamefont{and}
  \bibinfo{author}{\bibfnamefont{D.}~\bibnamefont{Pollney}},
  \bibinfo{journal}{Class. Quant. Grav.} \textbf{\bibinfo{volume}{28}},
  \bibinfo{pages}{195015} (\bibinfo{year}{2011}), \eprint{1006.1632}.

\bibitem[{\citenamefont{Purrer et~al.}(2012)\citenamefont{Purrer, Husa, and
  Hannam}}]{Purrer:2012wy}
\bibinfo{author}{\bibfnamefont{M.}~\bibnamefont{Purrer}},
  \bibinfo{author}{\bibfnamefont{S.}~\bibnamefont{Husa}}, \bibnamefont{and}
  \bibinfo{author}{\bibfnamefont{M.}~\bibnamefont{Hannam}},
  \bibinfo{journal}{Phys. Rev.} \textbf{\bibinfo{volume}{D85}},
  \bibinfo{pages}{124051} (\bibinfo{year}{2012}), \eprint{1203.4258}.

\bibitem[{\citenamefont{Hannam et~al.}(2010)\citenamefont{Hannam, Husa, Ohme,
  Muller, and Bruegmann}}]{Hannam:2010ec}
\bibinfo{author}{\bibfnamefont{M.}~\bibnamefont{Hannam}},
  \bibinfo{author}{\bibfnamefont{S.}~\bibnamefont{Husa}},
  \bibinfo{author}{\bibfnamefont{F.}~\bibnamefont{Ohme}},
  \bibinfo{author}{\bibfnamefont{D.}~\bibnamefont{Muller}}, \bibnamefont{and}
  \bibinfo{author}{\bibfnamefont{B.}~\bibnamefont{Bruegmann}},
  \bibinfo{journal}{Phys. Rev.} \textbf{\bibinfo{volume}{D82}},
  \bibinfo{pages}{124008} (\bibinfo{year}{2010}), \eprint{1007.4789}.

\bibitem[{\citenamefont{Schmidt et~al.}(2012)\citenamefont{Schmidt, Hannam, and
  Husa}}]{Schmidt:2012rh}
\bibinfo{author}{\bibfnamefont{P.}~\bibnamefont{Schmidt}},
  \bibinfo{author}{\bibfnamefont{M.}~\bibnamefont{Hannam}}, \bibnamefont{and}
  \bibinfo{author}{\bibfnamefont{S.}~\bibnamefont{Husa}},
  \bibinfo{journal}{Phys. Rev.} \textbf{\bibinfo{volume}{D86}},
  \bibinfo{pages}{104063} (\bibinfo{year}{2012}), \eprint{1207.3088}.

\bibitem[{\citenamefont{Schmidt et~al.}(2015)\citenamefont{Schmidt, Ohme, and
  Hannam}}]{Schmidt:2014iyl}
\bibinfo{author}{\bibfnamefont{P.}~\bibnamefont{Schmidt}},
  \bibinfo{author}{\bibfnamefont{F.}~\bibnamefont{Ohme}}, \bibnamefont{and}
  \bibinfo{author}{\bibfnamefont{M.}~\bibnamefont{Hannam}},
  \bibinfo{journal}{Phys. Rev.} \textbf{\bibinfo{volume}{D91}},
  \bibinfo{pages}{024043} (\bibinfo{year}{2015}), \eprint{1408.1810}.

\bibitem[{\citenamefont{Ramos-Buades et~al.}(2019)\citenamefont{Ramos-Buades,
  Husa, and Pratten}}]{Ramos-Buades:2018azo}
\bibinfo{author}{\bibfnamefont{A.}~\bibnamefont{Ramos-Buades}},
  \bibinfo{author}{\bibfnamefont{S.}~\bibnamefont{Husa}}, \bibnamefont{and}
  \bibinfo{author}{\bibfnamefont{G.}~\bibnamefont{Pratten}},
  \bibinfo{journal}{Phys. Rev. D} \textbf{\bibinfo{volume}{99}},
  \bibinfo{pages}{023003} (\bibinfo{year}{2019}), \eprint{1810.00036}.

\bibitem[{\citenamefont{Husa et~al.}(2008{\natexlab{b}})\citenamefont{Husa,
  Hannam, Gonzalez, Sperhake, and Bruegmann}}]{Husa:2007rh}
\bibinfo{author}{\bibfnamefont{S.}~\bibnamefont{Husa}},
  \bibinfo{author}{\bibfnamefont{M.}~\bibnamefont{Hannam}},
  \bibinfo{author}{\bibfnamefont{J.~A.} \bibnamefont{Gonzalez}},
  \bibinfo{author}{\bibfnamefont{U.}~\bibnamefont{Sperhake}}, \bibnamefont{and}
  \bibinfo{author}{\bibfnamefont{B.}~\bibnamefont{Bruegmann}},
  \bibinfo{journal}{Phys. Rev.} \textbf{\bibinfo{volume}{D77}},
  \bibinfo{pages}{044037} (\bibinfo{year}{2008}{\natexlab{b}}),
  \eprint{0706.0904}.

\bibitem[{\citenamefont{Hannam et~al.}(2014)\citenamefont{Hannam, Schmidt,
  Bohé, Haegel, Husa, Ohme, Pratten, and Pürrer}}]{Hannam:2013oca}
\bibinfo{author}{\bibfnamefont{M.}~\bibnamefont{Hannam}},
  \bibinfo{author}{\bibfnamefont{P.}~\bibnamefont{Schmidt}},
  \bibinfo{author}{\bibfnamefont{A.}~\bibnamefont{Bohé}},
  \bibinfo{author}{\bibfnamefont{L.}~\bibnamefont{Haegel}},
  \bibinfo{author}{\bibfnamefont{S.}~\bibnamefont{Husa}},
  \bibinfo{author}{\bibfnamefont{F.}~\bibnamefont{Ohme}},
  \bibinfo{author}{\bibfnamefont{G.}~\bibnamefont{Pratten}}, \bibnamefont{and}
  \bibinfo{author}{\bibfnamefont{M.}~\bibnamefont{Pürrer}},
  \bibinfo{journal}{Phys. Rev. Lett.} \textbf{\bibinfo{volume}{113}},
  \bibinfo{pages}{151101} (\bibinfo{year}{2014}), \eprint{1308.3271}.

\bibitem[{\citenamefont{London and Fauchon-Jones}(2019)}]{London:2018nxs}
\bibinfo{author}{\bibfnamefont{L.}~\bibnamefont{London}} \bibnamefont{and}
  \bibinfo{author}{\bibfnamefont{E.}~\bibnamefont{Fauchon-Jones}},
  \bibinfo{journal}{Class. Quant. Grav.} \textbf{\bibinfo{volume}{36}},
  \bibinfo{pages}{235015} (\bibinfo{year}{2019}), \eprint{1810.03550}.

\bibitem[{\citenamefont{Christodoulou}(1970)}]{Christodoulou:1970wf}
\bibinfo{author}{\bibfnamefont{D.}~\bibnamefont{Christodoulou}},
  \bibinfo{journal}{Phys. Rev. Lett.} \textbf{\bibinfo{volume}{25}},
  \bibinfo{pages}{1596} (\bibinfo{year}{1970}).

\bibitem[{\citenamefont{Cutler and Flanagan}(1994)}]{Cutler:1994ys}
\bibinfo{author}{\bibfnamefont{C.}~\bibnamefont{Cutler}} \bibnamefont{and}
  \bibinfo{author}{\bibfnamefont{E.~E.} \bibnamefont{Flanagan}},
  \bibinfo{journal}{Phys. Rev. D} \textbf{\bibinfo{volume}{49}},
  \bibinfo{pages}{2658} (\bibinfo{year}{1994}), \eprint{gr-qc/9402014}.

\bibitem[{\citenamefont{Poisson and Will}(1995)}]{Poisson:1995ef}
\bibinfo{author}{\bibfnamefont{E.}~\bibnamefont{Poisson}} \bibnamefont{and}
  \bibinfo{author}{\bibfnamefont{C.~M.} \bibnamefont{Will}},
  \bibinfo{journal}{Phys. Rev.} \textbf{\bibinfo{volume}{D52}},
  \bibinfo{pages}{848} (\bibinfo{year}{1995}), \eprint{gr-qc/9502040}.

\bibitem[{\citenamefont{{LIGO Scientific Collaboration}}(2018)}]{lalsuite}
\bibinfo{author}{\bibnamefont{{LIGO Scientific Collaboration}}},
  \emph{\bibinfo{title}{{LIGO} {A}lgorithm {L}ibrary - {LALS}uite}},
  \bibinfo{howpublished}{free software (GPL)} (\bibinfo{year}{2018}).

\bibitem[{\citenamefont{Ajith et~al.}(2011)}]{Ajith:2009bn}
\bibinfo{author}{\bibfnamefont{P.}~\bibnamefont{Ajith}} \bibnamefont{et~al.},
  \bibinfo{journal}{Phys. Rev. Lett.} \textbf{\bibinfo{volume}{106}},
  \bibinfo{pages}{241101} (\bibinfo{year}{2011}), \eprint{0909.2867}.

\bibitem[{\citenamefont{Baird et~al.}(2013)\citenamefont{Baird, Fairhurst,
  Hannam, and Murphy}}]{Baird:2012cu}
\bibinfo{author}{\bibfnamefont{E.}~\bibnamefont{Baird}},
  \bibinfo{author}{\bibfnamefont{S.}~\bibnamefont{Fairhurst}},
  \bibinfo{author}{\bibfnamefont{M.}~\bibnamefont{Hannam}}, \bibnamefont{and}
  \bibinfo{author}{\bibfnamefont{P.}~\bibnamefont{Murphy}},
  \bibinfo{journal}{Phys. Rev.} \textbf{\bibinfo{volume}{D87}},
  \bibinfo{pages}{024035} (\bibinfo{year}{2013}), \eprint{1211.0546}.

\bibitem[{\citenamefont{Apostolatos et~al.}(1994)\citenamefont{Apostolatos,
  Cutler, Sussman, and Thorne}}]{Apostolatos:1994mx}
\bibinfo{author}{\bibfnamefont{T.~A.} \bibnamefont{Apostolatos}},
  \bibinfo{author}{\bibfnamefont{C.}~\bibnamefont{Cutler}},
  \bibinfo{author}{\bibfnamefont{G.~J.} \bibnamefont{Sussman}},
  \bibnamefont{and} \bibinfo{author}{\bibfnamefont{K.~S.}
  \bibnamefont{Thorne}}, \bibinfo{journal}{Phys. Rev. D}
  \textbf{\bibinfo{volume}{49}}, \bibinfo{pages}{6274} (\bibinfo{year}{1994}).

\bibitem[{\citenamefont{Kidder}(1995)}]{Kidder:1995zr}
\bibinfo{author}{\bibfnamefont{L.~E.} \bibnamefont{Kidder}},
  \bibinfo{journal}{Phys. Rev. D} \textbf{\bibinfo{volume}{52}},
  \bibinfo{pages}{821} (\bibinfo{year}{1995}), \eprint{gr-qc/9506022}.

\bibitem[{\citenamefont{Berti et~al.}(2006)\citenamefont{Berti, Cardoso, and
  Will}}]{PhysRevD.73.064030}
\bibinfo{author}{\bibfnamefont{E.}~\bibnamefont{Berti}},
  \bibinfo{author}{\bibfnamefont{V.}~\bibnamefont{Cardoso}}, \bibnamefont{and}
  \bibinfo{author}{\bibfnamefont{C.~M.} \bibnamefont{Will}},
  \bibinfo{journal}{Phys. Rev. D} \textbf{\bibinfo{volume}{73}},
  \bibinfo{pages}{064030} (\bibinfo{year}{2006}),
  \urlprefix\url{https://link.aps.org/doi/10.1103/PhysRevD.73.064030}.

\bibitem[{\citenamefont{Tichy and Bruegmann}(2004)}]{Tichy:2003qi}
\bibinfo{author}{\bibfnamefont{W.}~\bibnamefont{Tichy}} \bibnamefont{and}
  \bibinfo{author}{\bibfnamefont{B.}~\bibnamefont{Bruegmann}},
  \bibinfo{journal}{Phys. Rev. D} \textbf{\bibinfo{volume}{69}},
  \bibinfo{pages}{024006} (\bibinfo{year}{2004}), \eprint{gr-qc/0307027}.

\bibitem[{\citenamefont{Alcubierre}(2008)}]{Alcubierre:1138167}
\bibinfo{author}{\bibfnamefont{M.}~\bibnamefont{Alcubierre}},
  \emph{\bibinfo{title}{{Introduction to 3+1 numerical relativity}}},
  International series of monographs on physics (\bibinfo{publisher}{Oxford
  Univ. Press}, \bibinfo{address}{Oxford}, \bibinfo{year}{2008}),
  \urlprefix\url{https://cds.cern.ch/record/1138167}.

\bibitem[{\citenamefont{Leaver}(1985)}]{leaver85}
\bibinfo{author}{\bibfnamefont{E.}~\bibnamefont{Leaver}},
  \bibinfo{journal}{Proc. Roy. Soc. Lond. A} \textbf{\bibinfo{volume}{A402}},
  \bibinfo{pages}{285} (\bibinfo{year}{1985}).

\bibitem[{\citenamefont{London}(2015)}]{London_nrutils_2015}
\bibinfo{author}{\bibfnamefont{L.}~\bibnamefont{London}},
  \emph{\bibinfo{title}{{nrutils}}} (\bibinfo{year}{2015}).

\bibitem[{\citenamefont{London et~al.}(2014)\citenamefont{London, Shoemaker,
  and Healy}}]{London:2014cma}
\bibinfo{author}{\bibfnamefont{L.}~\bibnamefont{London}},
  \bibinfo{author}{\bibfnamefont{D.}~\bibnamefont{Shoemaker}},
  \bibnamefont{and} \bibinfo{author}{\bibfnamefont{J.}~\bibnamefont{Healy}},
  \bibinfo{journal}{Phys. Rev.} \textbf{\bibinfo{volume}{D90}},
  \bibinfo{pages}{124032} (\bibinfo{year}{2014}), \eprint{1404.3197}.

\bibitem[{\citenamefont{Schmidt et~al.}(2017)\citenamefont{Schmidt, Harry, and
  Pfeiffer}}]{Schmidt:2017btt}
\bibinfo{author}{\bibfnamefont{P.}~\bibnamefont{Schmidt}},
  \bibinfo{author}{\bibfnamefont{I.~W.} \bibnamefont{Harry}}, \bibnamefont{and}
  \bibinfo{author}{\bibfnamefont{H.~P.} \bibnamefont{Pfeiffer}}
  (\bibinfo{year}{2017}), \eprint{1703.01076}.

\bibitem[{\citenamefont{Calder\'on~Bustillo
  et~al.}(2015)\citenamefont{Calder\'on~Bustillo, Boh\'e, Husa, Sintes, Hannam,
  and P\"urrer}}]{CalderonBustillo:2015lrg}
\bibinfo{author}{\bibfnamefont{J.}~\bibnamefont{Calder\'on~Bustillo}},
  \bibinfo{author}{\bibfnamefont{A.}~\bibnamefont{Boh\'e}},
  \bibinfo{author}{\bibfnamefont{S.}~\bibnamefont{Husa}},
  \bibinfo{author}{\bibfnamefont{A.~M.} \bibnamefont{Sintes}},
  \bibinfo{author}{\bibfnamefont{M.}~\bibnamefont{Hannam}}, \bibnamefont{and}
  \bibinfo{author}{\bibfnamefont{M.}~\bibnamefont{P\"urrer}}
  (\bibinfo{year}{2015}), \eprint{1501.00918}.

\bibitem[{\citenamefont{O'Shaughnessy et~al.}(2011)\citenamefont{O'Shaughnessy,
  Vaishnav, Healy, Meeks, and Shoemaker}}]{OShaughnessy:2011pmr}
\bibinfo{author}{\bibfnamefont{R.}~\bibnamefont{O'Shaughnessy}},
  \bibinfo{author}{\bibfnamefont{B.}~\bibnamefont{Vaishnav}},
  \bibinfo{author}{\bibfnamefont{J.}~\bibnamefont{Healy}},
  \bibinfo{author}{\bibfnamefont{Z.}~\bibnamefont{Meeks}}, \bibnamefont{and}
  \bibinfo{author}{\bibfnamefont{D.}~\bibnamefont{Shoemaker}},
  \bibinfo{journal}{Phys. Rev. D} \textbf{\bibinfo{volume}{84}},
  \bibinfo{pages}{124002} (\bibinfo{year}{2011}), \eprint{1109.5224}.

\bibitem[{\citenamefont{Boyle et~al.}(2011)\citenamefont{Boyle, Owen, and
  Pfeiffer}}]{Boyle:2011gg}
\bibinfo{author}{\bibfnamefont{M.}~\bibnamefont{Boyle}},
  \bibinfo{author}{\bibfnamefont{R.}~\bibnamefont{Owen}}, \bibnamefont{and}
  \bibinfo{author}{\bibfnamefont{H.~P.} \bibnamefont{Pfeiffer}},
  \bibinfo{journal}{Phys. Rev. D} \textbf{\bibinfo{volume}{84}},
  \bibinfo{pages}{124011} (\bibinfo{year}{2011}), \eprint{1110.2965}.

\bibitem[{\citenamefont{Ohme et~al.}(2011)\citenamefont{Ohme, Hannam, and
  Husa}}]{Ohme:2011zm}
\bibinfo{author}{\bibfnamefont{F.}~\bibnamefont{Ohme}},
  \bibinfo{author}{\bibfnamefont{M.}~\bibnamefont{Hannam}}, \bibnamefont{and}
  \bibinfo{author}{\bibfnamefont{S.}~\bibnamefont{Husa}},
  \bibinfo{journal}{Phys. Rev. D} \textbf{\bibinfo{volume}{84}},
  \bibinfo{pages}{064029} (\bibinfo{year}{2011}), \eprint{1107.0996}.

\bibitem[{\citenamefont{Lindblom et~al.}(2008)\citenamefont{Lindblom, Owen, and
  Brown}}]{Lindblom:2008cm}
\bibinfo{author}{\bibfnamefont{L.}~\bibnamefont{Lindblom}},
  \bibinfo{author}{\bibfnamefont{B.~J.} \bibnamefont{Owen}}, \bibnamefont{and}
  \bibinfo{author}{\bibfnamefont{D.~A.} \bibnamefont{Brown}},
  \bibinfo{journal}{Phys. Rev. D} \textbf{\bibinfo{volume}{78}},
  \bibinfo{pages}{124020} (\bibinfo{year}{2008}), \eprint{0809.3844}.

\bibitem[{\citenamefont{Ferguson et~al.}(2021)\citenamefont{Ferguson, Jani,
  Laguna, and Shoemaker}}]{Ferguson:2020xnm}
\bibinfo{author}{\bibfnamefont{D.}~\bibnamefont{Ferguson}},
  \bibinfo{author}{\bibfnamefont{K.}~\bibnamefont{Jani}},
  \bibinfo{author}{\bibfnamefont{P.}~\bibnamefont{Laguna}}, \bibnamefont{and}
  \bibinfo{author}{\bibfnamefont{D.}~\bibnamefont{Shoemaker}},
  \bibinfo{journal}{Phys. Rev. D} \textbf{\bibinfo{volume}{104}},
  \bibinfo{pages}{044037} (\bibinfo{year}{2021}), \eprint{2006.04272}.

\bibitem[{\citenamefont{Abbott
  et~al.}(2021{\natexlab{g}})}]{LIGOScientific:2021psn}
\bibinfo{author}{\bibfnamefont{R.}~\bibnamefont{Abbott}} \bibnamefont{et~al.}
  (\bibinfo{collaboration}{LIGO Scientific, VIRGO, KAGRA})
  (\bibinfo{year}{2021}{\natexlab{g}}), \eprint{2111.03634}.

\bibitem[{\citenamefont{P\"urrer et~al.}(2016)\citenamefont{P\"urrer, Hannam,
  and Ohme}}]{Purrer:2015nkh}
\bibinfo{author}{\bibfnamefont{M.}~\bibnamefont{P\"urrer}},
  \bibinfo{author}{\bibfnamefont{M.}~\bibnamefont{Hannam}}, \bibnamefont{and}
  \bibinfo{author}{\bibfnamefont{F.}~\bibnamefont{Ohme}},
  \bibinfo{journal}{Phys. Rev. D} \textbf{\bibinfo{volume}{93}},
  \bibinfo{pages}{084042} (\bibinfo{year}{2016}), \eprint{1512.04955}.

\bibitem[{\citenamefont{Ramos-Buades et~al.}(2020)\citenamefont{Ramos-Buades,
  Husa, Pratten, Estell\'es, Garc\'\i{}a-Quir\'os, Mateu-Lucena, Colleoni, and
  Jaume}}]{Ramos-Buades:2019uvh}
\bibinfo{author}{\bibfnamefont{A.}~\bibnamefont{Ramos-Buades}},
  \bibinfo{author}{\bibfnamefont{S.}~\bibnamefont{Husa}},
  \bibinfo{author}{\bibfnamefont{G.}~\bibnamefont{Pratten}},
  \bibinfo{author}{\bibfnamefont{H.}~\bibnamefont{Estell\'es}},
  \bibinfo{author}{\bibfnamefont{C.}~\bibnamefont{Garc\'\i{}a-Quir\'os}},
  \bibinfo{author}{\bibfnamefont{M.}~\bibnamefont{Mateu-Lucena}},
  \bibinfo{author}{\bibfnamefont{M.}~\bibnamefont{Colleoni}}, \bibnamefont{and}
  \bibinfo{author}{\bibfnamefont{R.}~\bibnamefont{Jaume}},
  \bibinfo{journal}{Phys. Rev. D} \textbf{\bibinfo{volume}{101}},
  \bibinfo{pages}{083015} (\bibinfo{year}{2020}), \eprint{1909.11011}.

\bibitem[{\citenamefont{Huerta et~al.}(2018)}]{Huerta:2017kez}
\bibinfo{author}{\bibfnamefont{E.~A.} \bibnamefont{Huerta}}
  \bibnamefont{et~al.}, \bibinfo{journal}{Phys. Rev. D}
  \textbf{\bibinfo{volume}{97}}, \bibinfo{pages}{024031}
  (\bibinfo{year}{2018}), \eprint{1711.06276}.

\bibitem[{\citenamefont{Hinder et~al.}(2018)\citenamefont{Hinder, Kidder, and
  Pfeiffer}}]{Hinder:2017sxy}
\bibinfo{author}{\bibfnamefont{I.}~\bibnamefont{Hinder}},
  \bibinfo{author}{\bibfnamefont{L.~E.} \bibnamefont{Kidder}},
  \bibnamefont{and} \bibinfo{author}{\bibfnamefont{H.~P.}
  \bibnamefont{Pfeiffer}}, \bibinfo{journal}{Phys. Rev. D}
  \textbf{\bibinfo{volume}{98}}, \bibinfo{pages}{044015}
  (\bibinfo{year}{2018}), \eprint{1709.02007}.

\bibitem[{\citenamefont{Nagar et~al.}(2021)\citenamefont{Nagar, Bonino, and
  Rettegno}}]{Nagar:2021gss}
\bibinfo{author}{\bibfnamefont{A.}~\bibnamefont{Nagar}},
  \bibinfo{author}{\bibfnamefont{A.}~\bibnamefont{Bonino}}, \bibnamefont{and}
  \bibinfo{author}{\bibfnamefont{P.}~\bibnamefont{Rettegno}},
  \bibinfo{journal}{Phys. Rev. D} \textbf{\bibinfo{volume}{103}},
  \bibinfo{pages}{104021} (\bibinfo{year}{2021}), \eprint{2101.08624}.

\bibitem[{\citenamefont{Islam et~al.}(2021)\citenamefont{Islam, Varma, Lodman,
  Field, Khanna, Scheel, Pfeiffer, Gerosa, and Kidder}}]{Islam:2021mha}
\bibinfo{author}{\bibfnamefont{T.}~\bibnamefont{Islam}},
  \bibinfo{author}{\bibfnamefont{V.}~\bibnamefont{Varma}},
  \bibinfo{author}{\bibfnamefont{J.}~\bibnamefont{Lodman}},
  \bibinfo{author}{\bibfnamefont{S.~E.} \bibnamefont{Field}},
  \bibinfo{author}{\bibfnamefont{G.}~\bibnamefont{Khanna}},
  \bibinfo{author}{\bibfnamefont{M.~A.} \bibnamefont{Scheel}},
  \bibinfo{author}{\bibfnamefont{H.~P.} \bibnamefont{Pfeiffer}},
  \bibinfo{author}{\bibfnamefont{D.}~\bibnamefont{Gerosa}}, \bibnamefont{and}
  \bibinfo{author}{\bibfnamefont{L.~E.} \bibnamefont{Kidder}},
  \bibinfo{journal}{Phys. Rev. D} \textbf{\bibinfo{volume}{103}},
  \bibinfo{pages}{064022} (\bibinfo{year}{2021}), \eprint{2101.11798}.

\bibitem[{\citenamefont{Liu et~al.}(2022)\citenamefont{Liu, Cao, and
  Zhu}}]{Liu:2021pkr}
\bibinfo{author}{\bibfnamefont{X.}~\bibnamefont{Liu}},
  \bibinfo{author}{\bibfnamefont{Z.}~\bibnamefont{Cao}}, \bibnamefont{and}
  \bibinfo{author}{\bibfnamefont{Z.-H.} \bibnamefont{Zhu}},
  \bibinfo{journal}{Class. Quant. Grav.} \textbf{\bibinfo{volume}{39}},
  \bibinfo{pages}{035009} (\bibinfo{year}{2022}), \eprint{2102.08614}.

\bibitem[{\citenamefont{Woods and Williams}()}]{citc}
\bibinfo{author}{\bibfnamefont{C.}~\bibnamefont{Woods}} \bibnamefont{and}
  \bibinfo{author}{\bibfnamefont{M.}~\bibnamefont{Williams}},
  \emph{\bibinfo{title}{Cluster in the cloud}},
  \urlprefix\url{https://cluster-in-the-cloud.readthedocs.io/en/latest/}.

\bibitem[{\citenamefont{Richardson and
  Glazebrook}(1911)}]{doi:10.1098/rsta.1911.0009}
\bibinfo{author}{\bibfnamefont{L.~F.} \bibnamefont{Richardson}}
  \bibnamefont{and} \bibinfo{author}{\bibfnamefont{R.~T.}
  \bibnamefont{Glazebrook}}, \bibinfo{journal}{Philosophical Transactions of
  the Royal Society of London. Series A, Containing Papers of a Mathematical or
  Physical Character} \textbf{\bibinfo{volume}{210}}, \bibinfo{pages}{307}
  (\bibinfo{year}{1911}).

\end{thebibliography}
\end{document}